\RequirePackage{lineno}
\documentclass[12pt]{iopart}
\usepackage{iopams}

\usepackage{xcolor}
\usepackage{cite}
\usepackage{graphicx}
\usepackage{amssymb,amsfonts}
\usepackage{hyperref}
\usepackage{multirow}
\usepackage[font=small,labelfont=bf]{caption}
\usepackage{subcaption}
\usepackage{makecell}
\usepackage{booktabs}
\usepackage[normalem]{ulem}

\begin{document}
\title[Extracting Cross Sections from Swarm Data using DNN]{Extracting Electron Scattering Cross Sections from Swarm Data using Deep Neural Networks}

\author{Vishrut Jetly$^1$ and  Bhaskar Chaudhury$^1$}

\address{
$^1$ Group in Computational Science and HPC, DA-IICT, Gandhinagar, India}

\vspace{5pt}
\ead{bhaskar\_chaudhury@daiict.ac.in}
\vspace{10pt}

\begin{abstract}
Electron-neutral scattering cross sections are fundamental quantities in simulations of low temperature plasmas used for many technological applications today.  From these microscopic cross sections, several macro-scale quantities (called ``swarm" parameters) can be calculated. However, measurements as well as theoretical calculations of cross sections are challenging.  Since the 1960s researchers have attempted to solve the inverse swarm problem of obtaining cross sections from swarm data; but the solutions are not necessarily unique.  To address this issues, we examine the use of deep learning models which are trained using the previous determinations of elastic momentum transfer, ionization and excitation cross sections for different gases available on the LXCat website and their corresponding swarm parameters calculated using the BOLSIG+ solver for the numerical solution of the Boltzmann equation for electrons in weakly ionized gases. We implement artificial neural network (ANN), convolutional neural network (CNN) and densely connected convolutional network (DenseNet) for this investigation. To the best of our knowledge, there is no study exploring the use of CNN and DenseNet for the inverse swarm problem. We test the validity of predictions by all these trained networks for a broad range of gas species and we deduce that DenseNet effectively extracts both long and short term features from the swarm data and hence, it predicts cross sections with significantly higher accuracy compared to ANN. Further, we apply Monte Carlo dropout as Bayesian approximation to estimate the probability distribution of the cross sections to determine all plausible solutions of this inverse problem.

\end{abstract}

\vspace{1pc}
\begin{tabular}{r l}
    \textit{Keywords:} & \multicolumn{1}{p{12.2cm}}{Scattering cross sections, transport coefficients, inverse~swarm~problem, deep~convolutional~neural~networks, uncertainty~quantification}
\end{tabular}
%
%
\maketitle
%
%


\section{Introduction}
Plasma science has an admirable track record as an enabling technology that underpin our modern society and has the potential to make wide-ranging contributions to address many societal challenges~\cite{roadmap2012,roadmap2017}. Technologies based on low-temperature plasmas (LTPs) are ubiquitous in today's society. These include mature technologies such as fluorescent lamps and gas lasers, for example, as well as other more ``modern" technologies in use but still being developed, such as plasma reactors for processing of semiconductors, for fabrication of microelectronics to name a few~\cite{encyclopedia}. Today, there are extensive research activities and rapidly emerging applications of LTPs in medicine and in agriculture~\cite{plasmamedicine,plasmaagriculture}. LTPs are generated most simply by applying a sufficiently high voltage across a gas gap separated by two electrodes~\cite{ltpreview}. The properties of the plasmas so generated vary considerably with the experimental parameters -  gas pressure and composition, geometrical configuration, means of applying an electromagnetic field (e.g., application of a DC, AC and/or rf, pulsed or steady-state voltage across the electrodes, injection of microwave) and the specificities of the external circuit. For purposes of discussion here, we will consider only an applied electric field. 
LTPs consist of electrons and ions flowing through a neutral background gas in response to the applied electric field, and, for many applications, the number density of the neutral molecules exceeds that of the charged particles by many orders of magnitude.  Our knowledge of the electron and ion interactions with atomic and molecular species within the plasma, and evaluation of cross-sections and reaction rates for such collisions has played an important role in the exploitation of plasmas in several applications~\cite{encyclopedia}.

Being much lighter than ions, electrons are more easily accelerated in the electric field that sustains the plasma, and hence the electrons are the vector through which electrical energy is transferred to the neutrals through collisions. For a wide range of conditions in LTPs, electrons collide predominately with background neutral gas molecules in their ground state. In these conditions, the electron energy distribution function is generally non-Maxwellian and the electron ``temperature" is much higher than the temperature of the ions or that of the neutrals. Because of the huge range of realizable conditions, optimization of a LTP plasma for a particular application necessarily requires a combination of experiment and modeling.   
The data required for modeling LTPs depend on the level of description but are in all cases are extensive~\cite{plasmasimulation1,alvesltpmodeling}. Fluid models - where the electrons and the ions are treated as separate fluids because of their widely disparate temperatures and these are coupled to Maxwell's equations for the EM. In their simplest form, fluid models require electron and ion mobilities, diffusion coefficients, and electron ionization/attachment rate coefficients.  The product of the mobility and neutral density, $\mu N$; the product of the diffusion coefficient and neutral density, $DN$; and rate coefficients are dependent on $E/N$, the ratio of the electric field strength to the neutral density in the limit of a constant (in time and space)  electric field.  These transport and rate coefficients as functions of $E/N$ are commonly called ``swarm" parameters in analogy with a drifting spreading swarm of bees where the average kinetic energy is much higher than the directed – or drift – energy.  On the other hand, the more detailed kinetic models (such as Particle-In-Cell simulations with Monte Carlo Collisions) require electron-neutral and ion-neutral cross sections vs. energy for each possible outcome of the collision, whether it be elastic, excitation, or ionization.   Of course, there are many different possible excitation channels and a cross section for each as a function of energy is required in general. The future developments in the LTP areas will be based upon our ability in the manipulation of the plasma properties which requires a thorough understanding of plasma chemistry, and availability of accurate cross section data and swarm parameters~\cite{leanne-gec,pnascollisions}. Swarm parameters can be measured fairly easily and with very high accuracy (0.5$\%$ for the drift velocity, for example), and since the works of Ramsauer, Mayer, Townsend and Bailey in the early 1920's researchers have aimed to extract information about microscopic cross sections from measurements of macroscopic swarm parameters.  Cross sections, on the other hand, are much more difficult to measure and highly accurate quantum mechanical calculations for simple atomic target species are just now becoming available. However, despite their significance, many processes are not well understood because of the lack of availability of the required cross sections and its absence is a major impediment for experimentalists as well as computational investigators.

\begin{figure}
    \centering
    \includegraphics[width=\linewidth]{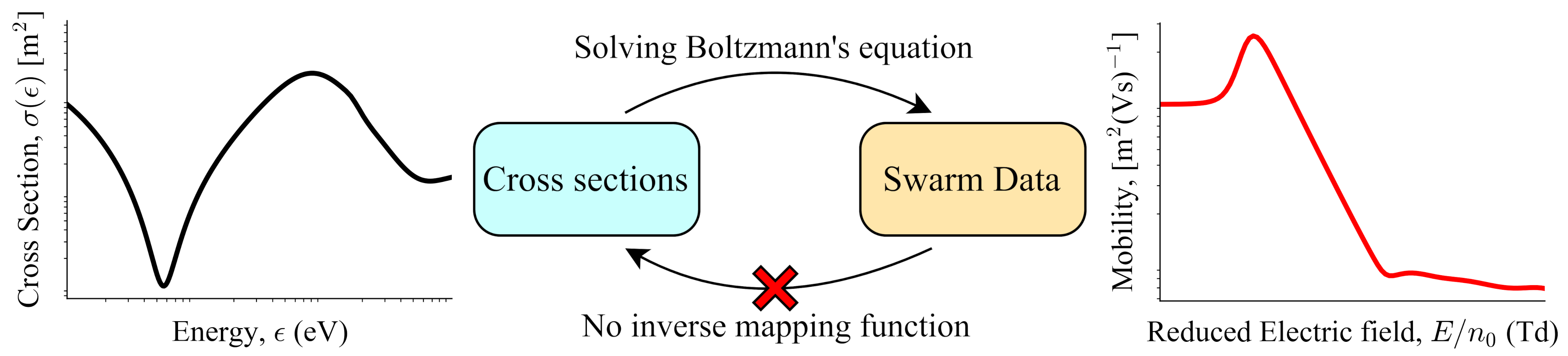}
    \caption{Forward problem of mapping from a set of cross sections to a set of corresponding swarm data is well-posed and can be solved numerically by solving Boltzmann’s equation. On the other hand, the inverse problem is ill-posed and this inverse mapping function does not exist.}
    \label{inverseproblem}
\end{figure}

The complete sets of cross sections include a momentum transfer cross section for elastic scattering, and cross sections for excitation and ionization processes for a given target species. A partial set includes a subset of the important scattering processes for that species. Complete sets of cross sections are needed as input to a Boltzmann equation solver to determine the electron or ion energy distribution function, subsequently from which swarm parameters can be computed (Forward mapping in Fig.~\ref{inverseproblem}). Therefore, complete sets of cross section data play an important role in designing new experiments as well as simulations. Obtaining cross sections from swarm parameters is a challenging task and was pioneered by Townsend and Ramsauer in the $1920$s. This inverse problem is ill-posed and the inverse mapping function does not exist, as depicted in Fig.~\ref{inverseproblem}. Thus, the method used in those early analyses involved inverting the integral relating the drift velocity and momentum transfer cross sections using a simplified expression of the electron energy distribution. This approach was significantly improved in the $1960$s by Phelps along with other collaborators, employing iterative methods to solve the Boltzmann's equation to obtain an accurate energy distribution of the electrons~\cite{frost1962rotational, engelhardt1963elastic, engelhardt1964determination, hake1967momentum}. This allowed accurate computation of the momentum transfer and lower energy inelastic cross sections from the available swarm data. The iterative process of inverting the swarm data to obtain cross sections involves solving the Boltzmann's equation, calculating the electron energy distribution and altering the model cross sections till a satisfactory match is found between the original and computed swarm parameters, making it a computationally expensive problem to solve. To address this issue,~\cite{suzuki1990momentum, morgan1991use} used numerical optimization algorithms to help speed up the process of obtaining cross sections from swarm data. But, the inverse swarm problem is ill-posed, especially when there is a lack of available swarm data. Therefore, these optimization algorithms would often get stuck in a local minima due to the non-uniqueness of the inverse swarm mapping.

Neural networks have been successfully used to learn the non-linear mappings between two sets of data, and once the network has been trained, it can give the outputs in roughly $\mathcal{O}(1)$. Also, it is relatively easier to avoid local minima during optimization using neural networks. Hoping to utilize these advantages, W. L. Morgan investigated the feasibility to use neural networks to solve the inverse swarm problem~\cite{morgan1991feasibility} and concluded that neural network were indeed useful to determine the cross sections from electron swarm data but couldn't achieve high accuracy levels due to the lack of quality cross section data available, along with various limitations of the commercially available neural net simulator of those times. Artificial neural networks have been also used to successfully predict the proton impact single ionization double differential cross sections of atoms and molecules~\cite{harris2013applications}.

Since Morgan's findings in $1991$, there has been an increase in the amount of available cross sections and swarm data (LXCat~\cite{lxcat}). Recently, study carried out by Stokes \textit{et al.}~\cite{stokes2019determining} verified Morgan's claims and their work~\cite{stokes2020self} demonstrated their automatic solution using artificial neural network had an accuracy comparable to that of a human expert in determining cross sections of the biomolecule tetrahydrofuran (THF) using experimentally measured swarm data. In~\cite{stokes2019determining}, they also showed that use of large amount of synthetic training data generated using the real cross sections available from LXCat indeed gave good results when used to predict elastic momentum transfer and ionization cross sections of Helium and Argon. However, the same needs to be verified for a number of different gas species to safely conclude the feasibility of this machine learning approach to solve the inverse swarm problem. Moreover, their study was limited only to the use of artificial neural network, which had minor improvements over the architecture proposed by Morgan to increase the parameter efficiency and training speed of the model. Additionally in the last decade, there has been drastic increase in computing power along with vast improvements of machine learning algorithms, allowing creation of large and powerful neural networks. There are numerous applications in computer vision and image processing where other neural network types, such as CNN, outperform ANN's predictions because of its ability to extract spatial information. In this problem too, the swarm data which is to be used as an input to the neural network is in the form of continuous series and thus, it becomes imperative to study performance of the CNN architectures in solving the inverse swarm problem. Additionally, since this inverse problem in itself is ill-posed in nature, it is more reasonable to find the entire distribution from which the plausible solutions can be sampled.

Thus, in this study, we explore the suitability of deep neural networks to identify the inverse relationship for a wide range of gas species and assess the efficacy of different neural network architectures in predicting scattering cross sections from simulated swarm data. Furthermore, we perform uncertainty quantification to estimate the distribution of all the plausible solutions of the inverse problem.  
To this end, study exploring the use of CNN and DenseNet for this inverse swarm problem has not yet been reported. In section 2, we describe our complete data-driven methodology starting from data preparation to the implementation of two new neural networks (CNN and DenseNet based) for the solution of this inverse swarm problem. In section 3, we present a detailed comparison of the performance of three neural networks (ANN, CNN and DenseNet) in determining the cross sections of seven gas species. Reliability of the predictions has been also evaluated by using an uncertainty quantification method in subsection 3.1. Finally we conclude the paper with a summary of our results and brief discussion on how accuracy of this data driven approach can be further improved.

\section{Methodology}

\begin{figure}
    \centering
    \includegraphics[width=\linewidth]{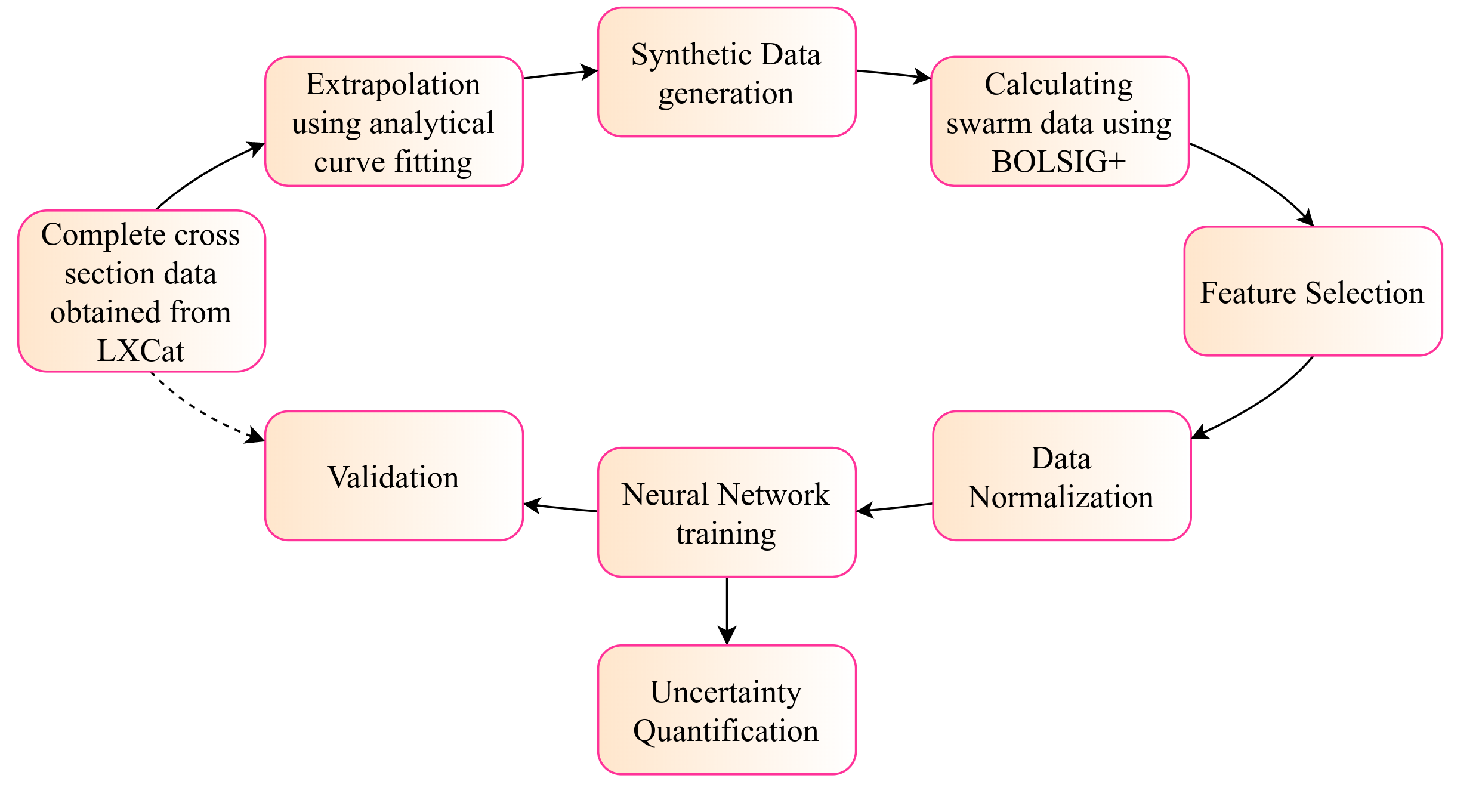}
    \caption{Complete workflow used in this study for solving the inverse swarm problem.}
    \label{workflow}
\end{figure}

Our data driven methodology for determining a set of cross sections consistent with swarm parameters involves several steps such as data collection and profiling, feature engineering, building the suitable Machine Learning (ML) models followed by training and evaluation. Figure~\ref{workflow} describes the complete workflow used in this study for solving the inverse swarm problem. Firstly, complete sets of cross sections for different gas species are obtained from the LXCat~\cite{lxcat} database, however since this data is limited, we generate abundant synthetic cross sections data. Secondly, using the cross section data, we compute the corresponding swarm coefficients using the freeware Boltzmann equation solver BOLSIG+~\cite{bolsig+}. Thirdly, we perform a feature selection followed by data normalization. Finally different neural network models are designed and are trained using the combination of cross section and swarm data. The predicted results are compared to the cross sections obtained from LXCat. We then also estimate the complete distribution of the plausible cross sections by quantifying the uncertainty in the solution using Monte Carlo Dropout~\cite{gal2016dropout}. In the following subsections we provide a detailed description of each of the above mentioned steps.

\subsection{Dataset}
\begin{figure}
    \centering
    \includegraphics[width=\linewidth]{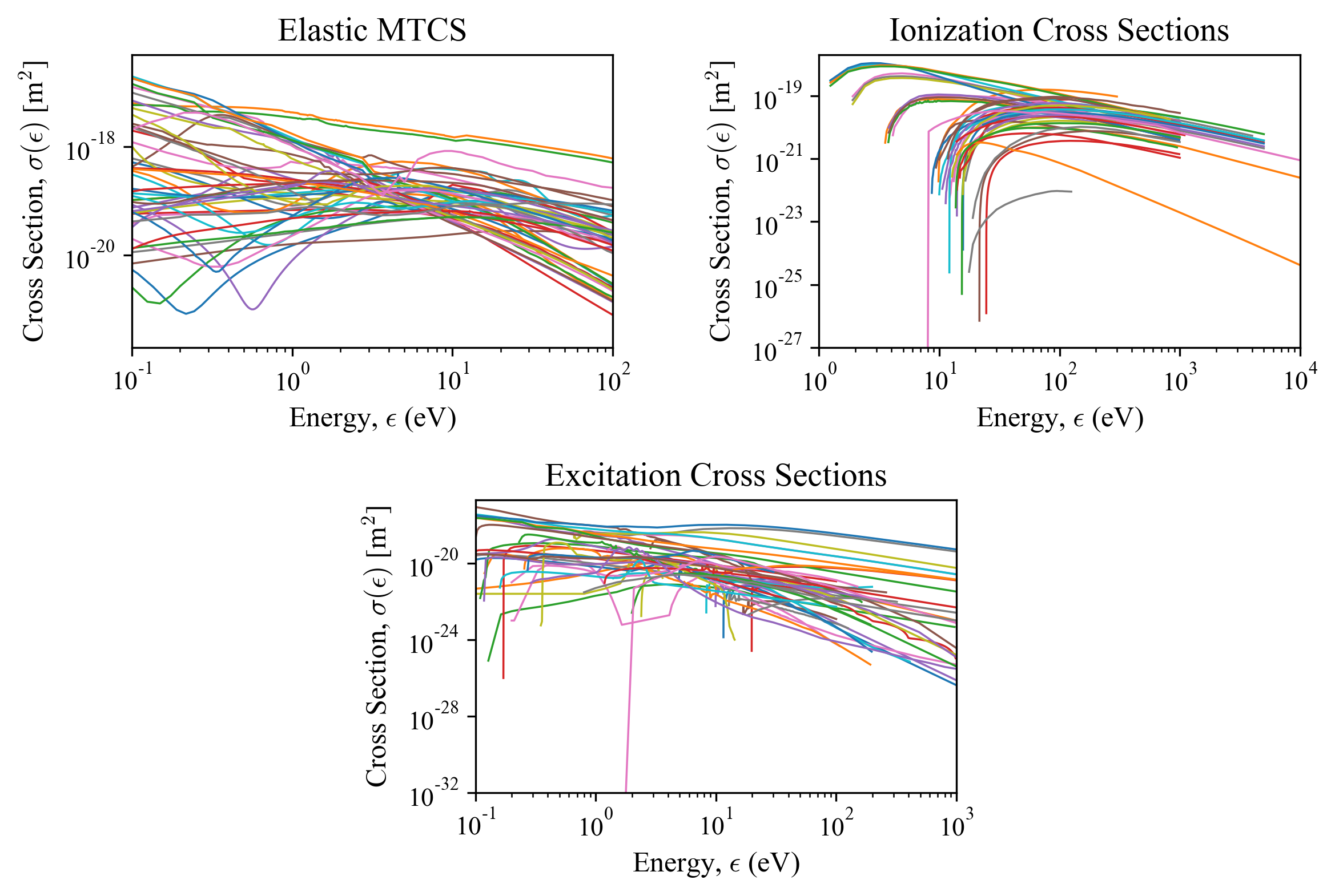}
    \caption{Complete cross section data of various gas species obtained from LXCat}
    \label{cs}
\end{figure}

Efficient training of the neural network to identify an inverse non-linear relationship between swarm data and cross sections requires abundant training data. Morgan generated these training cross sections using a power-law model of the form $\sigma(\epsilon) = \epsilon_0 / \epsilon^{p}$, where $\epsilon$ and $p$ are randomly chosen from $(10^{-17},\; 10^{-14})$ and $(0,\; 1)$ respectively~\cite{morgan1991feasibility}. This parameterized method allows to generate an infinite number of training examples and thus can be considered ideal for machine learning problems. However, this parameterized equation represents only a small subset of physically plausible cross sections. To expose our deep learning models to more realistic data, we use cross sections data for gas species compiled on the LXCat website, shown in Fig.~\ref{cs}. The cross sections include the energy-dependent momentum transfer cross section for elastic scattering, and total (angle integrated) cross sections for excitation and ionization processes for a given target species. In general, the probability of a collision of a particular type occurring depends on the relative velocity of the collision partners and the scattering angle. However, it has been shown that the additional detail regarding angular scattering has very little effect on the calculated swarm parameters. Note that there are many different excitation processes with different energy thresholds,  and predicting all (or even the most important) of them is a challenging task.  In this work, and for the sake of demonstrating the features of different ML algorithms, we consider only one excitation cross section; that is, the training data consider only the lowest excitation process from the compilations available on LXCat. The Boltzmann equation is solved using only these three input cross sections, and the swarm parameters so calculated are not be compared with those tabulated from experiments on the LXCat website, which are generally well predicted when a complete set of cross sections is used.   This procedure considerably simplifies the computational requirements and is expected to correctly demonstrate the capabilities of each of the ML algorithms studied here.   


\subsubsection{Extrapolation of inelastic cross sections}
In this work, we aim to predict elastic momentum transfer, ionization and excitation cross section for the energies in range [$10^{-1}$ eV, $10^2$ eV], [$10^{0}$ eV, $10^4$ eV], [$10^{-1}$ eV, $10^3$ eV] respectively, and as evident from Fig.~\ref{cs}, inelastic cross sections of many gas species in LXCat databases are not available for the entire energy domain under consideration. Thus, we use an analytical expression to extrapolate these cross sections to higher energies. For the ionization cross sections, we use the parameterization (Eq.~\ref{ionization_formula}) proposed  by Rost and Pattard~\cite{rost1997analytical}
\begin{equation}
    \sigma(E) =  \frac{kE^\alpha}{(E+E_M /\alpha)^{\alpha+1}}
    \label{ionization_formula}
\end{equation}
where $E$ is the excess energy of the system measured from the ionization threshold, $E_M$ corresponds to the energy where the cross section attains a maximum value, $k$ and $\alpha$ are the parameters which are computed to obtain the best fit. 

Various approximation from quantum mechanics could be used to extrapolate excitation cross sections to higher energies. However, we have chosen to simply use a power-law relationship, Eq.~\ref{excitation_formula}, to extrapolate the data.  
\begin{equation}
    \ln\sigma(E) =  k\ln E + C
    \label{excitation_formula}
\end{equation}

\subsubsection{Synthetic data generation for training}

\begin{figure}
    \centering
    \includegraphics[width=\linewidth]{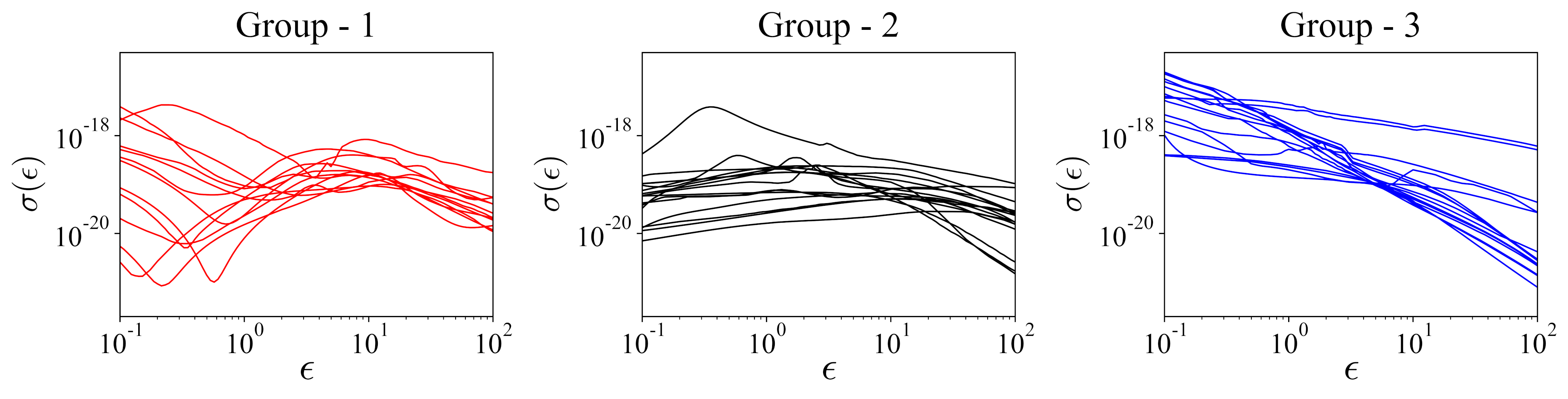}
    \caption{Gas species separated into three different classes based on the characteristics of their elastic momentum transfer cross sections Group 1 consists of Ar, Kr, SF\textsubscript{6}, Xe, CO\textsubscript{2}, Si(CH\textsubscript{3})\textsubscript{4}, CF\textsubscript{4}, CH\textsubscript{4}, H\textsubscript{2}O, HCl, SiH\textsubscript{4} and Cu, Group 2 consists of D\textsubscript{2}, H\textsubscript{2}, He, C, Be, C\textsubscript{2}H\textsubscript{2}, Ne, O\textsubscript{2}, N\textsubscript{2}, F, C(2p(2)\_1D), C(2p(2)\_1S), C\textsubscript{2}H\textsubscript{4}, O, N-elec, CO, C\textsubscript{3}H\textsubscript{6} and O\textsubscript{2}(0.98), and Group 3 consists of CHF\textsubscript{3}, Be(2s\_2p\_1P), N, C(2p3s\_1Po), C(2p3s\_3Po), H(1S), H(2P), H(2S), H(3D), H(3P), H(3S), H(4D), H(4F), H(4P), H(4S) and H~\cite{database1, database2, database3, database4, database5, database6, database7, database8, database9, database10, database11, database12, database13, database14, database15, database16, database17, zatsarinny2004b, allan2006near, bray1992convergent, fursa1995calculation, zammit2014electron, zammit2016complete, christophorou2000electron}}
    \label{classes}
\end{figure}

Deep neural networks require large training datasets for effective performance. The cross section data obtained from LXCat however, are very limited (complete cross sections of only 46 different gas species) and clearly insufficient to properly train the model. Therefore, we generate synthetic training examples by interpolating the actual cross sections. Firstly, all the 46 gas species for which complete sets of data exist on LXCat are manually classified into three different groups based on the characteristics of their elastic momentum transfer cross section as shown in Fig.~\ref{classes}. Group-1, Group-2 and Group-3 consists of 12, 18 and 16 different species respectively. To avoid generation of nonphysical cross sections, a new artificial cross section is calculated by taking a weighted geometric average~\cite{stokes2019determining} of two actual cross sections belonging to the same group : 
\begin{equation}
    \sigma_{\mathit{new}}(\epsilon) = \sigma_1^{1-r}(\epsilon + \epsilon_1 - \epsilon_1^{1-r}\epsilon_2^r)\;\sigma_2^{r}(\epsilon + \epsilon_2 - \epsilon_1^{1-r}\epsilon_2^r)
    \label{interpolate}
\end{equation}
where $\sigma_1(\epsilon)$ and $\sigma_2(\epsilon)$ are the cross sections of gas species belonging to the same group, $\epsilon_1$, $\epsilon_2$ and $\epsilon_1^{1-r}\epsilon_2^r$ are the threshold energies of $\sigma_1(\epsilon)$, $\sigma_2(\epsilon)$ and $\sigma_{\mathit{new}}(\epsilon) $ respectively, and $0\leq r\leq 1$ is a uniformly distributed random variable. 

\begin{figure}
    \centering
    \includegraphics[width=\linewidth]{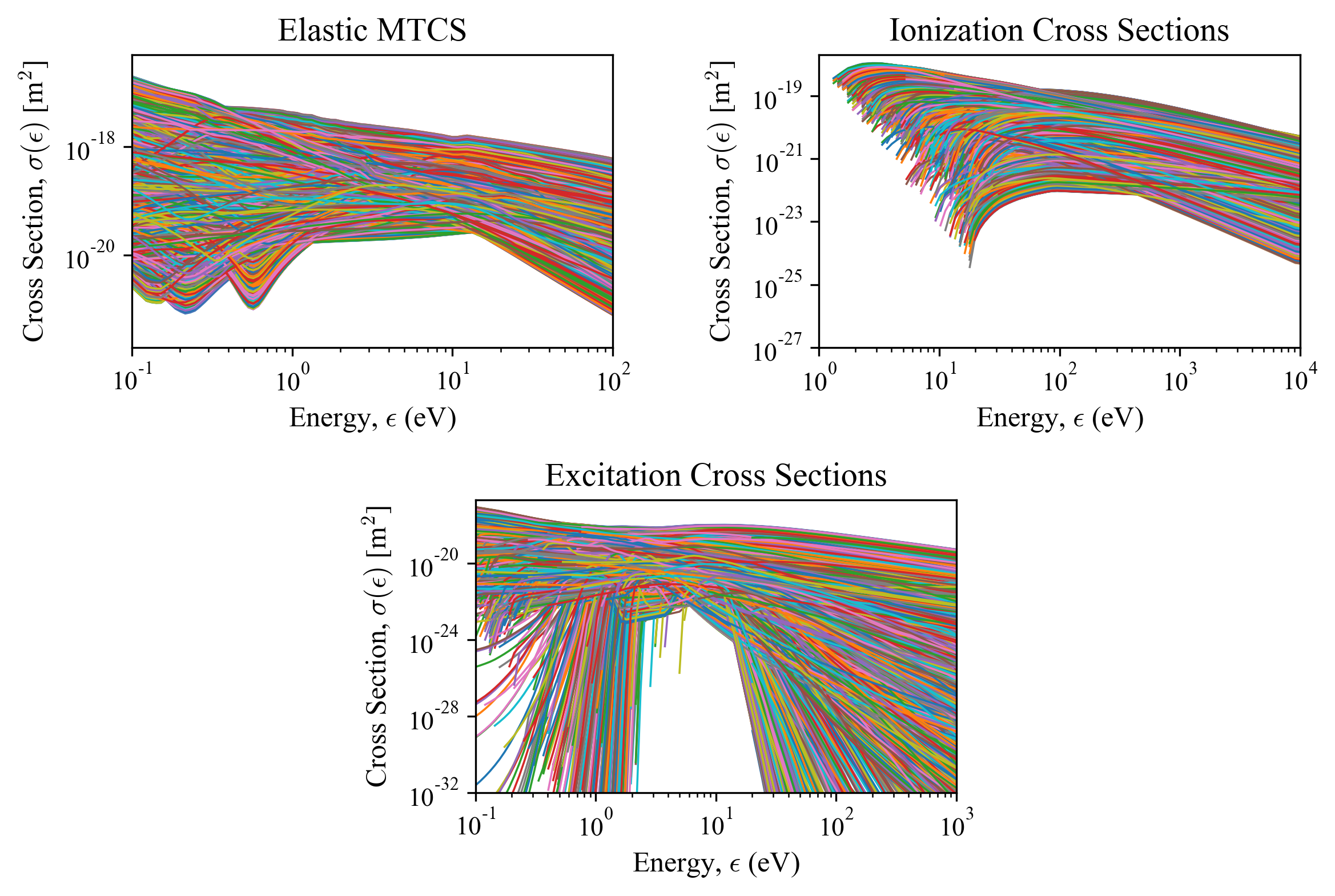}
    \caption{Synthetically generated cross sections data}
    \label{cs_interpolated}
\end{figure}

Out of the $46$ gas species available, we set apart one gas species so that we can later on use our deep learning model to predict its cross sections and compare that with the actual cross section from LxCAT to determine the accuracy of our model. Then for our training data, we use Eq.~\ref{interpolate} to generate a total of $10\,000$ different cross sections (Fig.~\ref{cs_interpolated}) including the actual complete cross section of 45 different gas species. Thus, only the cross sections of the gas species for which the model is to be tested is excluded, while all the other gas species contribute equally in generating these synthetic training cross sections. Subsequently, these cross sections are sampled at $100$ discrete log-spaced energy values, between the energy range considered for prediction. So, we have a total of $10^6$ energy -- cross section pairs in our training dataset.

\subsubsection{Swarm data calculation and Feature selection}
Finally, we complete the input-output training pairs by computing the swarm coefficients corresponding to the cross sections present in our training dataset. Swarm data are computed using the BOLSIG+~\cite{bolsig+} solver for the  numerical solution of the Boltzmann's equation~\cite{hagelaar2005solving}, with the cross sections data as input. Swarm data are calculated at temperature $T = 300$K for $100$ equally log-spaced reduced electric fields in the range $10^{-3}$ Td to $10^{3}$ Td ($1$ Td = $10^{-21}$ Vm$^2$). Note that BOLSIG+ can extrapolate cross sections to higher energies if needed for very high E/N. 

\begin{figure}
    \centering
    \includegraphics[width=\linewidth]{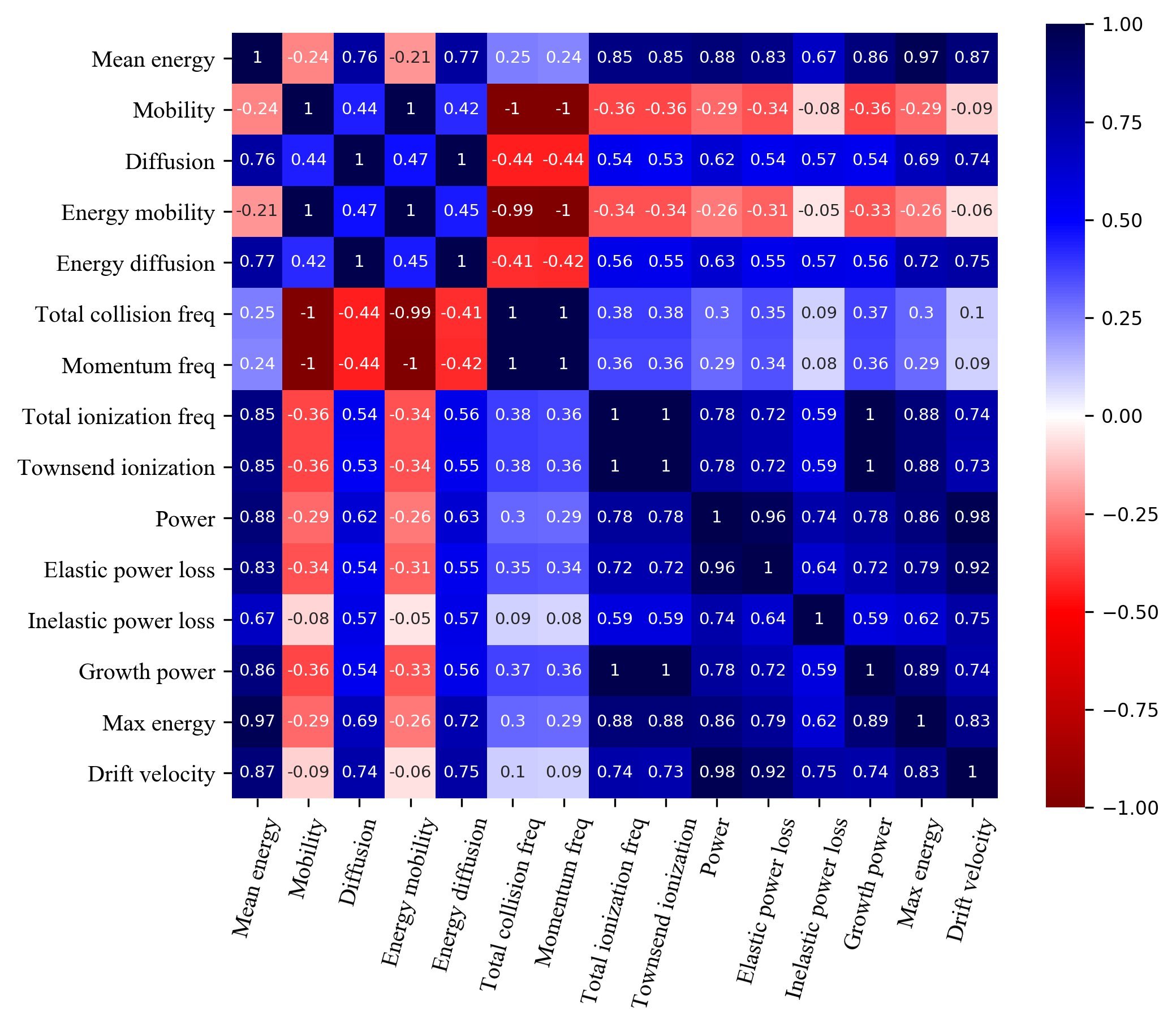}
    \caption{Pearson correlation coefficient of different swarm parameters obtained from BOLSIG+}
    \label{corr}
\end{figure} 

Mean energy, mobility, diffusion, energy mobility, energy diffusion, total collision frequency, momentum frequency, total ionization frequency, Townsend ionization coefficient, power, elastic power loss, inelastic power loss, growth power, maximum energy and drift velocity are 15 different quantities which are included in the output of the BOLSIG+ solver. However, unlike BOLSIG+, in most Boltzmann solvers, the max energy is input. Using all of these quantities as input to the neural network is not feasible as it would increase both the training time of the model and its memory requirements. It might even reduce the overall effectiveness of the model and hence, we use feature selection to reduce the input data by removing redundant variables. We compute the Pearson correlation coefficient between all these possible inputs as depicted in Fig.~\ref{corr}. Pearson correlation coefficient shown in Fig.~\ref{corr} is average of all the gas species except for helium (assuming He is the test species on which the model will be tested). Features with high correlation value ($>0.85$ or $<-0.85$) are more linearly dependent and hence have almost the same information content, thus we keep one and drop rest of these highly correlated variables.

Using this feature selection method, we are left with mean energy, mobility, diffusion, Townsend ionization, elastic power loss and inelastic power loss. However, mean energy, elastic power loss and inelastic power loss are the swarm coefficients whose data is not readily available on LXCat because these are generally not experimentally measured and thus we drop them from our feature set because in the long-term, we would like to apply our methodology to analysis of experimental swarm data. However, as a future study, it will be interesting to see how the results will get affected if we include these three parameters in the data.

\subsection{ML methods and Model Training}

\subsubsection{Data normalization}

Cross sections along with swarm data scale across many orders of magnitude. Directly using this data to train the neural network will severely impede neural network's ability to learn meaningful trends in the data. Also, large input values would result in large weight values of the neural networks, making them highly unstable. Small input values having zero mean and standard deviation of one are generally considered as ideal for neural networks, and thus we log transform everything (Eq.~\ref{log1}) and then subsequently normalize it to [-1, 1] range (Eq.~\ref{log2}). 
\begin{equation}
    y = \log(x)
    \label{log1}
\end{equation}
\begin{equation}
    z = 2\left(\frac{y-y_{min}}{y_{max}-y_{min}}\right) - 1
    \label{log2}
\end{equation}
If the data value is zero, then it is replaced by sufficiently small positive quantity ($\delta=10^{-50}$) before applying the log transformation.

\subsubsection{Neural network architecture}

Input to our network consists of different swarm parameters --- mobility ($\mu N$), diffusion coefficient ($ND$) and Townsend ionization coefficient ($\alpha/N$) --- measured at 100 distinct reduced electric fields $E_1/N$, $E_2/N$, $\ldots$, where $N$ (or $n_0$) is the number density of the background neutrals. We use neural network itself to estimate the cross sections as function of energy. Thus energy $\epsilon$ is also added to the input to the neural network and output is the single value of cross section corresponding to that energy, $\sigma(\epsilon)$
\begin{equation}
    x = \left[
    \begin{array}{cc}
    \epsilon \\
    N\mu(E_1/N) \\ 
    N\mu(E_2/N) \\ 
    \vdots\\ 
    ND(E_1/N) \\ 
    ND(E_2/N) \\ 
    \vdots\\ 
    \alpha/N(E_1/N) \\ 
    \alpha/N(E_2/N) \\ 
    \vdots
    \end{array}
    \right]
    , \;\;\; y = \sigma(\epsilon)
\end{equation}

Neural networks are composed of several artificial neurons. The structure of these neurons and its connections play an important role in inferring the function which maps the input to the output. Hence, we test different neural networks to study these performance changes. Artificial Neural Network (ANN) is the most basic form of neural network and its use to solve the inverse swarm problem has been proposed in~\cite{morgan1991feasibility}. Minor improvements were made to this architecture by~\cite{stokes2019determining}, which made the network simpler and faster to train. This ANN architecture has three hidden layers, each having 128 neurons, with $\mathit{swish}$ as non-linear activation function, where $\mathit{swish}(x) = x/(1+\exp{(-x)})$. We consider this as our benchmark architecture.

\begin{figure}
    \begin{subfigure}{\textwidth}
        \centering
        \includegraphics[width=0.95\linewidth]{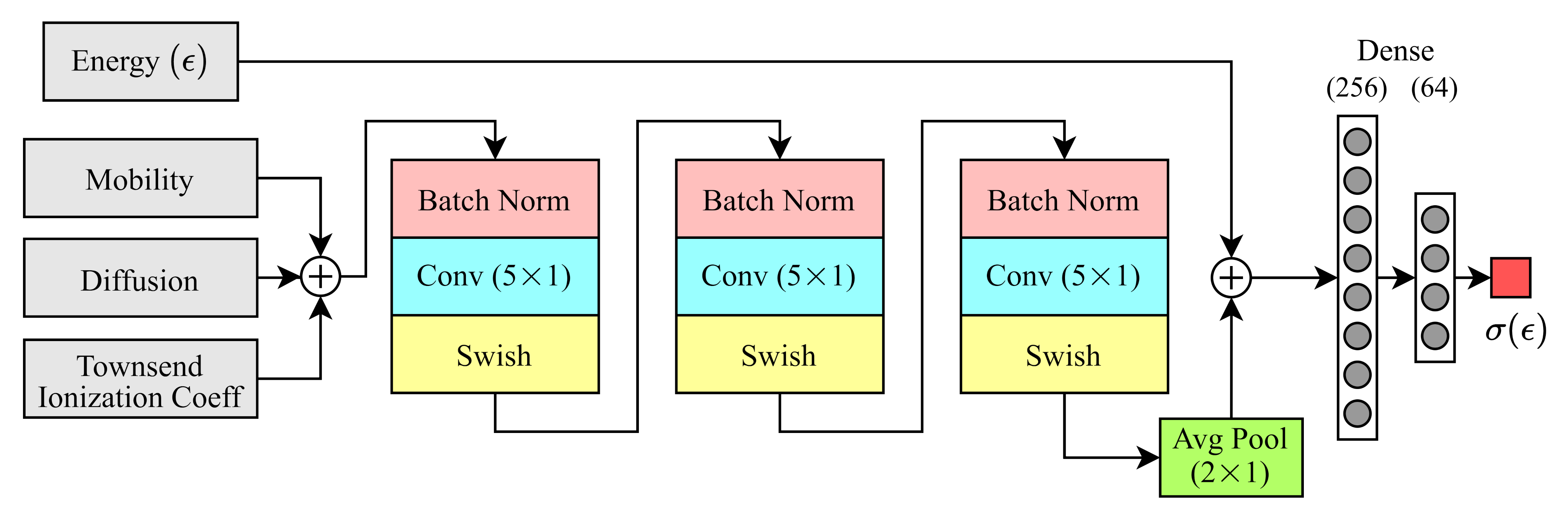}
        \caption{Convolutional Neural Network}
        \label{cnn_architecture}
    \end{subfigure}
    \newline
    \begin{subfigure}{\textwidth}
        \centering
        \includegraphics[width=\linewidth]{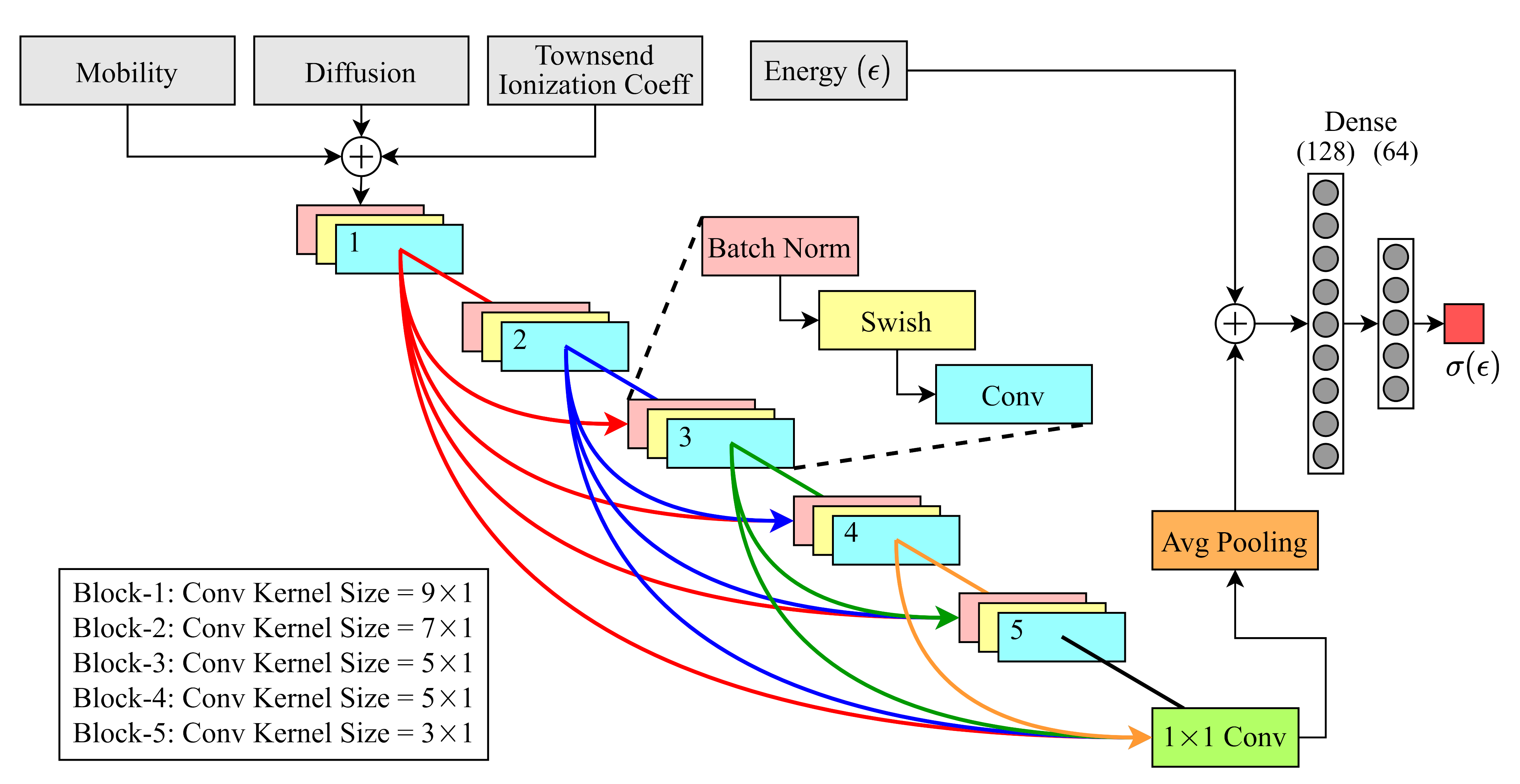}
        \caption{DenseNet}
        \label{densenet_architecture}
    \end{subfigure}
    \caption{Neural Network layouts used in this study. Various CNN and DenseNet architectures with different hyperparameters like number of convolutional filters, kernel size, number of hidden layers and the choice of activation function were implemented, and the layout having best performance has been finally chosen.}
    \label{architecture}
\end{figure}

For our other model, we implement a 1D Convolutional Neural Network~\cite{lecun1995convolutional} because of its ability to extract spatial information from the input data, which is in the form of continuous series. Various CNN architectures were trained to determine optimal hyper-parameters and Fig.~\ref{cnn_architecture} shows the one for which the best results were obtained. Features from the different swarm coefficients are extracted by three successive blocks, each consisting of batch normalization layer, convolutional layer with 64 filters and kernel size of $5\times1$, and a \textit{swish} activation layer. This is followed by an average pooling layer, which is then flattened and passed to two fully connected layers along with the energy input. FC layers have 256 and 64 neurons with \textit{swish} activation function. Finally, it is connected to the linearly activated single output neuron.

Dense Convolutional Neural Network (DenseNet) is extension of CNN, which provides substantial performance improvement with comparison to previous CNN architectures~\cite{huang2017densely} and hence we try 1D-DenseNet architecture as our third model. DenseNet improves the information flow between the layers by introducing direct connections from any layer to all subsequent layers. This also leads to feature reuse throughout the network and hence, it requires fewer parameters than a CNN architecture to achieve similar performance (Parameter efficiency). The concatenation of feature maps of all preceding layers, $x_0, x_1, \ldots, x_{l-1} $ are provided as input to the $l^{th}$ layer.
\begin{equation}
    x_l = H_l([x_0, x_1, \ldots, x_{l-1}])
\end{equation}
The composite block $H_l$ consists of three successive layers: batch normalization (BN), followed by \textit{swish} activation and a convolution layer.

We trained and tested different DenseNet layouts having 3-6 such composite blocks, where the number of convolution filters in each block were kept constant (32 filters) and zero padding was applied to each end of the input so as to keep the feature map's size fixed. Highest accuracy levels were achieved for DenseNet having five composite layers $H_l$ (Fig.~\ref{densenet_architecture}). We use longer convolution kernel to begin with, and gradually decrease its size in the subsequent layers. Concatenation of all these different length features allows the network to learn short-term as well as long-term trends from the swarm data but on the downside, these accumulated features substantially increases the model size. Hence, we use a $1\times1$ convolution followed by average pooling, to reduce the dimension of feature map and avoid overfitting before passing it, along with energy input, to two fully connected layers having size 128 and 64, with \textit{swish} activation.

The output layer in each of the architectures is a single neuron corresponding to the cross section being predicted. Hence, we need to train three separate models to predict elastic momentum transfer, ionization and excitation cross sections, for each architecture discussed above. This also allows the network to have feature maps pertinent to each cross section type. This separate training can be eliminated if the output layer is increased to 3 neurons, one for each cross section type. However, this would severely inhibit the network's capability as it would force the network to work with same feature maps even for different types of cross section. Thus, we avoid simultaneous prediction of different cross section types.


Purely from the machine learning perspective, neural networks are trained to improve their predictions by heavily penalizing large errors. Although this seems logical, Stokes~\textit{et al.}~\cite{stokes2019determining} found that using $L_2$ loss actually provided worse results than $L_1$ loss for the inverse swarm problem due to the inherent uncertainty in the solution of this inverse problem and consistently trying to fit these uncertain cross sections impeded the model's overall performance. Hence, we also choose mean absolute error ($L_1$-norm) as it is less sensitive to large errors, but make a slight modification to improve model's performance. As discussed earlier, zero-valued cross sections are replaced with a small threshold value $\delta=10^{-50}$ before performing data normalization. This is just an approximate value of $\delta$ and clearly it would be wrong to penalize the network if the predicted value lies in the range [$0$, $\delta$]. Thus, we use a custom $L_1$ loss function 
\begin{equation}
    L(y, \,\hat{y}) = \frac{1}{N}\sum_{i=1}^{N}| \mathit{max}(y_i, \,\Delta) - \hat{y_i}|
    \label{loss}
\end{equation}
where $N$ is the number of training examples, $y_i$ is the model's prediction, $\hat{y_i}$ is the target value and $\Delta$ is log normalized value of $\delta$ calculated using Eq.~\ref{log1}~\&~\ref{log2} . This loss function clips the predicted output if it is less than $\Delta$, allowing the network's final prediction to be less than $\delta$ without any penalty. This slight modification significantly improves the prediction results of ionization and excitation cross sections.

The training dataset is divided into batches containing $10^3$ samples and all the models are trained by minimizing Eq.~\ref{loss} using Adam optimizer~\cite{kingma2014adam} with learning rate of $10^{-4}$ and exponential decay rates of the first moment ($\beta_1$) and second moment ($\beta_2$) as $0.9$ and $0.999$ respectively. The models were implemented using Keras~(2.3.0)~\cite{chollet2015keras} with Tensorflow~(2.2.0) backend~\cite{tensorflow2015-whitepaper} having GPU support.

\subsubsection{Determining training duration}
During the iterative training of our neural networks, its error on the training set continuously decreases. However, the same does not apply on its generalization error (errors on unseen data), which actually begins to increase after a point in training (overfitting) and hence, ideally we must stop training our network when the generalization error is the least. Since it is not possible to calculate the generalization error explicitly, we try to roughly determine it using $k$-fold cross-validation as dividing our data simply into training and validation dataset is not feasible due to the insufficient availability of the actual cross sections. Each of the 3 groups of gas species divided earlier (Fig.~\ref{classes}) are subdivided in two separate parts (randomly) to form a total of 6 parts. Of these 6 parts, one part is kept as validation data and the remaining 5 parts will be present in training data set. The synthetic cross sections, which are generated using Eq.~\ref{interpolate} from two cross sections $\sigma_1$ and $\sigma_2$ will be split as per the following criteria -- if both $\sigma_1$ and $\sigma_2$ belong to the newly formed training dataset, then this artificial cross section too will be added to the training dataset whereas if both $\sigma_1$ and $\sigma_2$ belong to the newly formed validation dataset, then it will be added to the validation dataset. Then, we train the networks on this newly formed training dataset and monitor the changes in validation error at each epoch. This process is repeated 6 times, with each of the 6 parts used exactly once in the validation data. This ensures that no data is wasted and our models get the opportunity to train on multiple train-validate splits. All the six validation errors are averaged at each epoch and this averaged validation error can be considered as a close substitute for the generalization error. Thus, we determine the optimal number of epochs when this averaged validation error reaches a minimum value. Later while testing our models, we will train them again on all the $10^6$ examples (no division into training-validation dataset) for this optimal number of epochs.

\section{Results}

All the architectures -- ANN, CNN and DenseNet -- were trained to separately predict elastic momentum transfer, ionization and excitation cross sections using a total of $10^6$ examples which were generated using the process described earlier. These trained models were used to predict the unseen cross sections of nitrogen~(N\textsubscript{2}), argon~(Ar), helium~(He), fluorine~(F), methane~(CH\textsubscript{4}), oxygen~(O\textsubscript{2}) and sulphur hexafluoride~(SF\textsubscript{6}). Here, we would like to explicitly state that even though only one cross section type was predicted at a time, no assumption was made whatsoever regarding the values of other cross section types while predicting a particular cross section--i.e. while estimating the elastic MTCS of a gas species, we do not provide any details about the values of its ionization or excitation cross sections. We test the trained models for such a wide range of gas species, having different physical and chemical properties, to ensure robust performance.

\begin{figure}
    \begin{subfigure}{0.5\textwidth}
        \centering
        \includegraphics[width=0.875\linewidth]{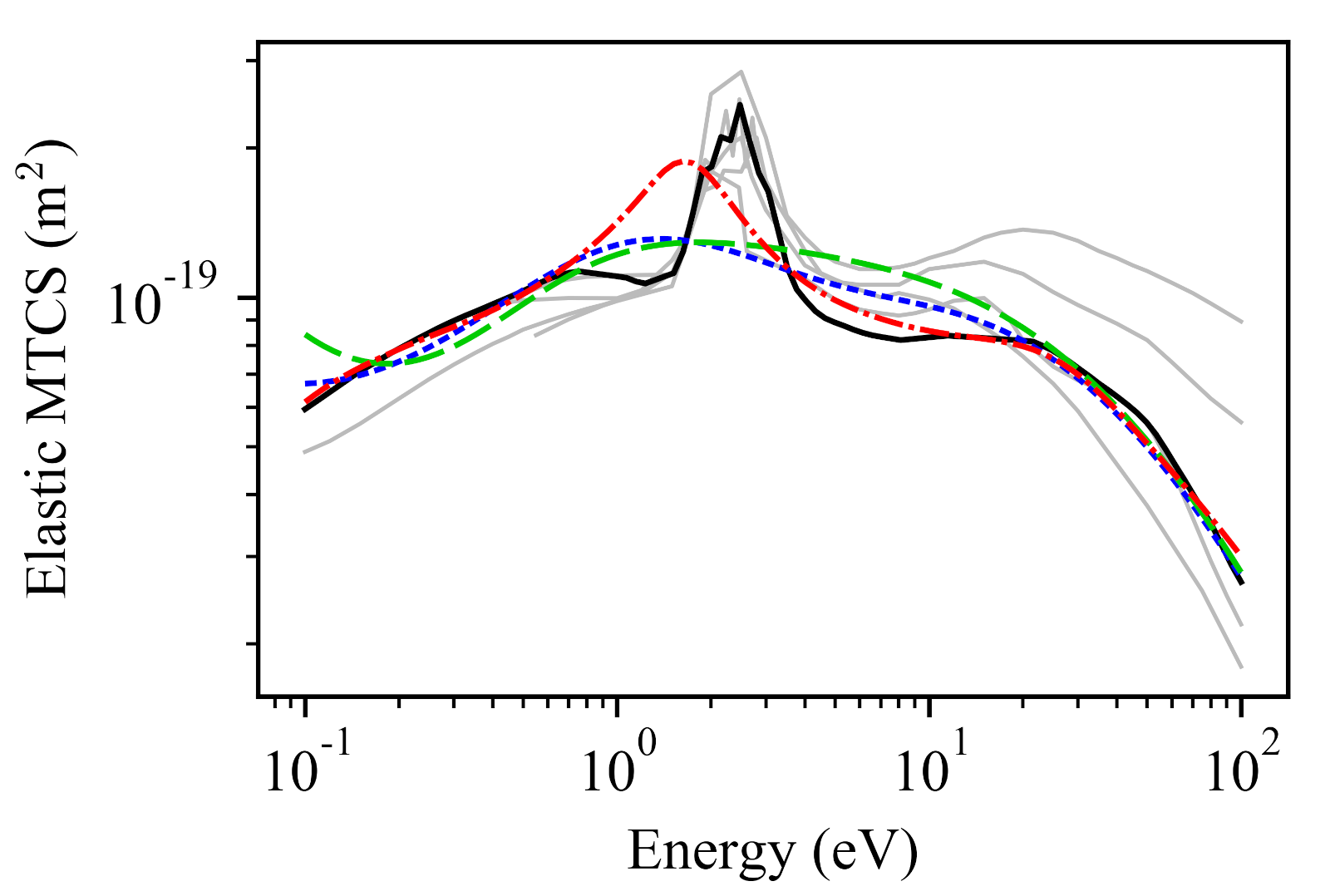}
        \caption{Nitrogen (N\textsubscript{2})}
        \label{elastic_N2}
    \end{subfigure}
    \begin{subfigure}{0.5\textwidth}
        \centering
        \includegraphics[width=0.875\linewidth]{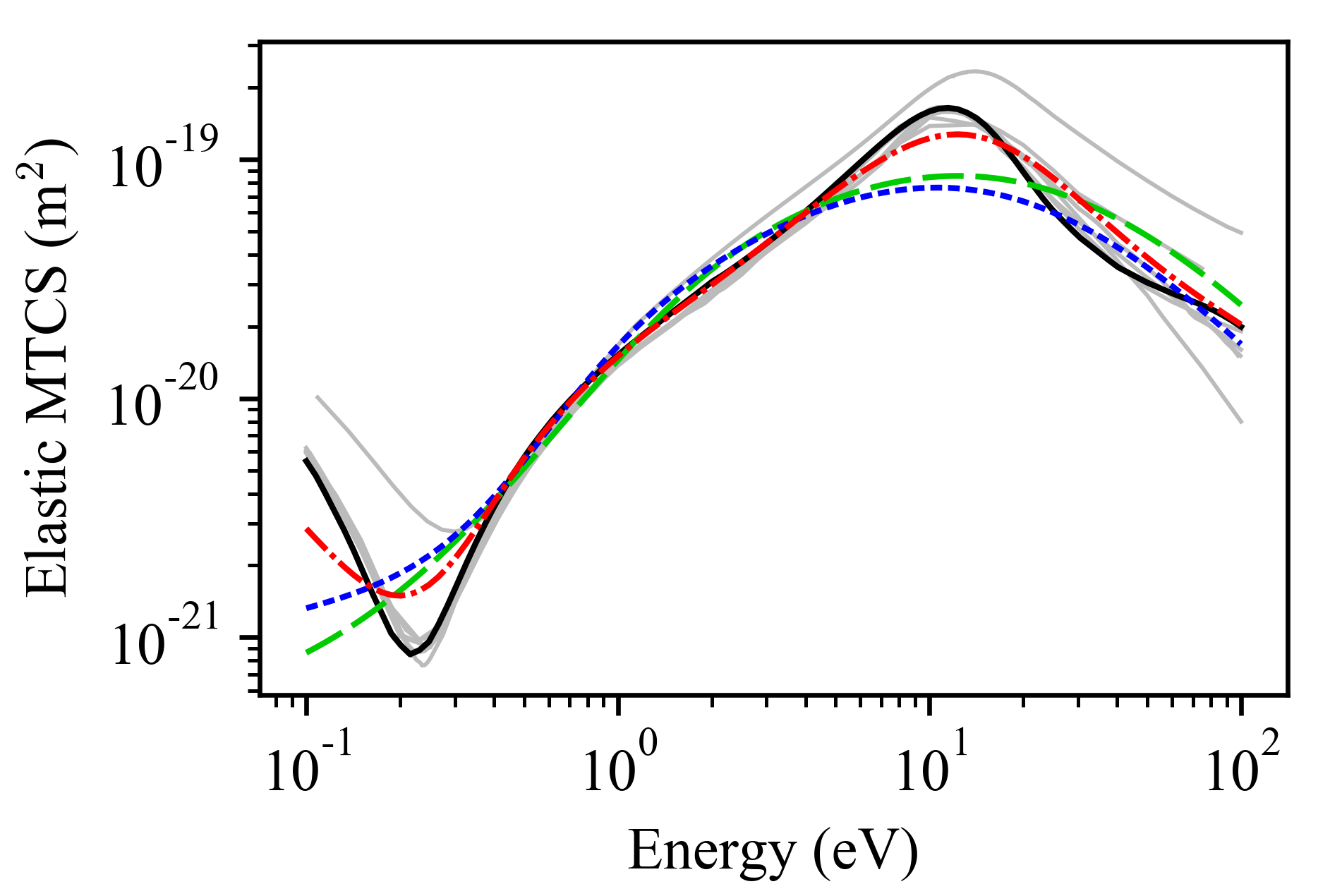}
        \caption{Argon (Ar)}
        \label{elastic_Ar}
    \end{subfigure}
    \begin{subfigure}{0.5\textwidth}
        \centering
        \includegraphics[width=0.875\linewidth]{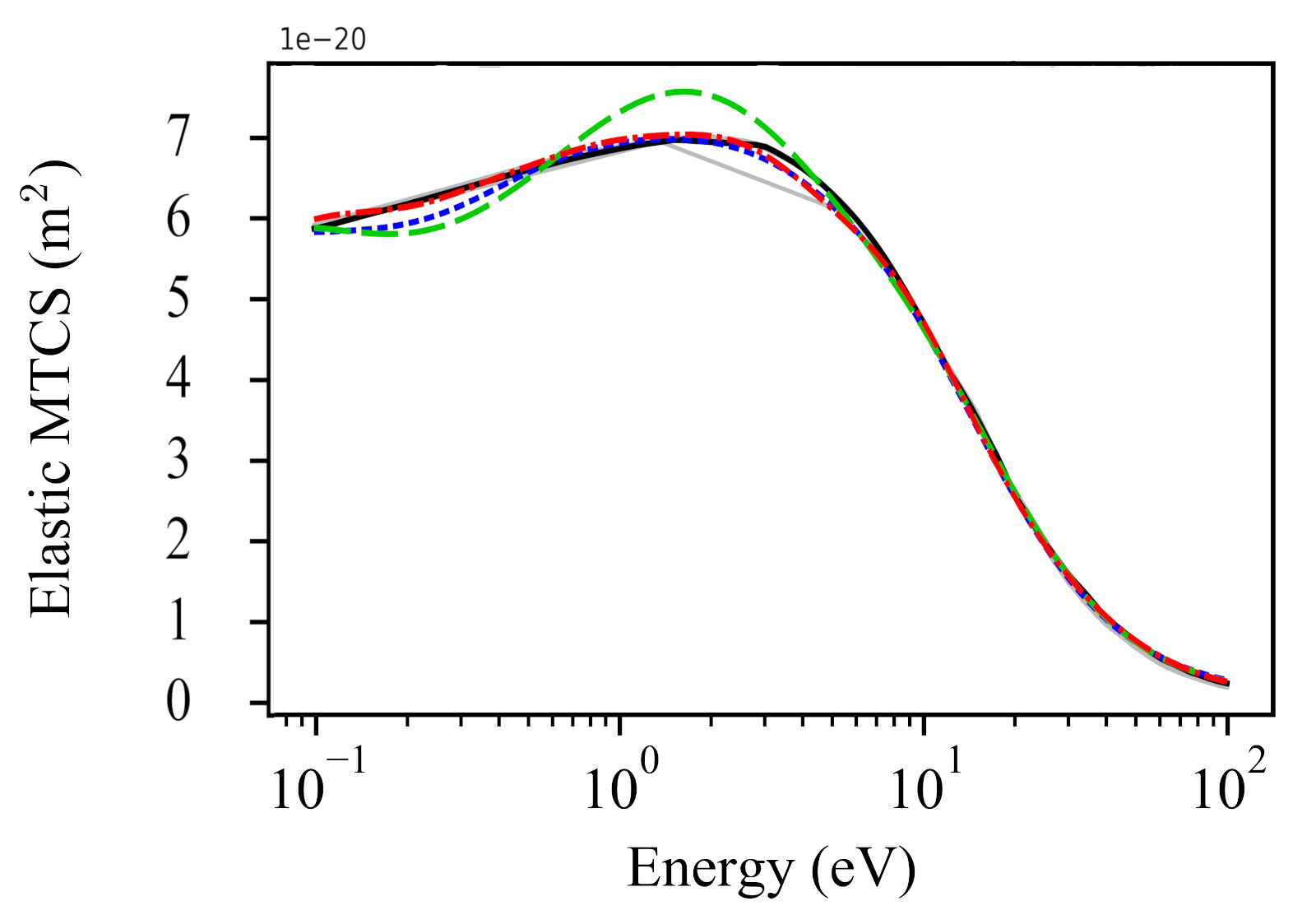}
        \caption{Helium (He)}
        \label{elastic_He}
    \end{subfigure}
    \begin{subfigure}{0.5\textwidth}
        \centering
        \includegraphics[width=0.875\linewidth]{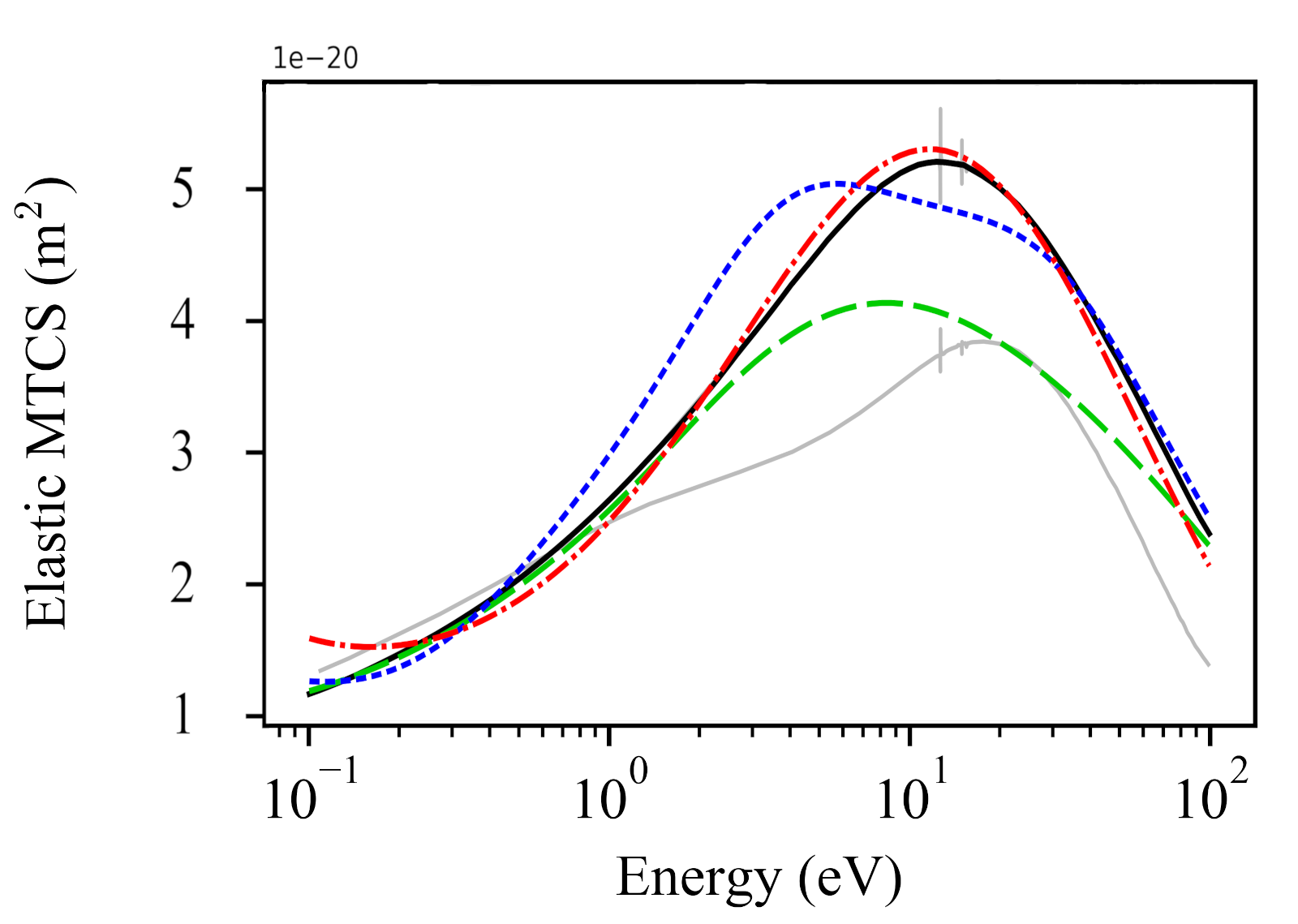}
        \caption{Fluorine (F)}
        \label{elastic_F}
    \end{subfigure}
    \begin{subfigure}{0.5\textwidth}
        \centering
        \includegraphics[width=0.875\linewidth]{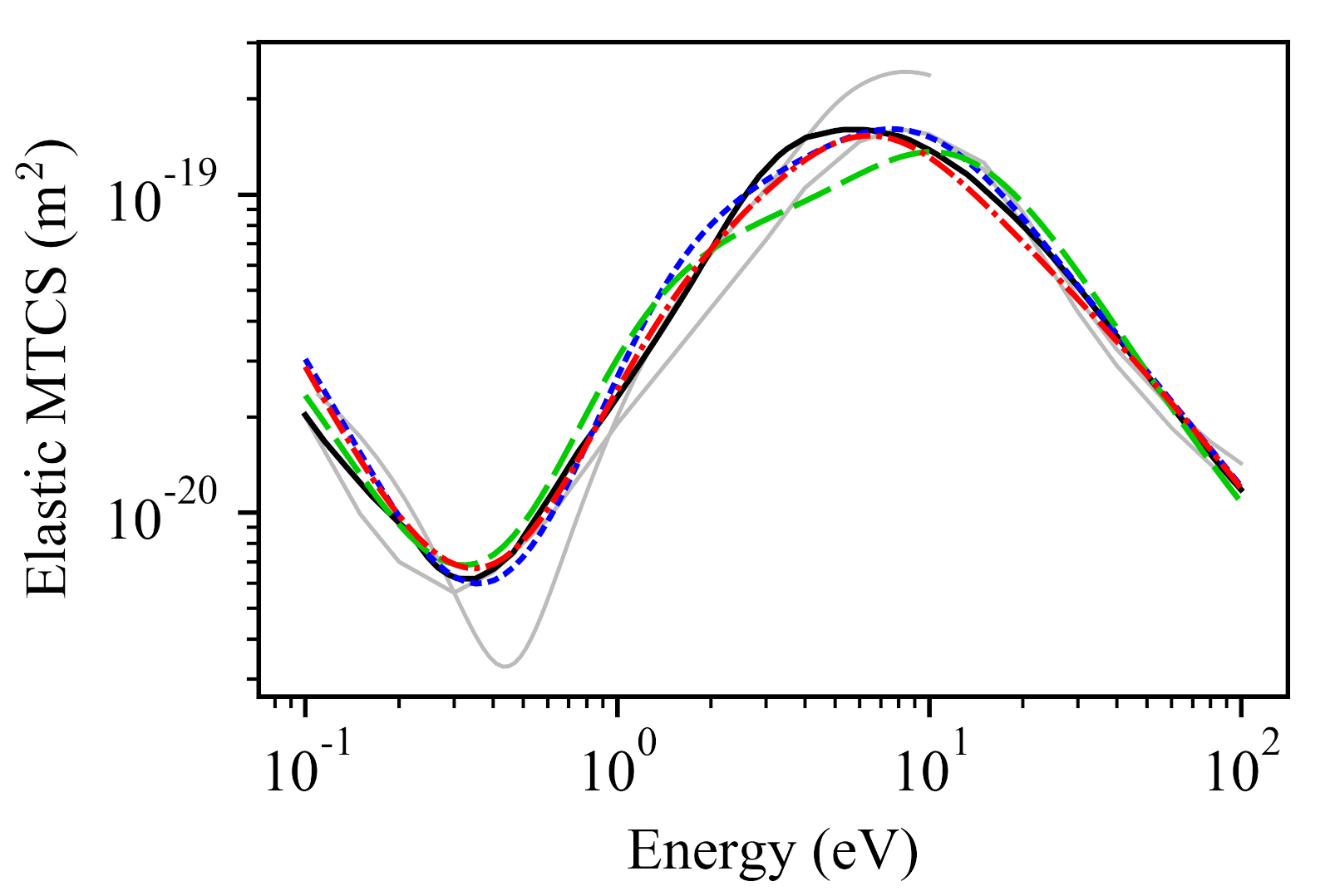}
        \caption{Methane (CH\textsubscript{4})}
        \label{elastic_CH4}
    \end{subfigure}
    \begin{subfigure}{0.5\textwidth}
        \centering
        \includegraphics[width=0.875\linewidth]{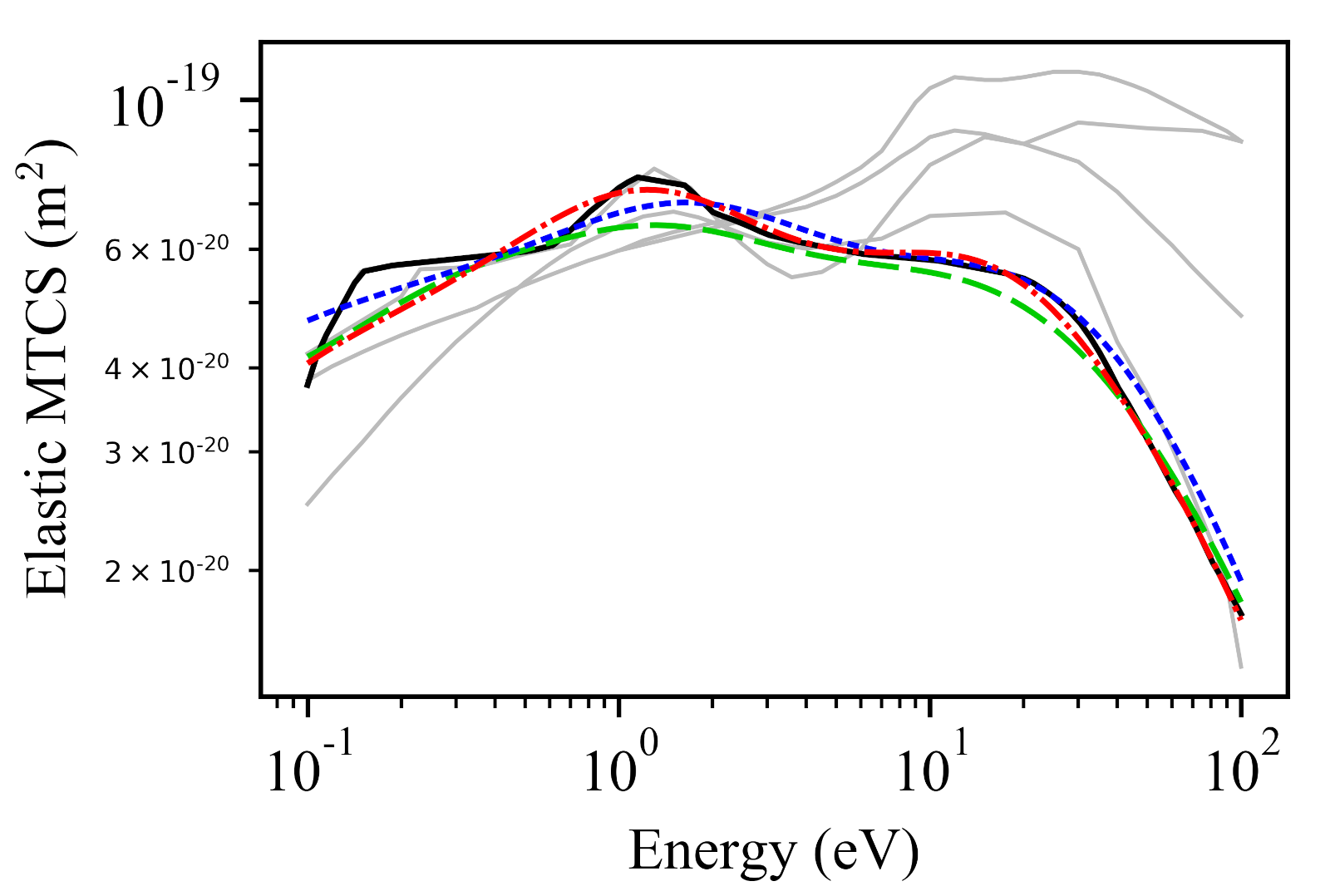}
        \caption{Oxygen (O\textsubscript{2})}
        \label{elastic_O2}
    \end{subfigure}
    \begin{subfigure}{0.5\textwidth}
        \centering
        \includegraphics[width=0.875\linewidth]{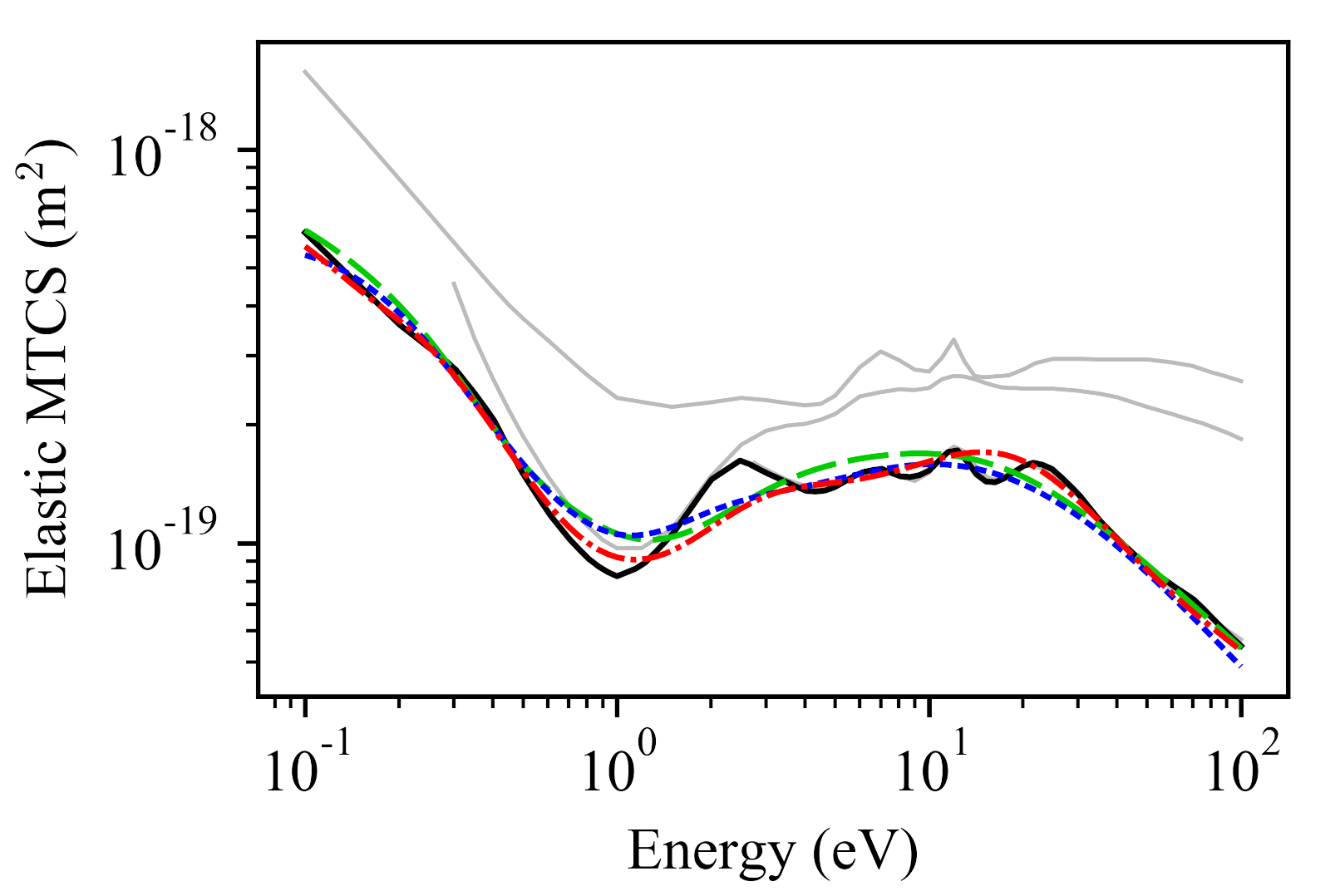}
        \caption{Sulfur hexafluoride (SF\textsubscript{6})}
        \label{elastic_SF6}
    \end{subfigure}
    \begin{subfigure}{0.5\textwidth}
        \centering
        \includegraphics[width=0.875\linewidth]{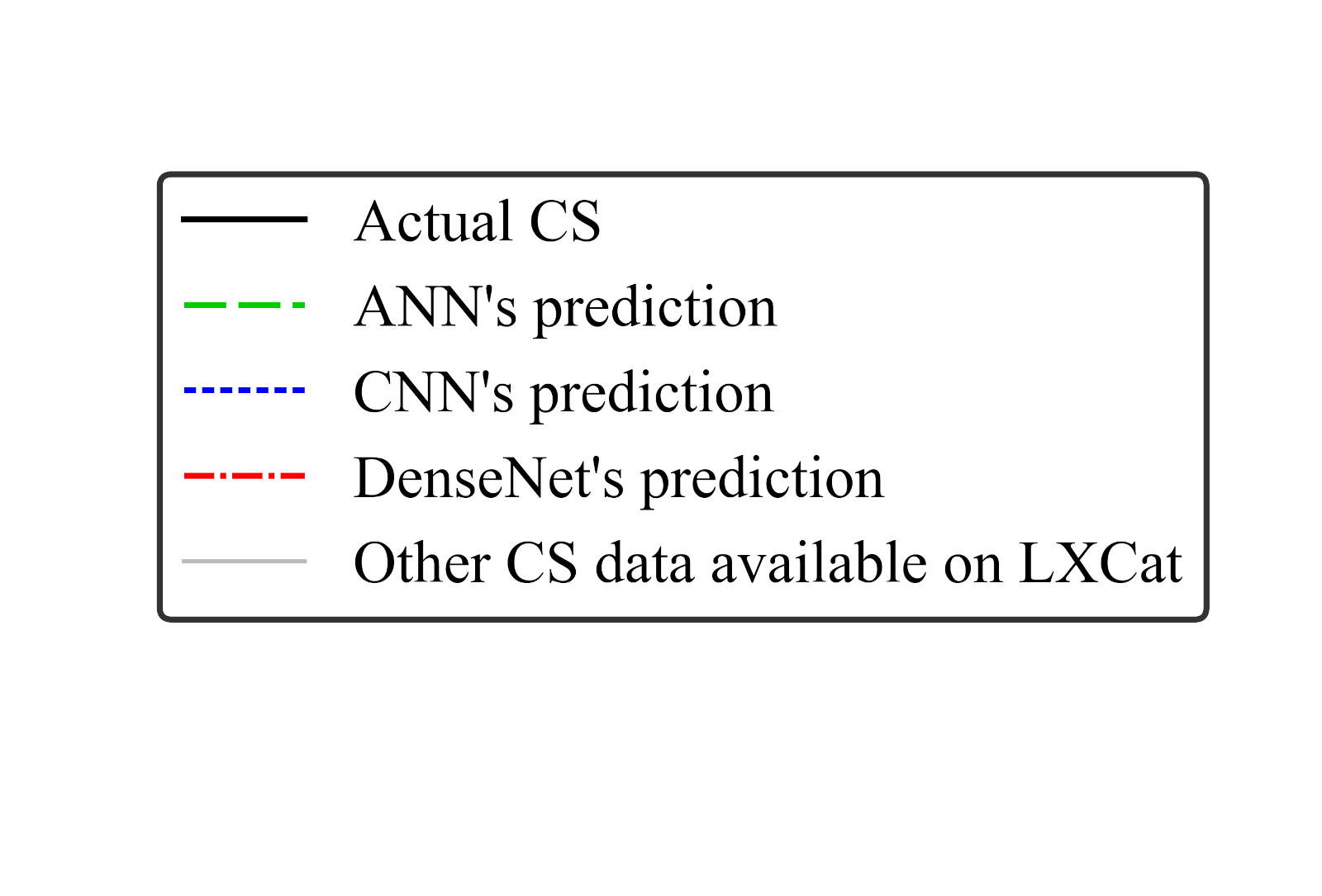}
        \label{elastic_legend}
    \end{subfigure}
    \caption{Prediction of elastic momentum transfer cross sections of various gas species. It is to be noted that in some cases ``Other CS data available on LXCat" (shown in grey color) consists of both elastic momentum transfer and total elastic scattering cross sections. The grey lines simply provides some estimation of the inherent variations in determination of cross sections already available in the literature.}
    \label{results_elastic}
\end{figure}

\begin{figure}
    \begin{subfigure}{0.5\textwidth}
        \centering
        \includegraphics[width=0.87\linewidth]{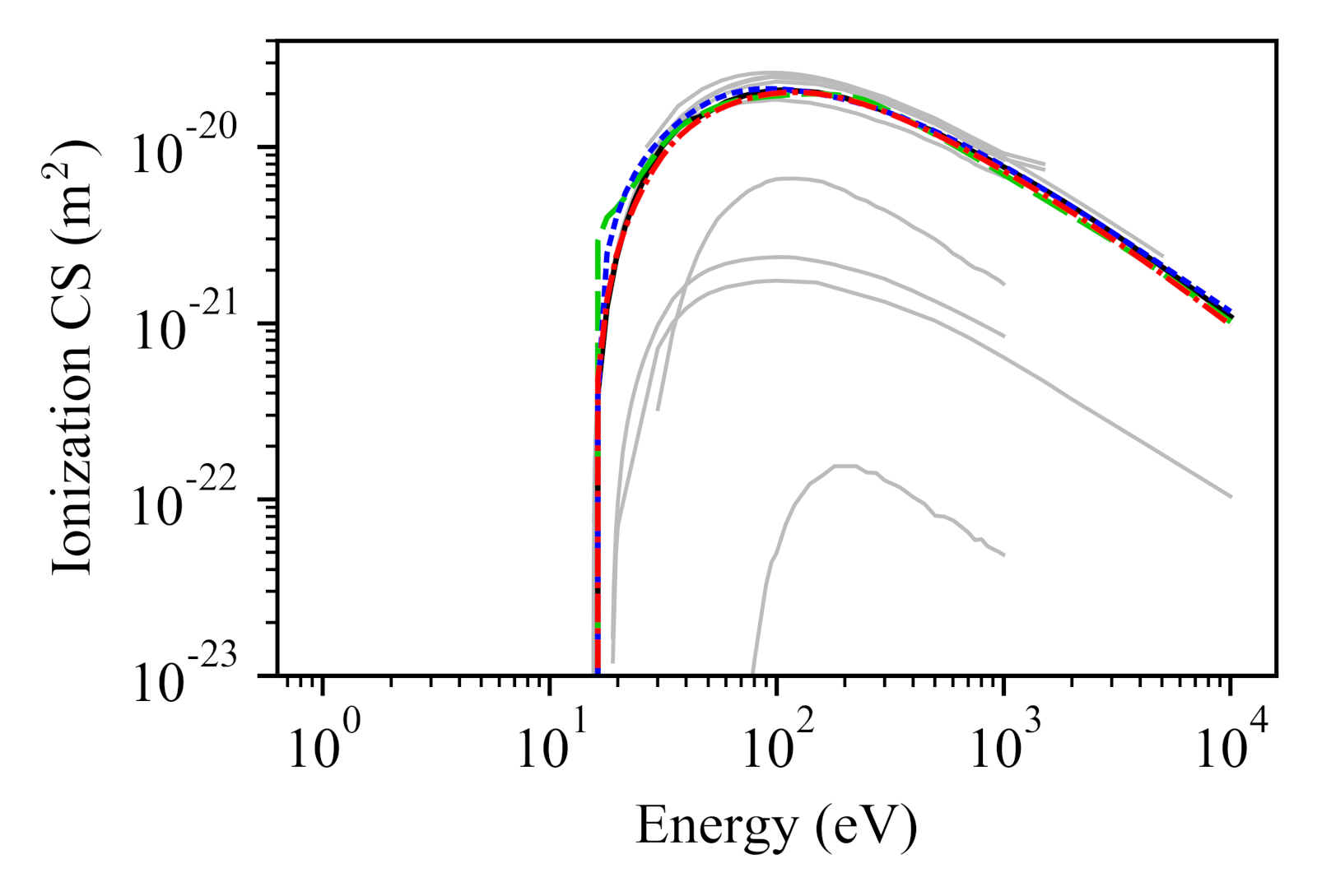}
        \caption{Nitrogen (N\textsubscript{2})}
        \label{ionization_N2}
    \end{subfigure}
    \begin{subfigure}{0.5\textwidth}
        \centering
        \includegraphics[width=0.87\linewidth]{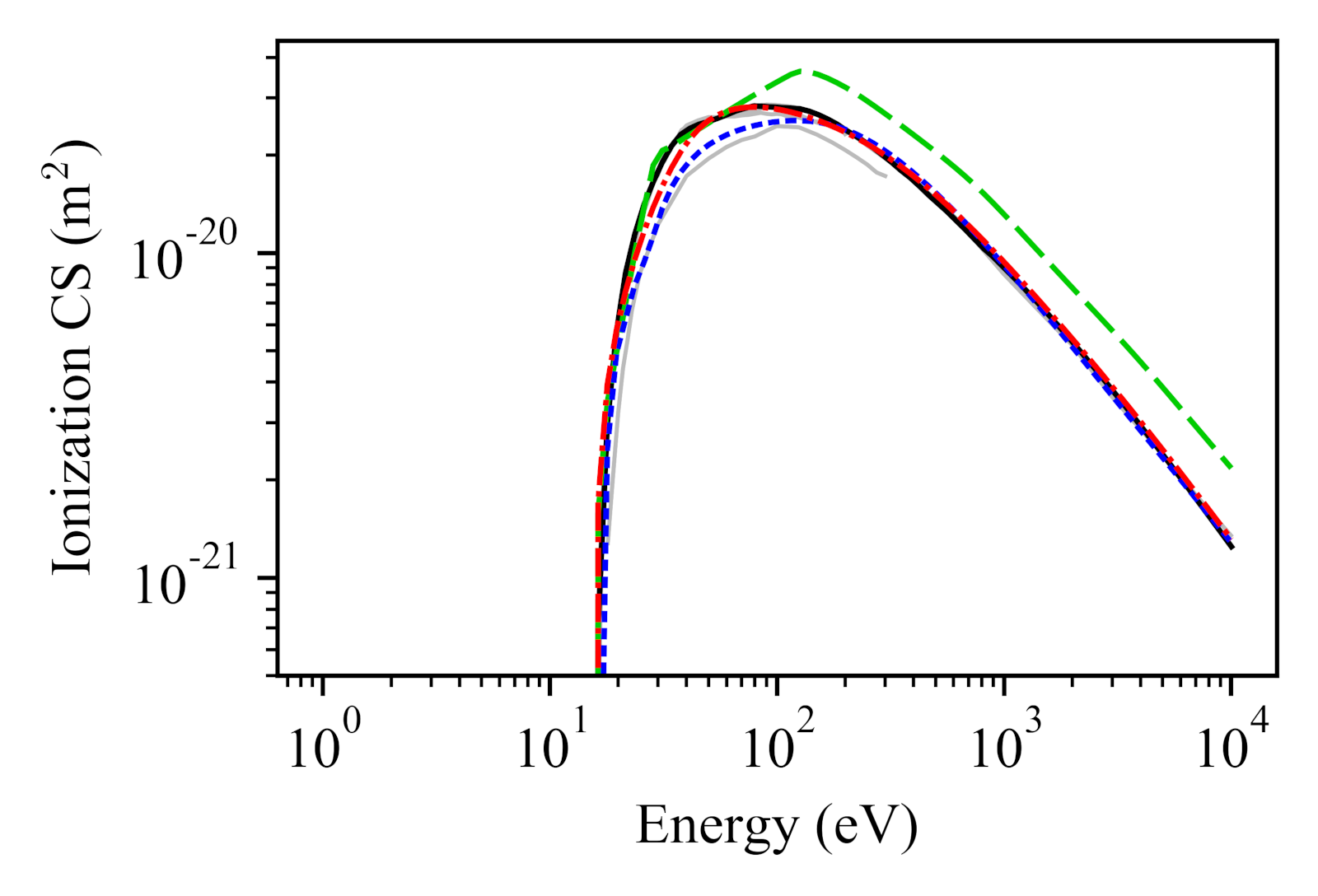}
        \caption{Argon (Ar)}
        \label{ionization_Ar}
    \end{subfigure}
    \begin{subfigure}{0.5\textwidth}
        \centering
        \includegraphics[width=0.87\linewidth]{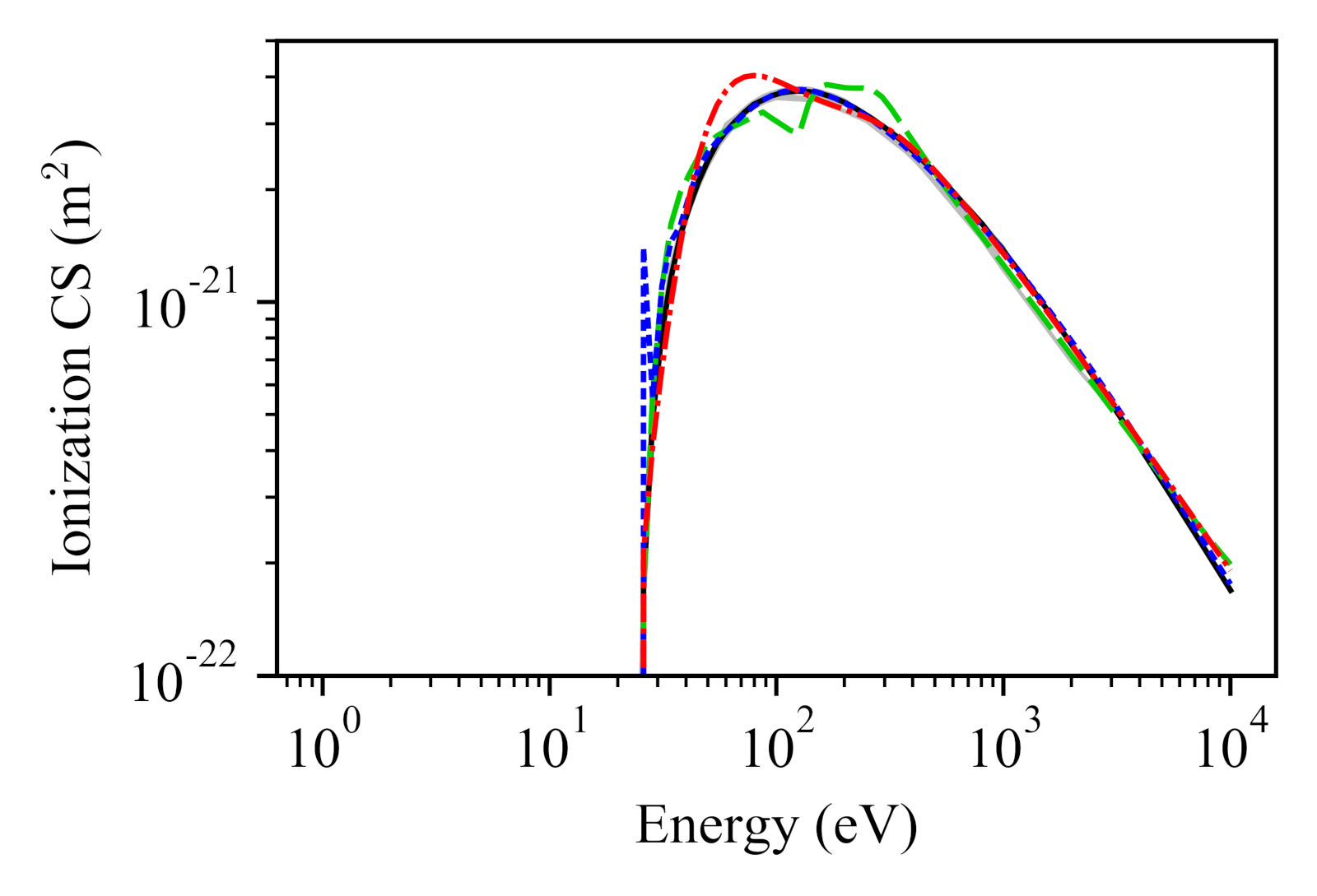}
        \caption{Helium (He)}
        \label{ionization_He}
    \end{subfigure}
    \begin{subfigure}{0.5\textwidth}
        \centering
        \includegraphics[width=0.87\linewidth]{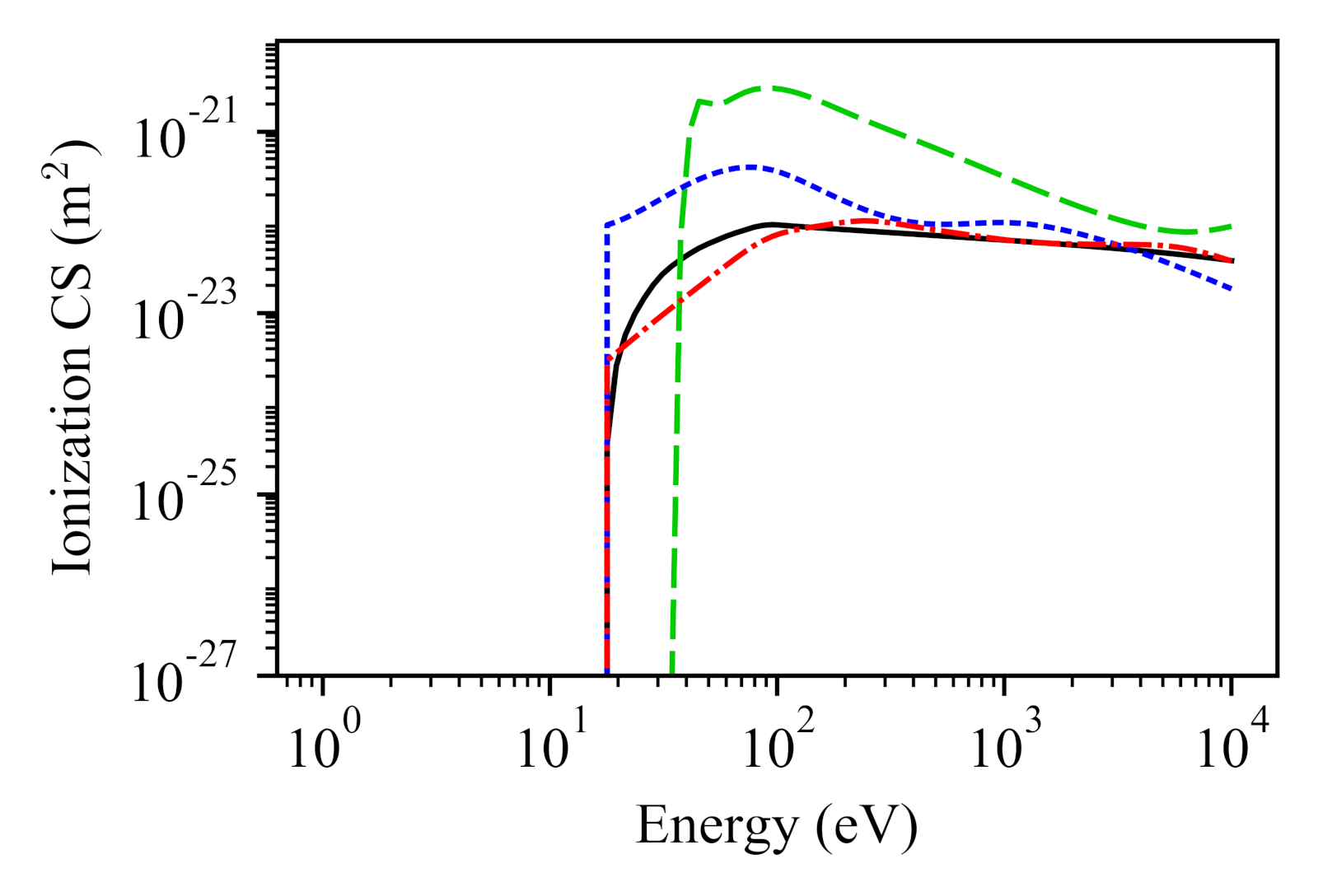}
        \caption{Fluorine (F)}
        \label{ionization_F}
    \end{subfigure}
    \begin{subfigure}{0.5\textwidth}
        \centering
        \includegraphics[width=0.87\linewidth]{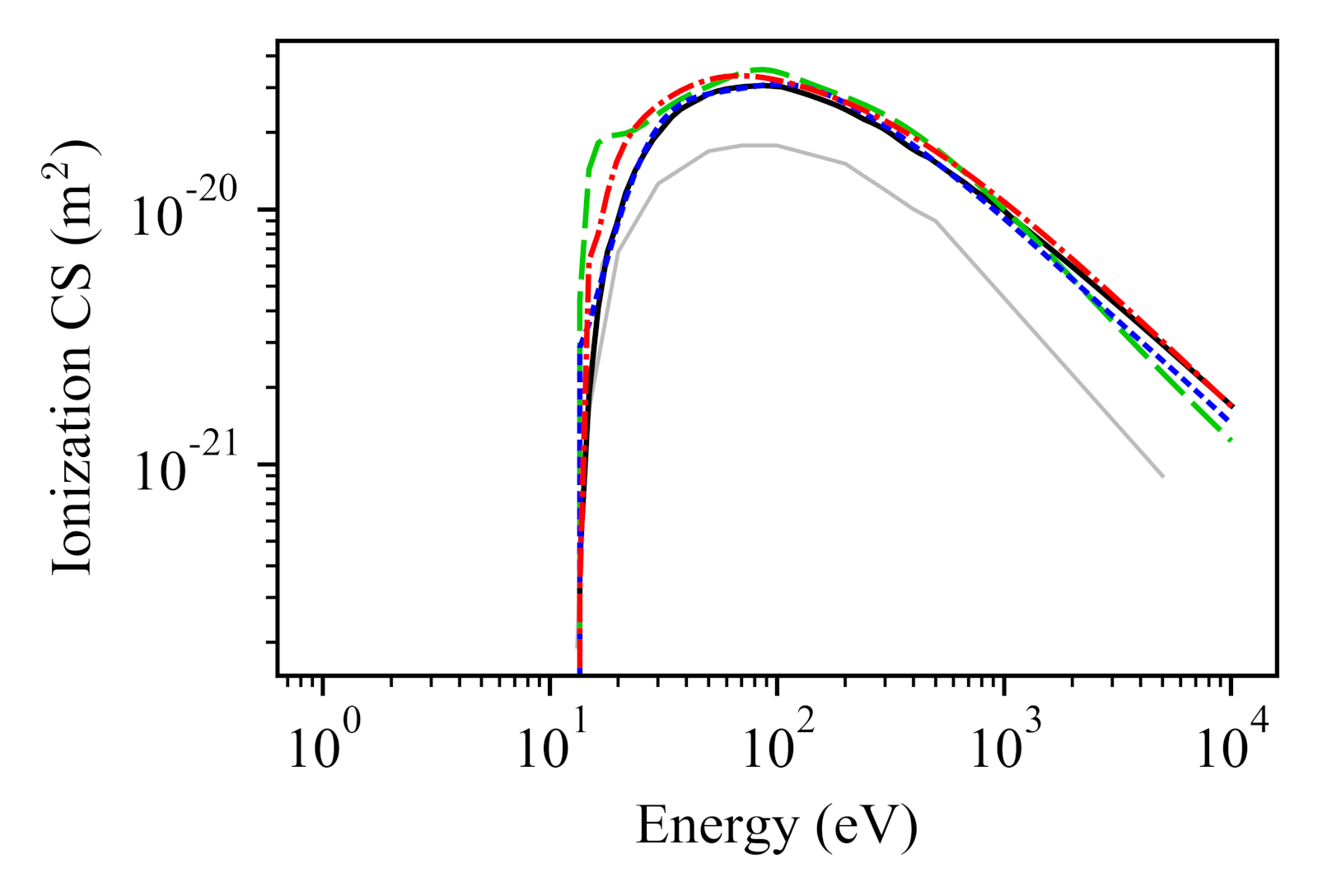}
        \caption{Methane (CH\textsubscript{4})}
        \label{ionization_CH4}
    \end{subfigure}
    \begin{subfigure}{0.5\textwidth}
        \centering
        \includegraphics[width=0.87\linewidth]{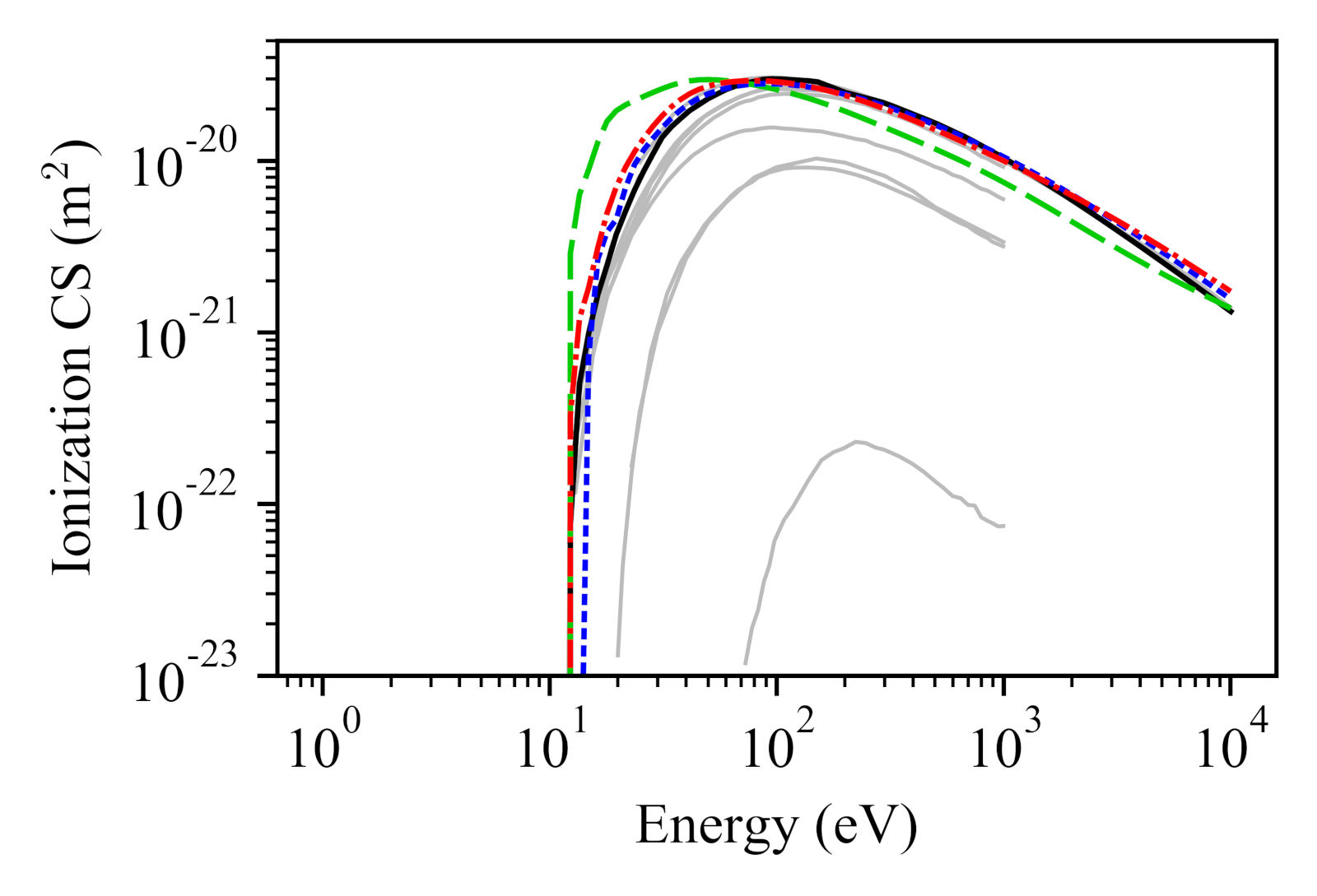}
        \caption{Oxygen (O\textsubscript{2})}
        \label{ionization_O2}
    \end{subfigure}
    \begin{subfigure}{0.5\textwidth}
        \centering
        \includegraphics[width=0.87\linewidth]{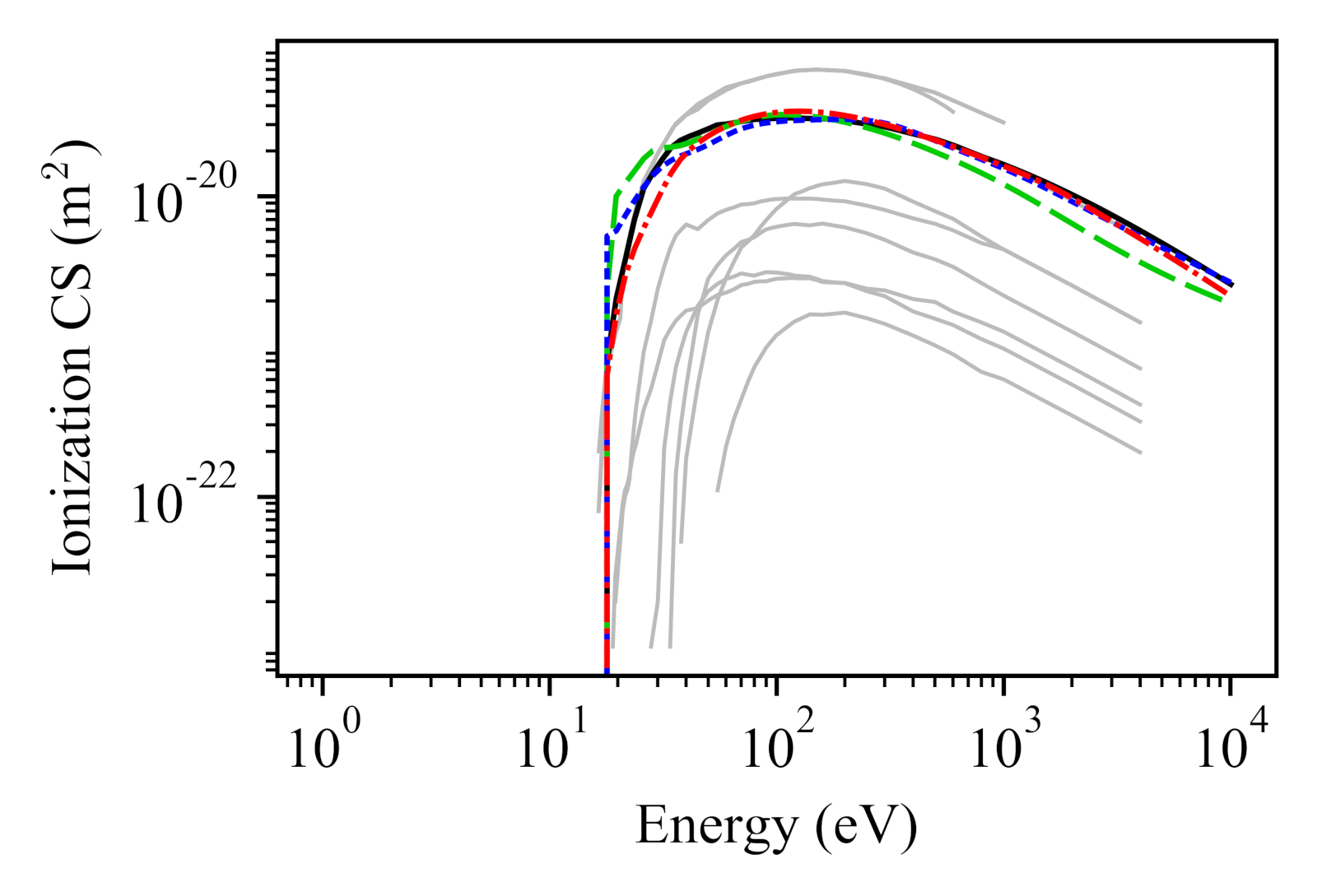}
        \caption{Sulfur hexafluoride (SF\textsubscript{6})}
        \label{ionization_SF6}
    \end{subfigure}
    \begin{subfigure}{0.5\textwidth}
        \centering
        \includegraphics[width=0.875\linewidth]{images/results/legend_vertical_extension.png}
        \label{ionization_legend}
    \end{subfigure}
    \caption{Prediction of ionization cross sections of various gas species. It is to be noted that in some cases ``Other CS data available on LXCat" (shown in grey color) consists of both individual ionization processes as well as sums of all ionization processes. The grey lines simply provides some estimation of the inherent variations in determination of cross sections already available in the literature.}
    \label{results_ionization}
\end{figure}

\begin{figure}
    \begin{subfigure}{0.5\textwidth}
        \centering
        \includegraphics[width=0.927\linewidth]{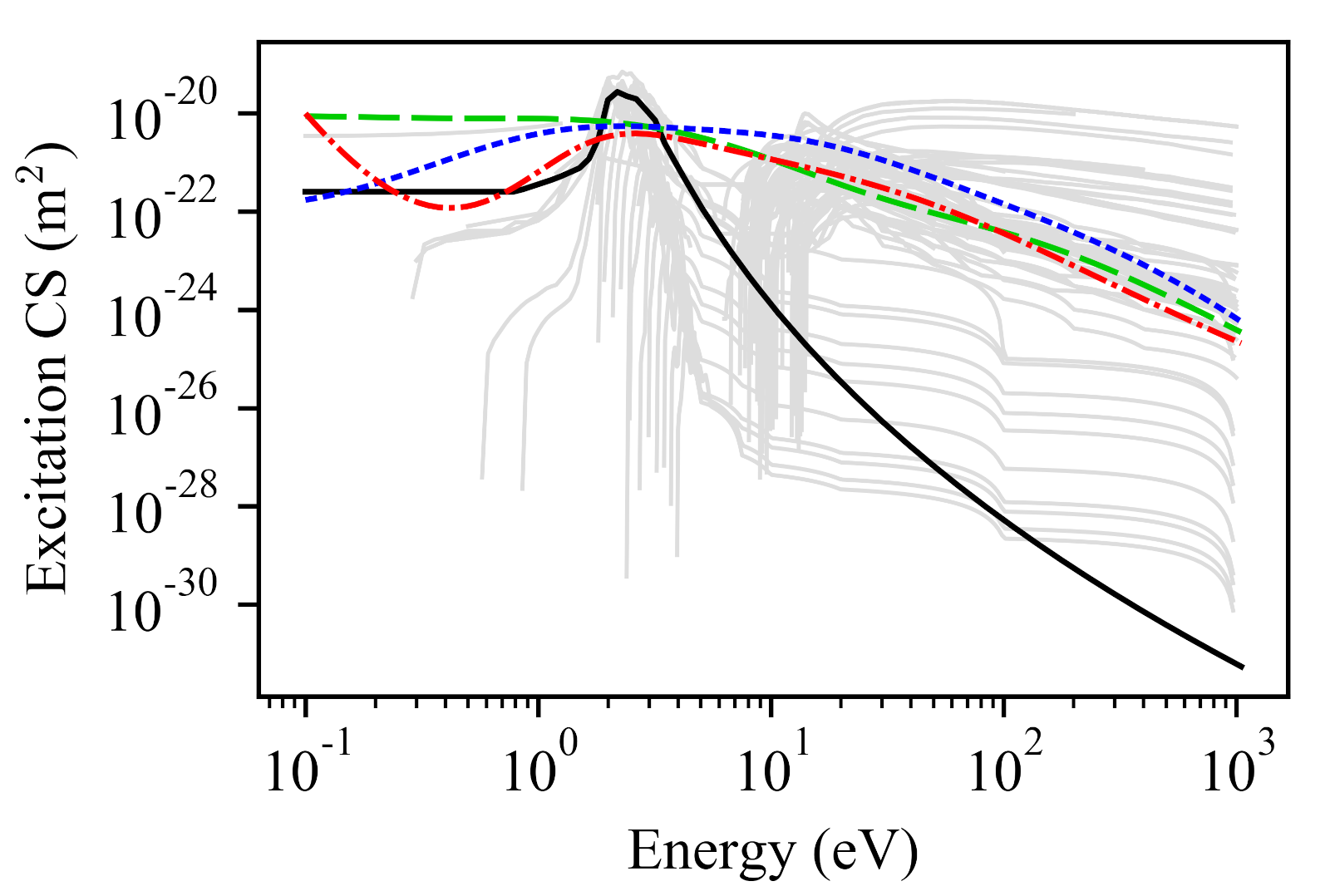}
        \caption{Nitrogen (N\textsubscript{2})}
        \label{excitation_N2}
    \end{subfigure}
    \begin{subfigure}{0.5\textwidth}
        \centering
        \includegraphics[width=0.927\linewidth]{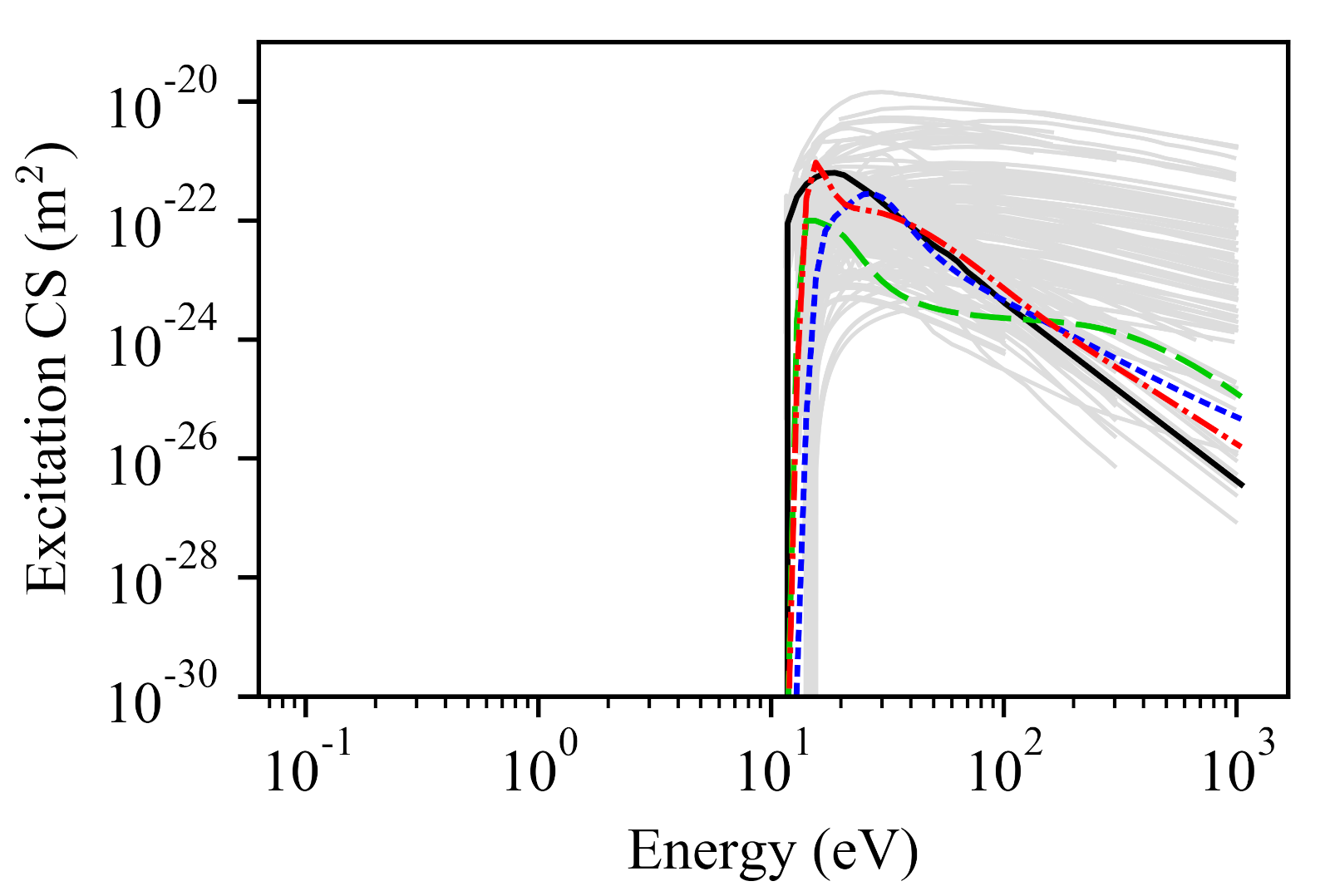}
        \caption{Argon (Ar)}
        \label{excitation_Ar}
    \end{subfigure}
    \begin{subfigure}{0.5\textwidth}
        \centering
        \includegraphics[width=0.927\linewidth]{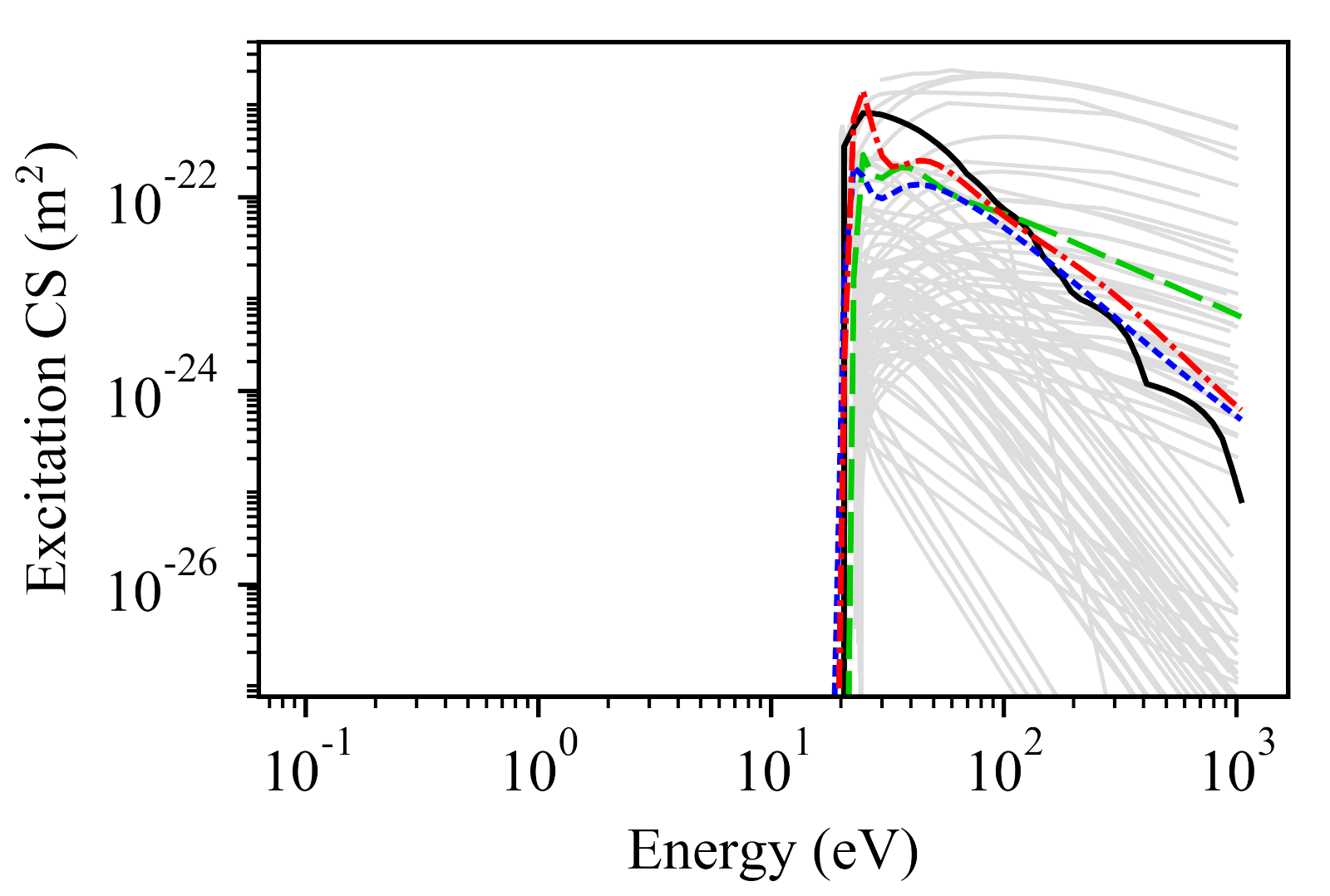}
        \caption{Helium (He)}
        \label{excitation_He}
    \end{subfigure}
    \begin{subfigure}{0.5\textwidth}
        \centering
        \includegraphics[width=0.927\linewidth]{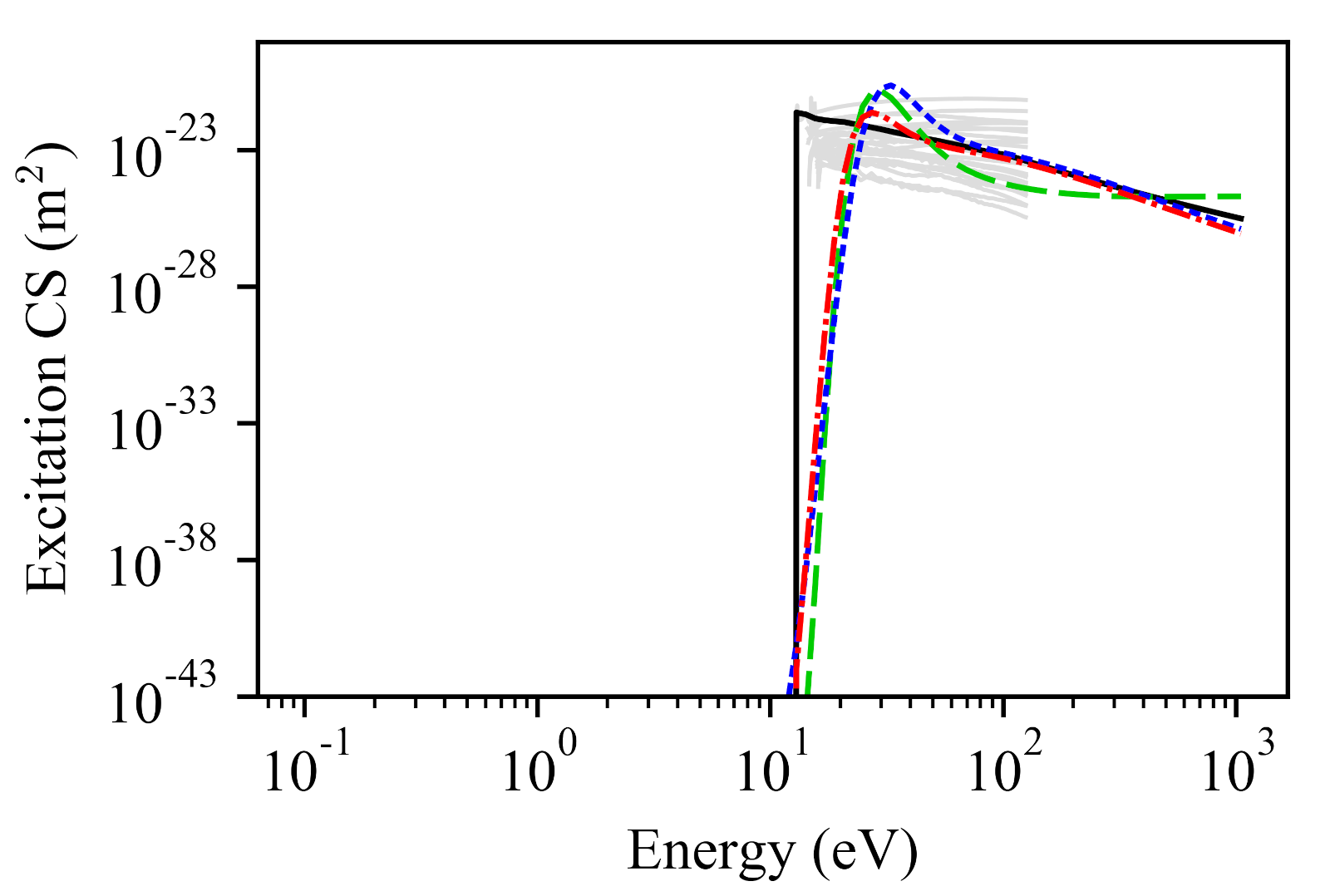}
        \caption{Fluorine (F)}
        \label{excitation_F}
    \end{subfigure}
    \begin{subfigure}{0.5\textwidth}
        \centering
        \includegraphics[width=0.927\linewidth]{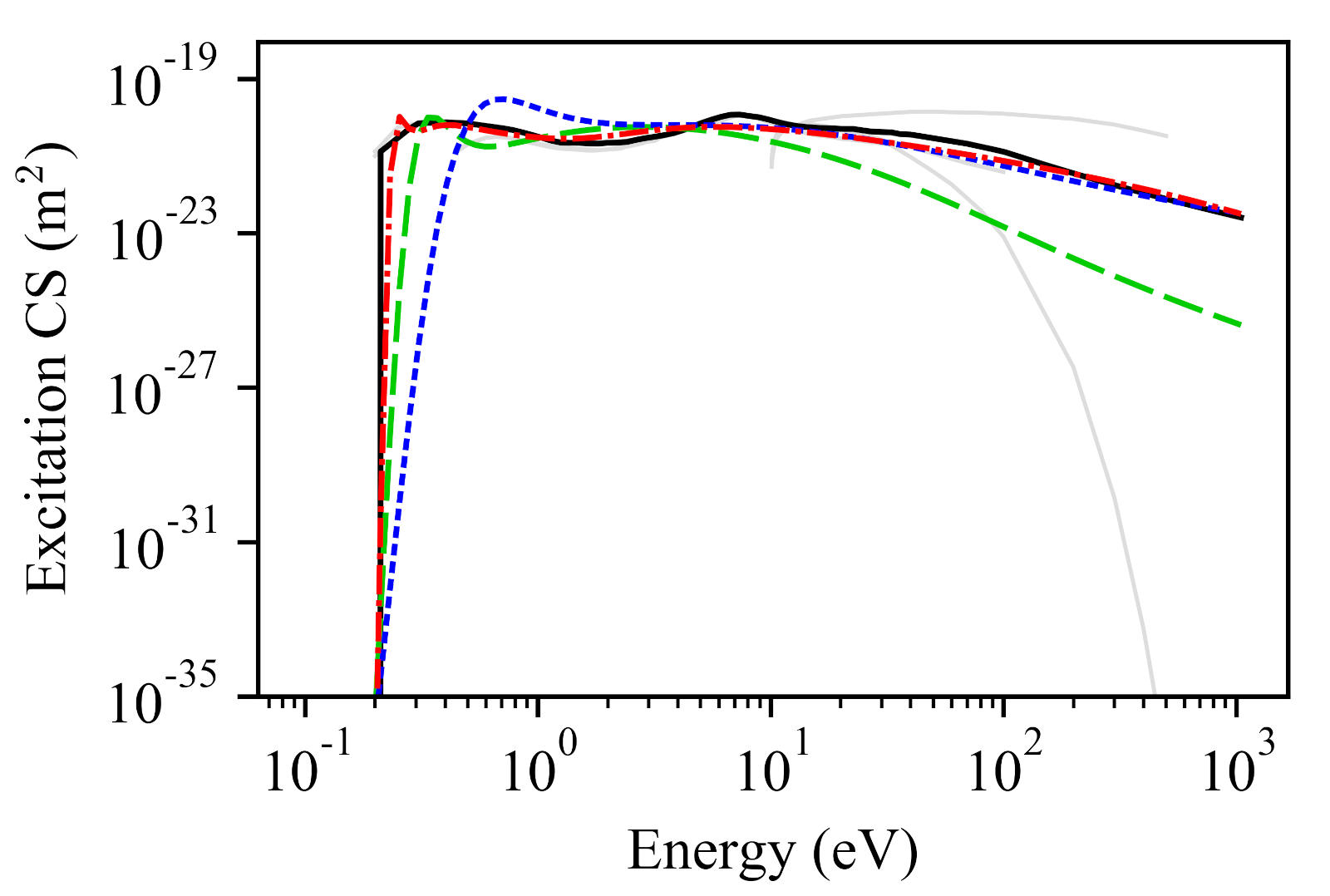}
        \caption{Methane (CH\textsubscript{4})}
        \label{excitation_CH4}
    \end{subfigure}
    \begin{subfigure}{0.5\textwidth}
        \centering
        \includegraphics[width=0.927\linewidth]{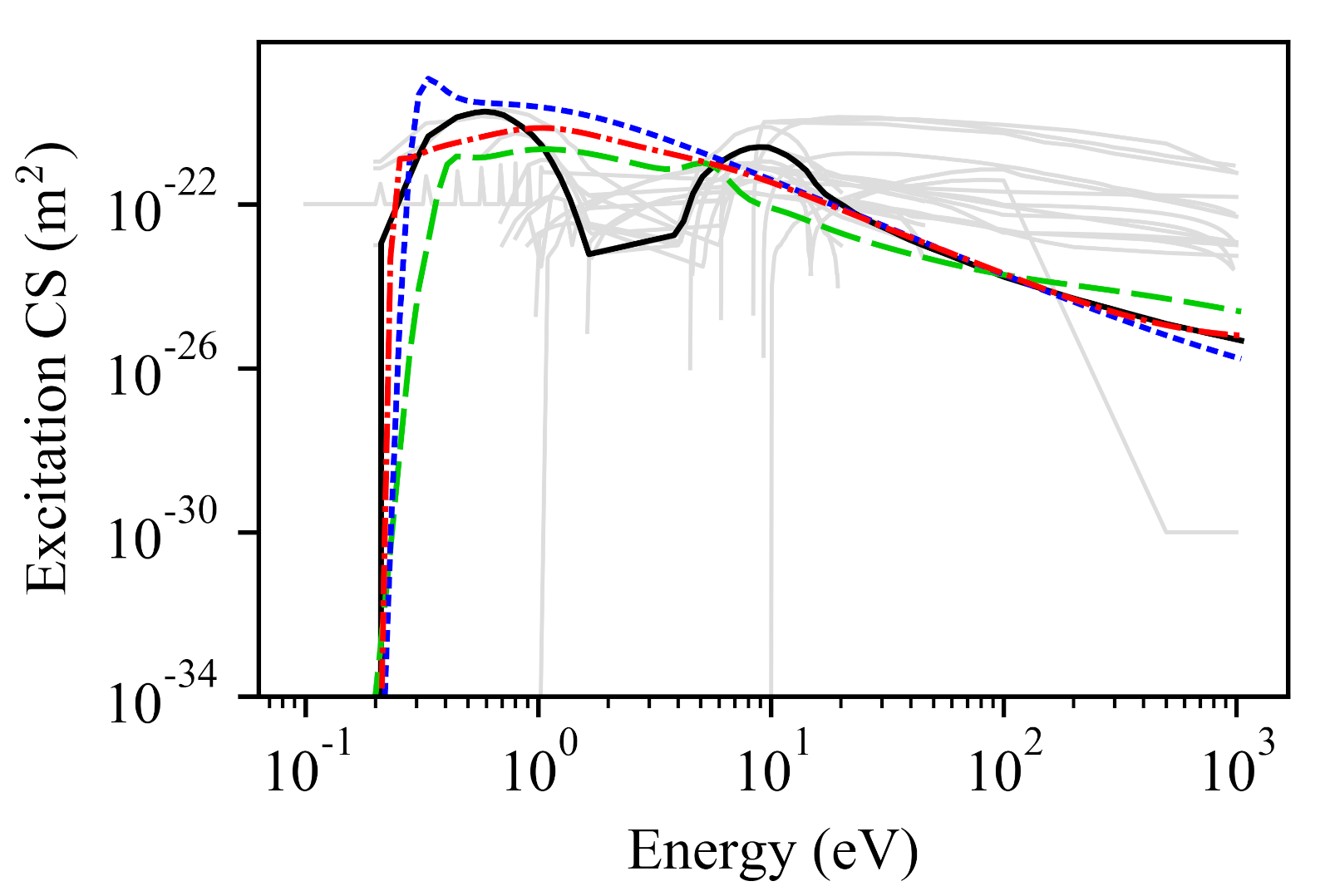}
        \caption{Oxygen (O\textsubscript{2})}
        \label{excitation_O2}
    \end{subfigure}
    \begin{subfigure}{0.5\textwidth}
        \centering
        \includegraphics[width=0.927\linewidth]{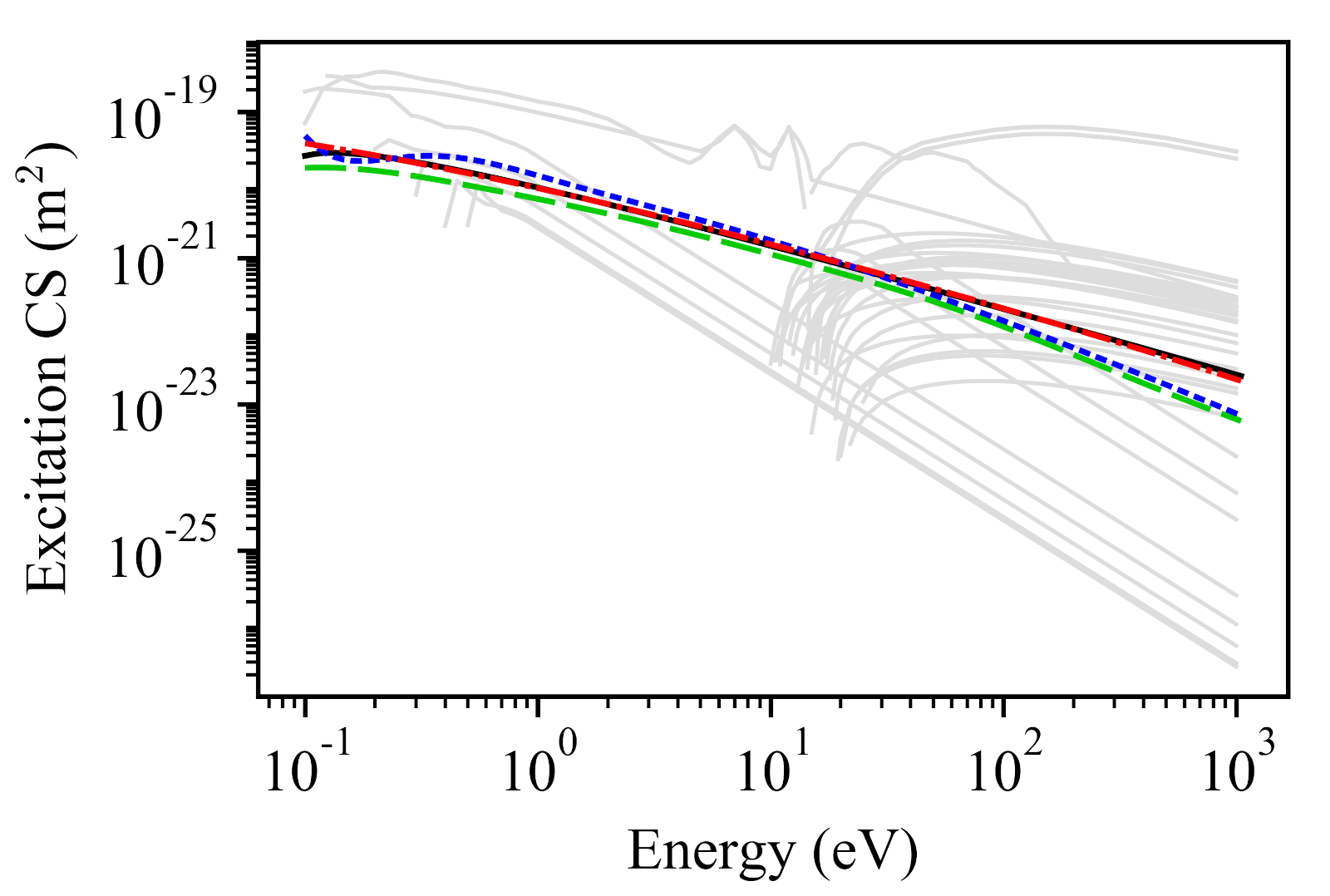}
        \caption{Sulfur hexafluoride (SF\textsubscript{6})}
        \label{excitation_SF6}
    \end{subfigure}
    \begin{subfigure}{0.5\textwidth}
        \centering
        \includegraphics[width=0.875\linewidth]{images/results/legend_vertical_extension.png}
        \label{excitation_legend}
    \end{subfigure}
    \caption{Prediction of excitation cross sections of various gas species}
    \label{results_excitation}
\end{figure}

\begin{figure}
    \begin{subfigure}{0.5\textwidth}
        \centering
        \includegraphics[width=0.927\linewidth]{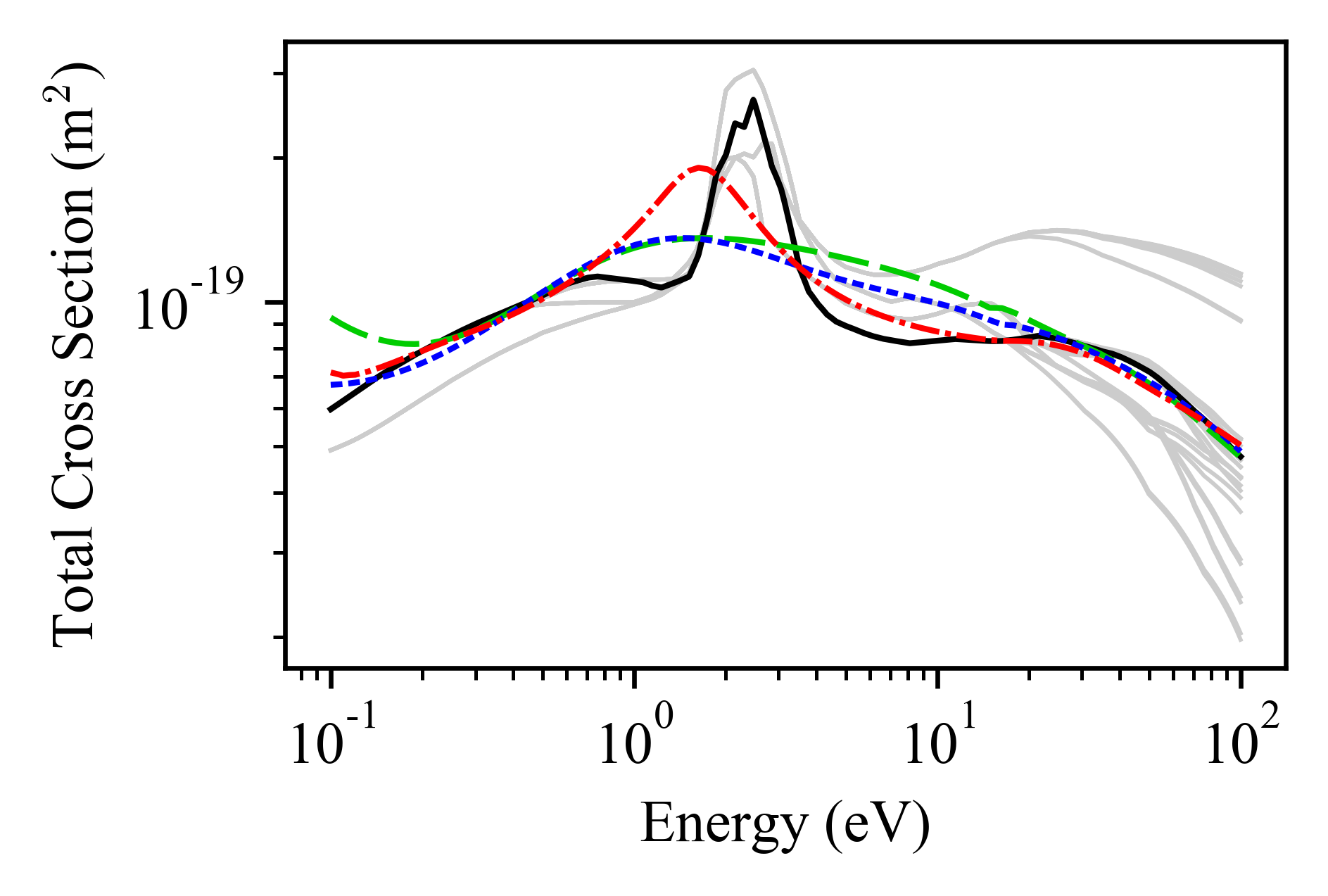}
        \caption{Nitrogen (N\textsubscript{2})}
        \label{total_N2}
    \end{subfigure}
    \begin{subfigure}{0.5\textwidth}
        \centering
        \includegraphics[width=0.927\linewidth]{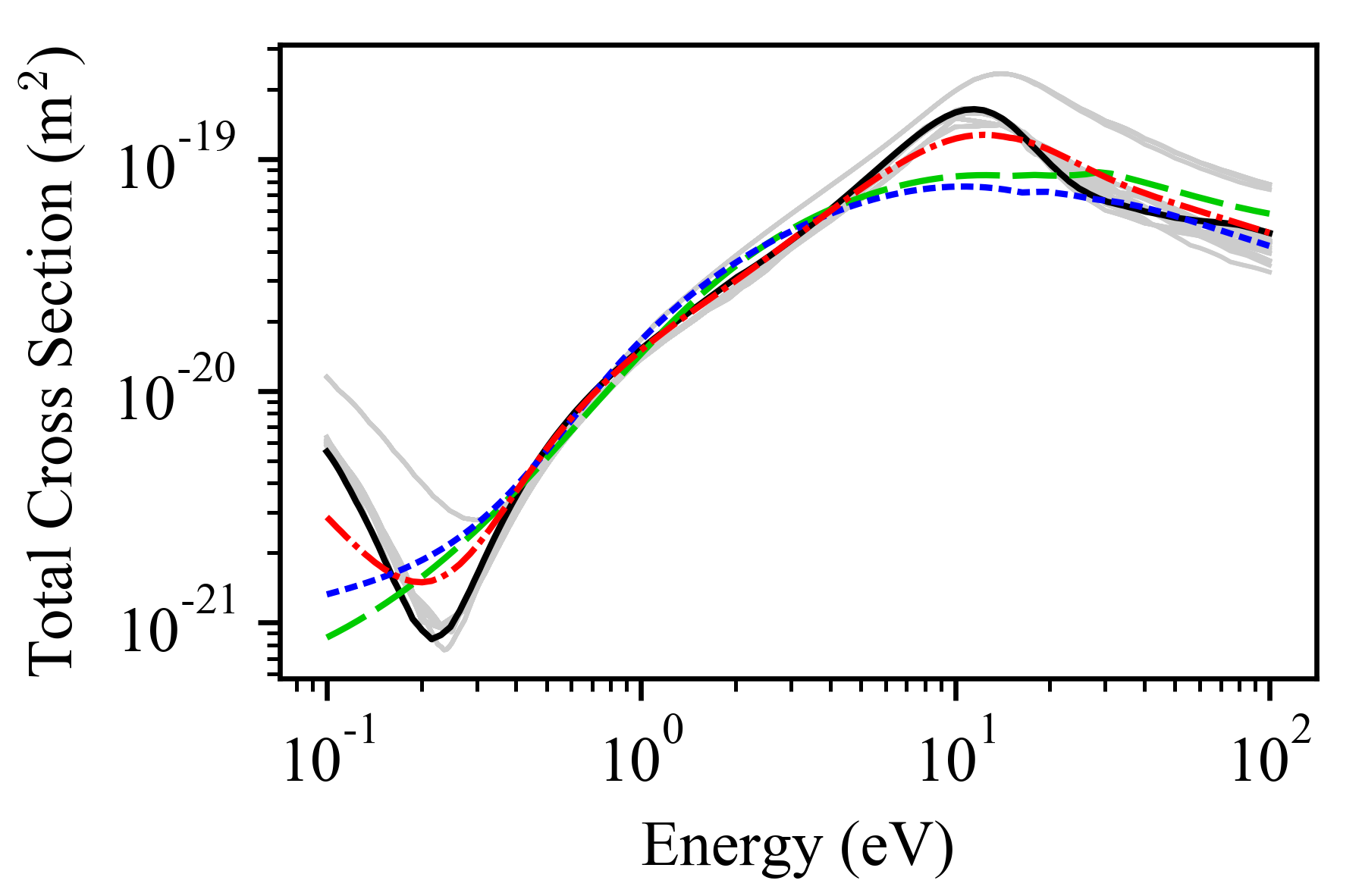}
        \caption{Argon (Ar)}
        \label{total_Ar}
    \end{subfigure}
    \begin{subfigure}{0.5\textwidth}
        \centering
        \includegraphics[width=0.927\linewidth]{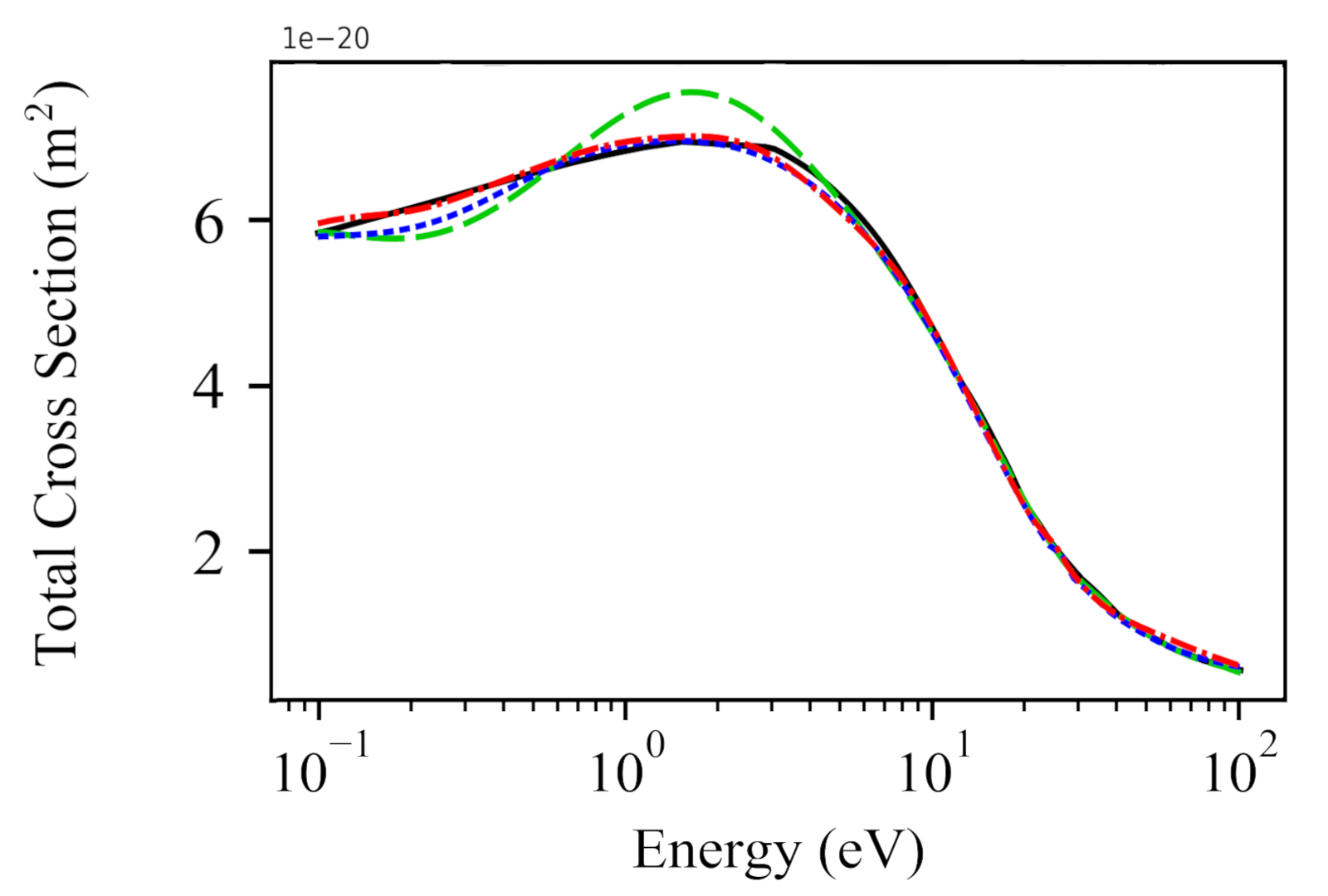}
        \caption{Helium (He)}
        \label{total_He}
    \end{subfigure}
    \begin{subfigure}{0.5\textwidth}
        \centering
        \includegraphics[width=0.927\linewidth]{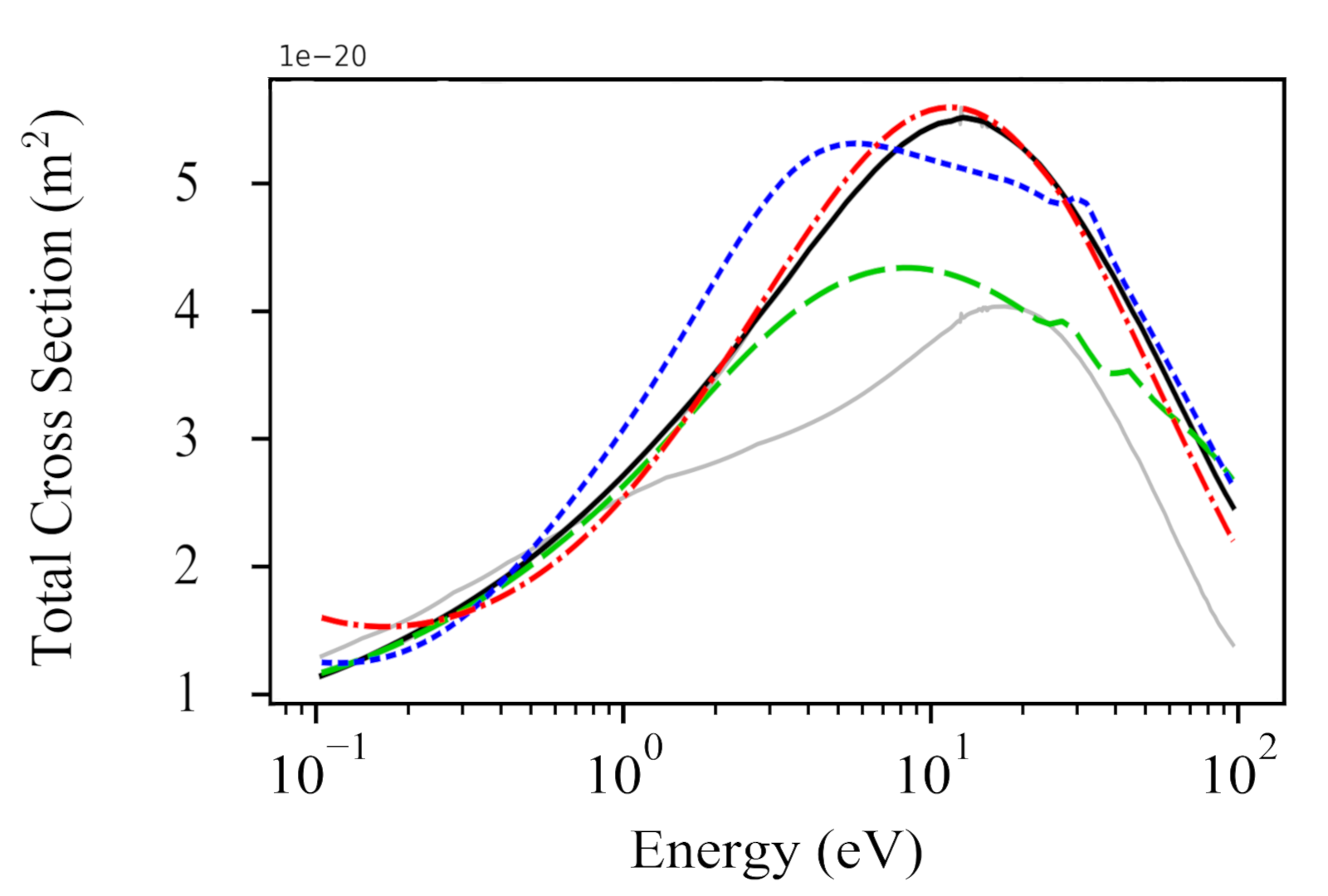}
        \caption{Fluorine (F)}
        \label{total_F}
    \end{subfigure}
    \begin{subfigure}{0.5\textwidth}
        \centering
        \includegraphics[width=0.927\linewidth]{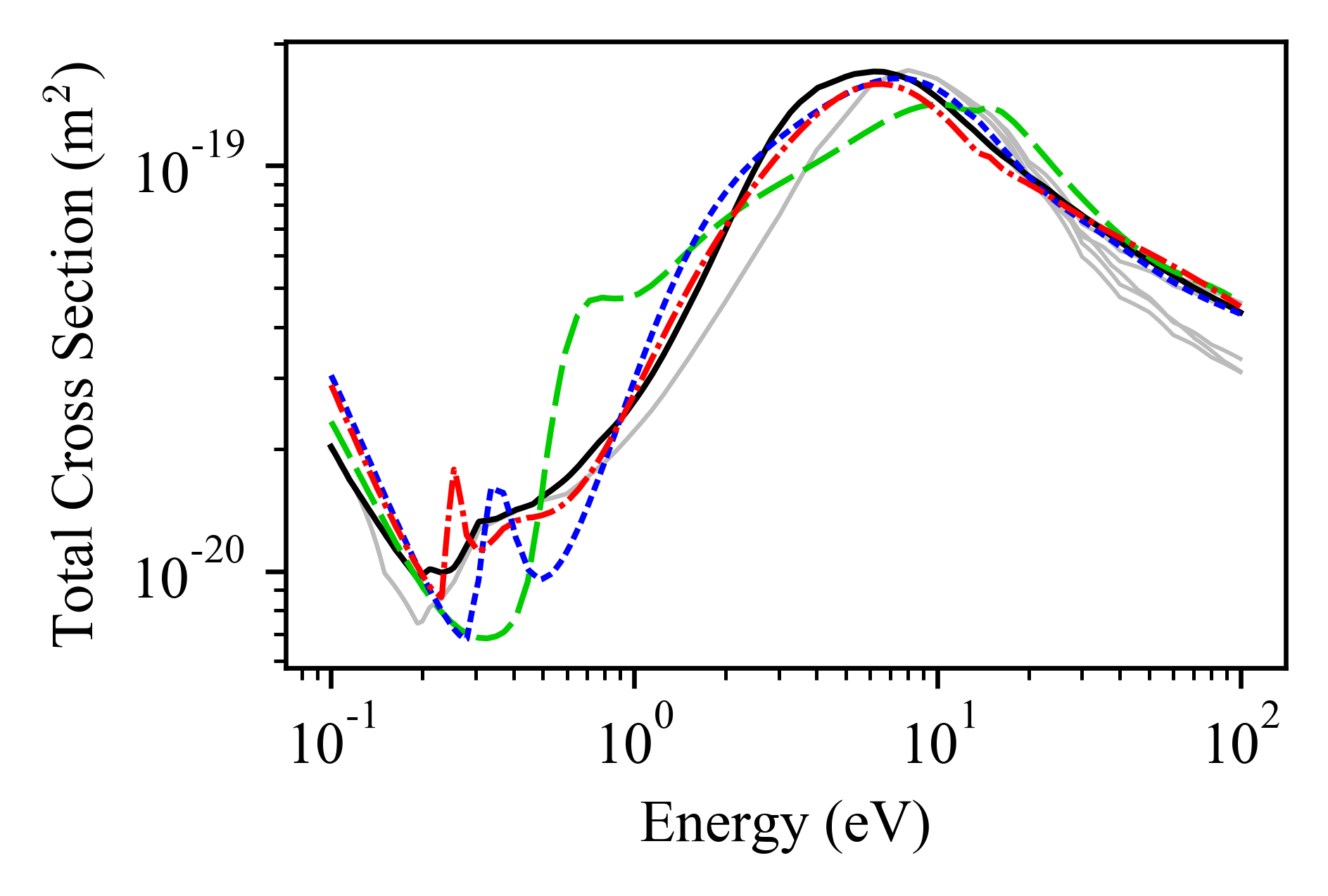}
        \caption{Methane (CH\textsubscript{4})}
        \label{total_CH4}
    \end{subfigure}
    \begin{subfigure}{0.5\textwidth}
        \centering
        \includegraphics[width=0.927\linewidth]{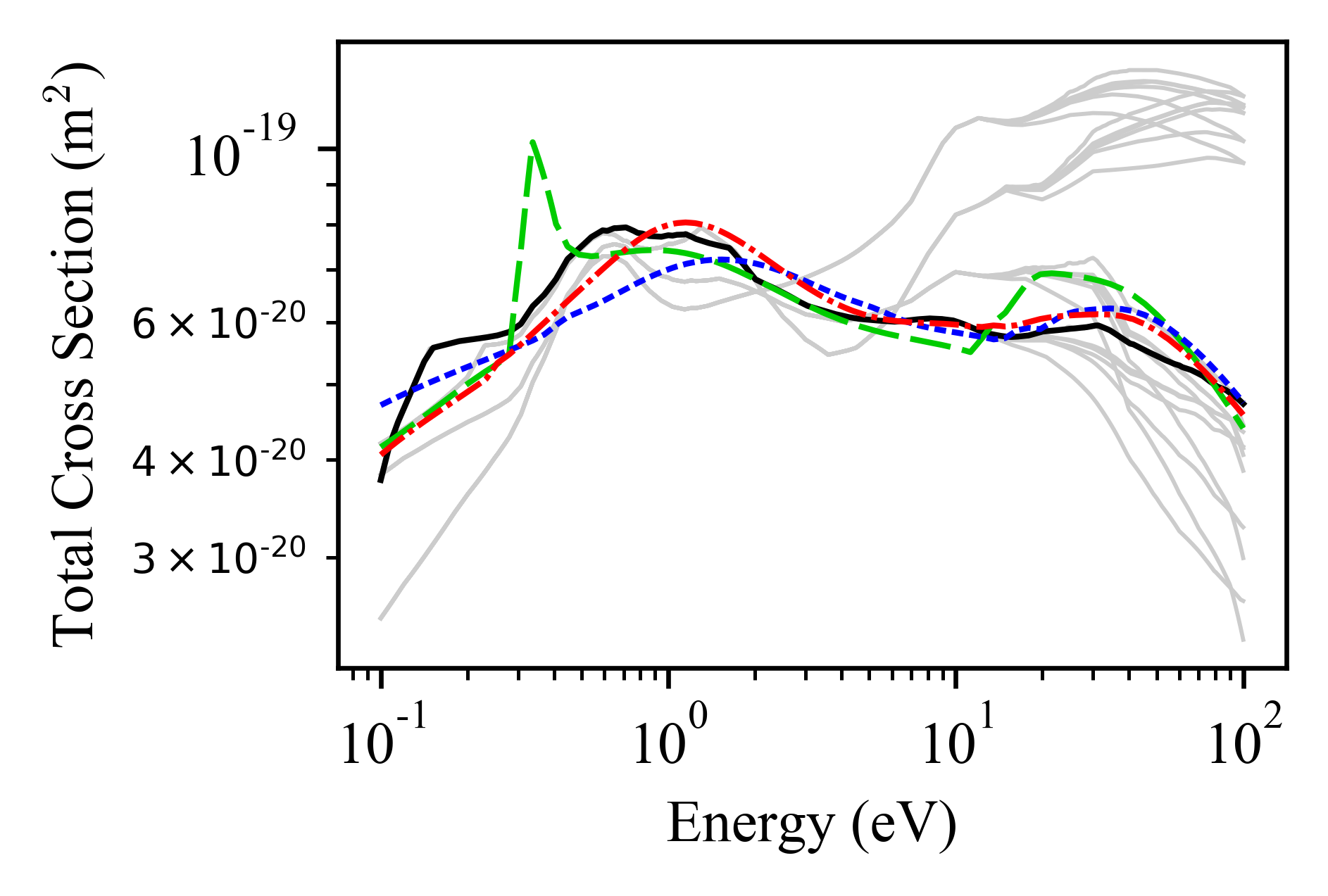}
        \caption{Oxygen (O\textsubscript{2})}
        \label{total_O2}
    \end{subfigure}
    \begin{subfigure}{0.5\textwidth}
        \centering
        \includegraphics[width=0.927\linewidth]{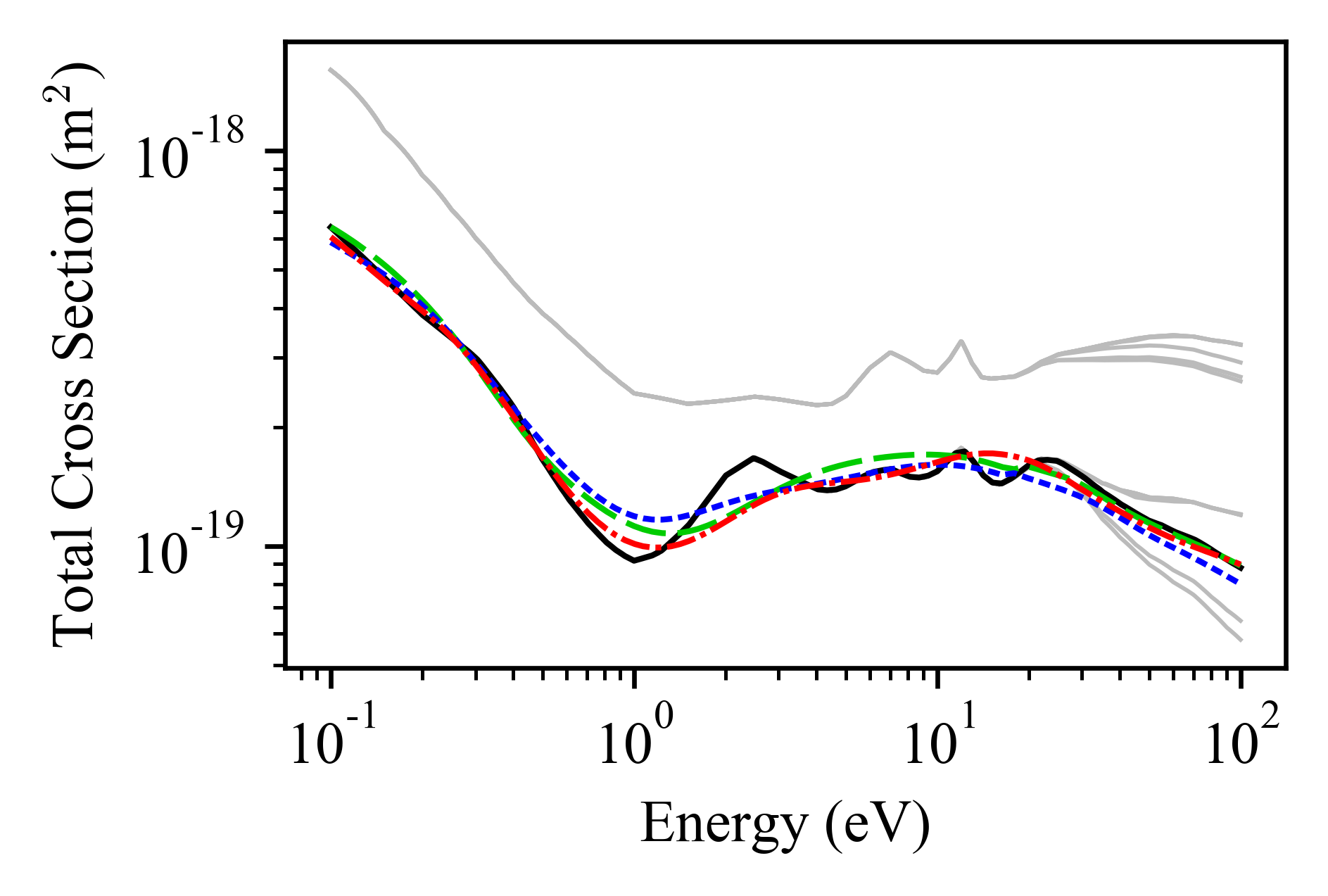}
        \caption{Sulfur hexafluoride (SF\textsubscript{6})}
        \label{total_SF6}
    \end{subfigure}
    \begin{subfigure}{0.5\textwidth}
        \centering
        \includegraphics[width=0.875\linewidth]{images/results/legend_vertical_extension.png}
        \label{total_legend}
    \end{subfigure}
    \caption{Predicted total cross sections of various gas species}
    \label{results_total}
\end{figure}

\begin{figure}
    \begin{subfigure}{0.5\textwidth}
        \centering
        \includegraphics[width=0.985\linewidth]{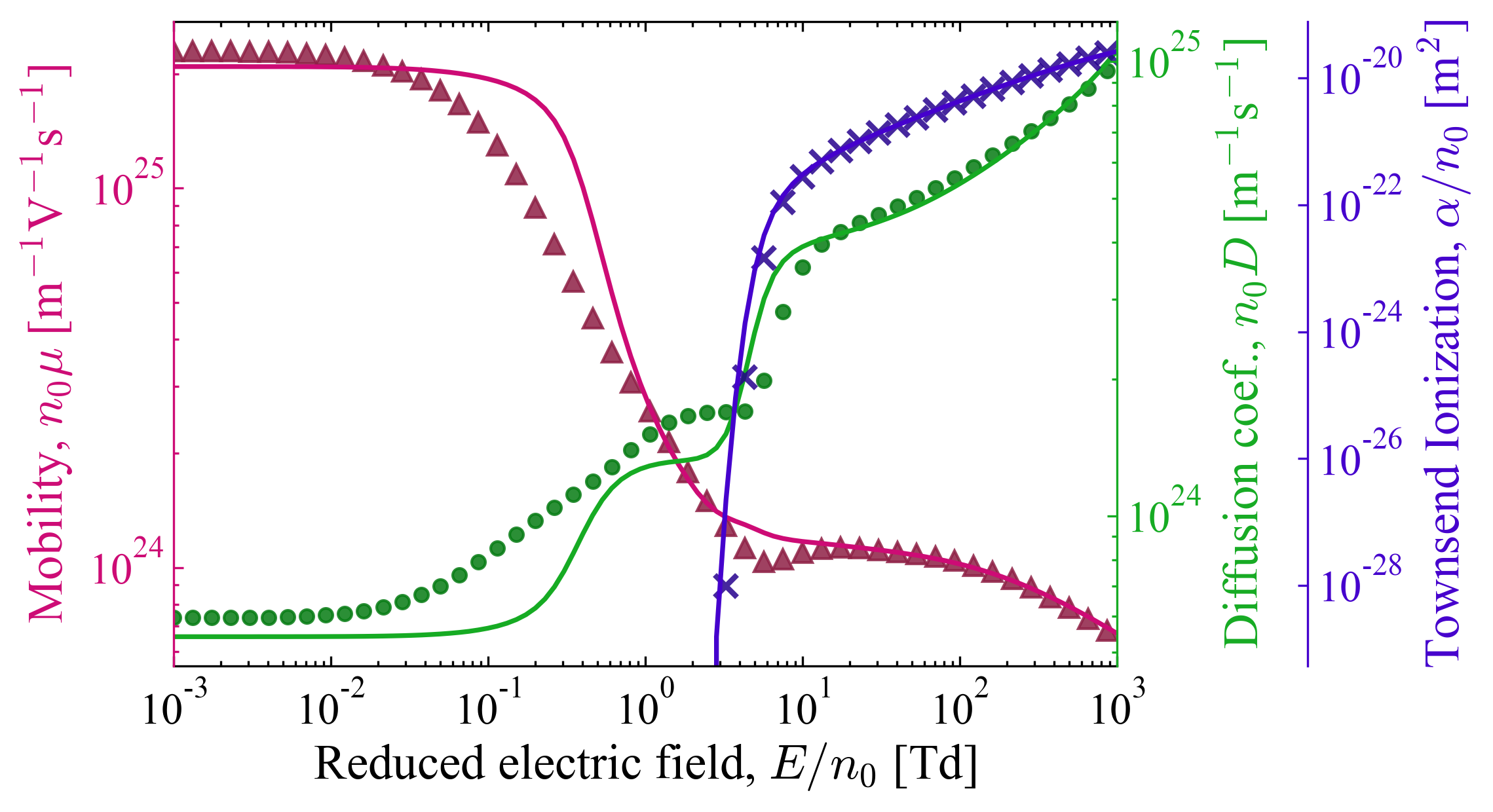}
        \caption{Nitrogen (N\textsubscript{2})}
        \label{swarm_N2}
    \end{subfigure}
    \begin{subfigure}{0.5\textwidth}
        \centering
        \includegraphics[width=0.985\linewidth]{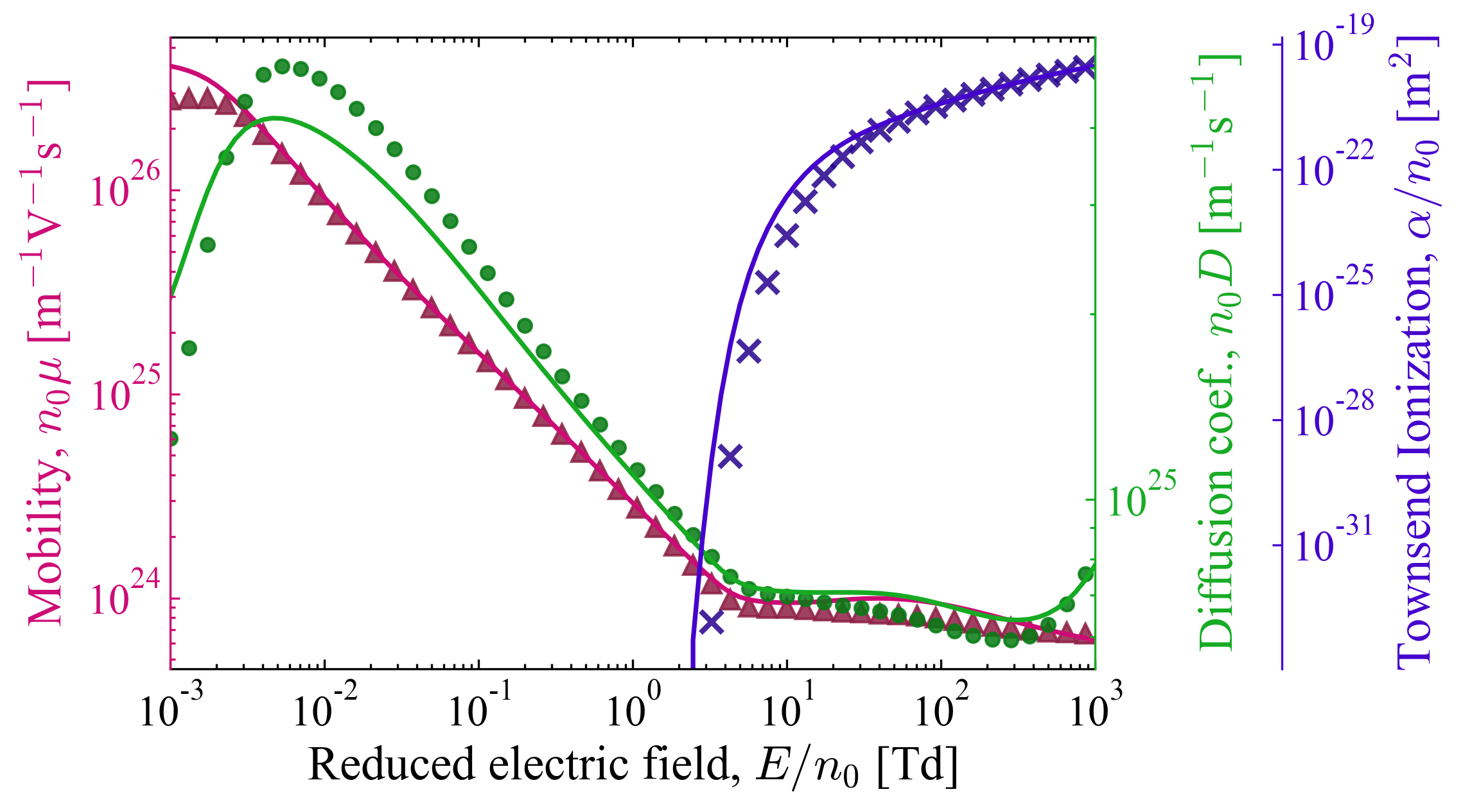}
        \caption{Argon (Ar)}
        \label{swarm_Ar}
    \end{subfigure}
    \begin{subfigure}{0.5\textwidth}
        \centering
        \includegraphics[width=0.985\linewidth]{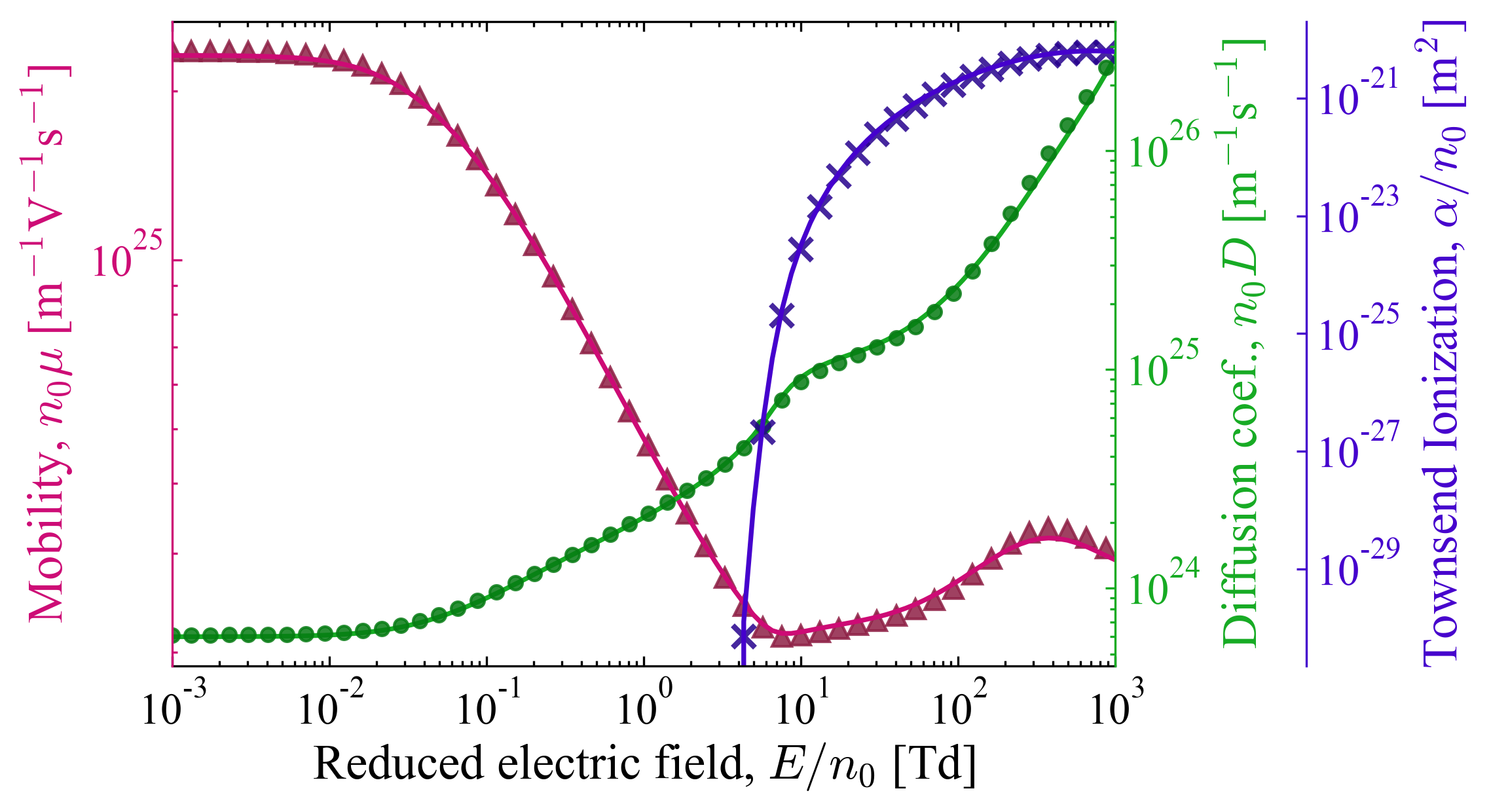}
        \caption{Helium (He)}
        \label{swarm_He}
    \end{subfigure}
    \begin{subfigure}{0.5\textwidth}
        \centering
        \includegraphics[width=0.985\linewidth]{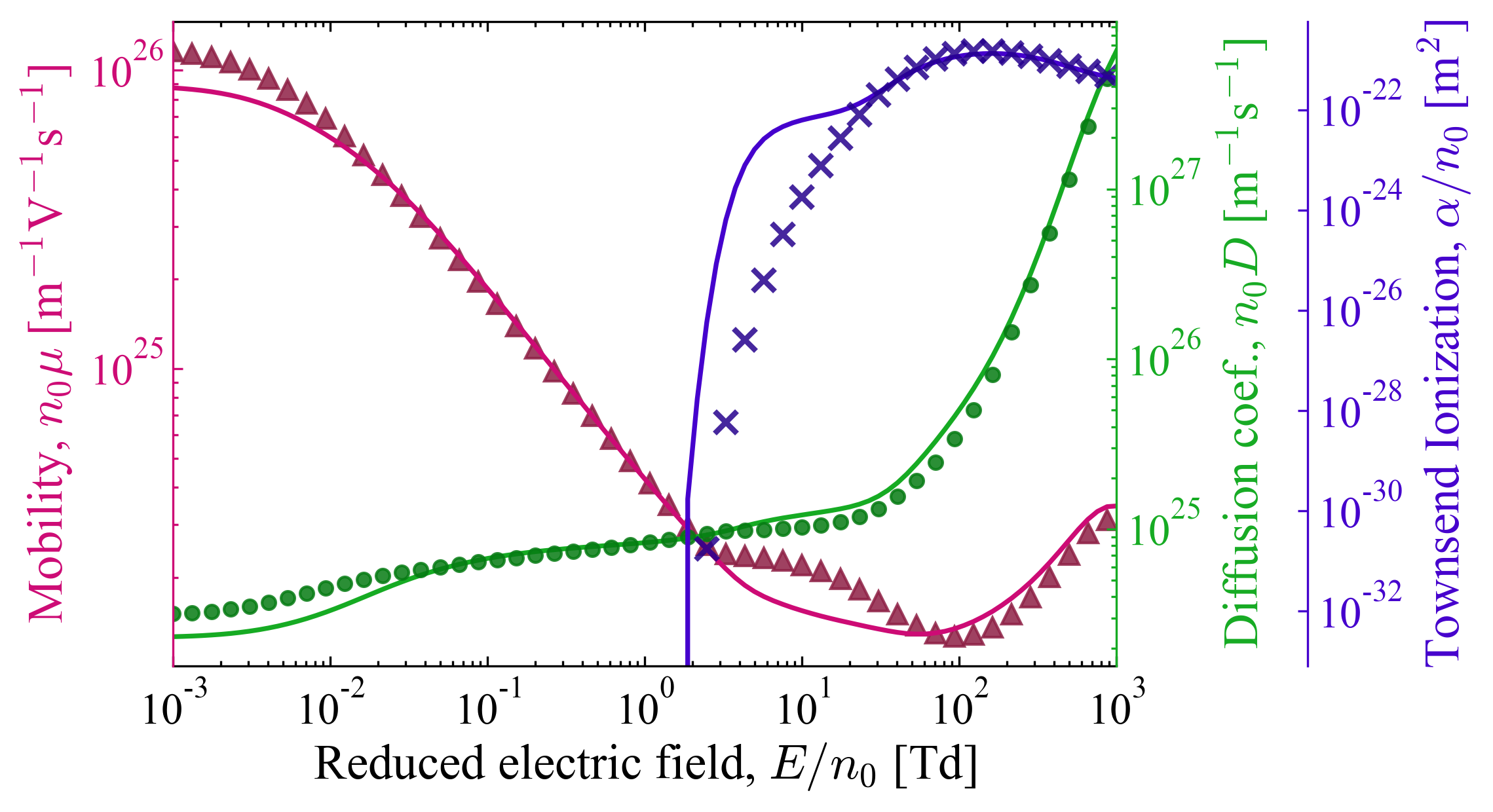}
        \caption{Fluorine (F)}
        \label{swarm_F}
    \end{subfigure}
    \begin{subfigure}{0.5\textwidth}
        \centering
        \includegraphics[width=0.985\linewidth]{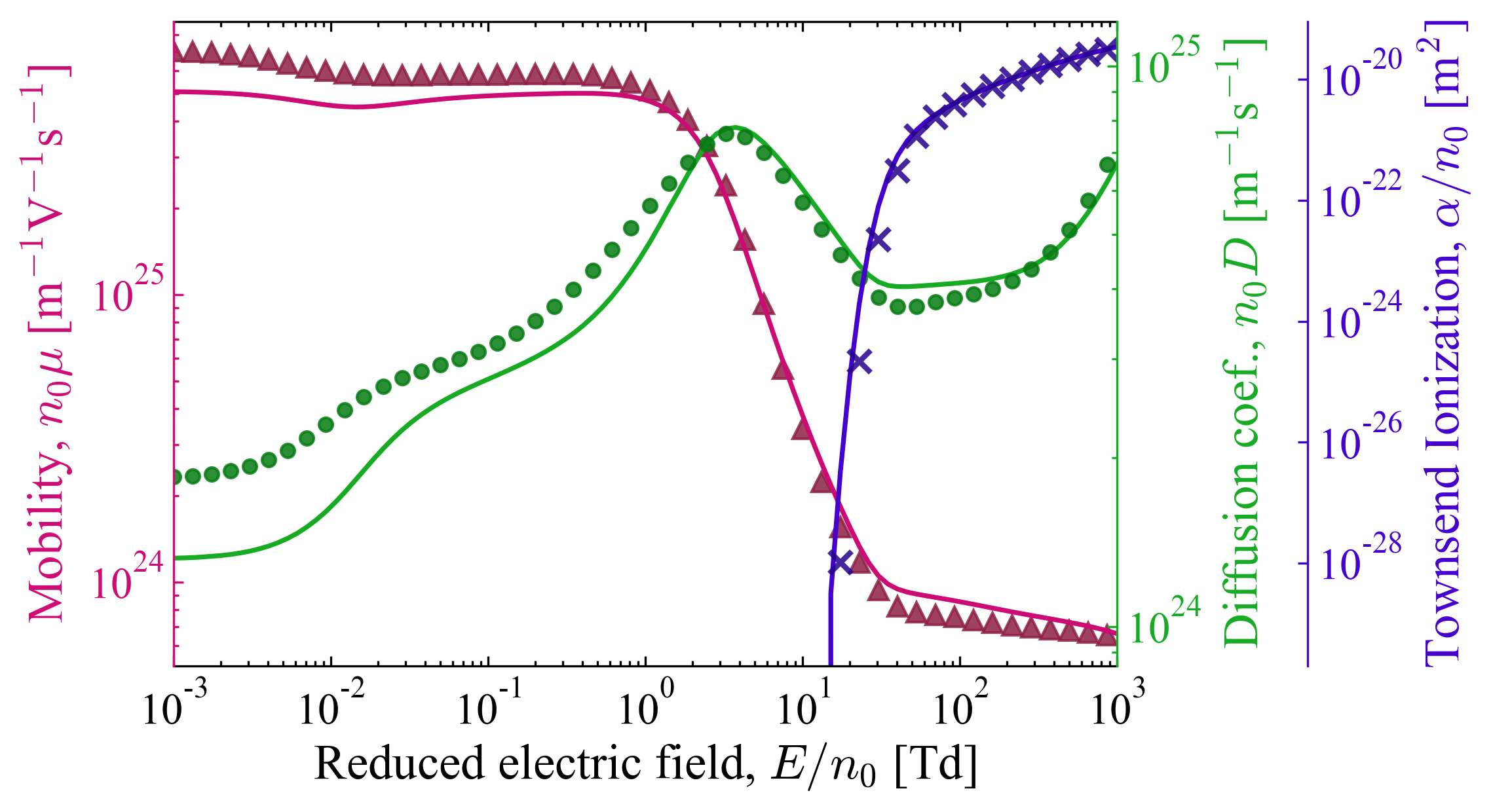}
        \caption{Methane (CH\textsubscript{4})}
        \label{swarm_CH4}
    \end{subfigure}
    \begin{subfigure}{0.5\textwidth}
        \centering
        \includegraphics[width=0.985\linewidth]{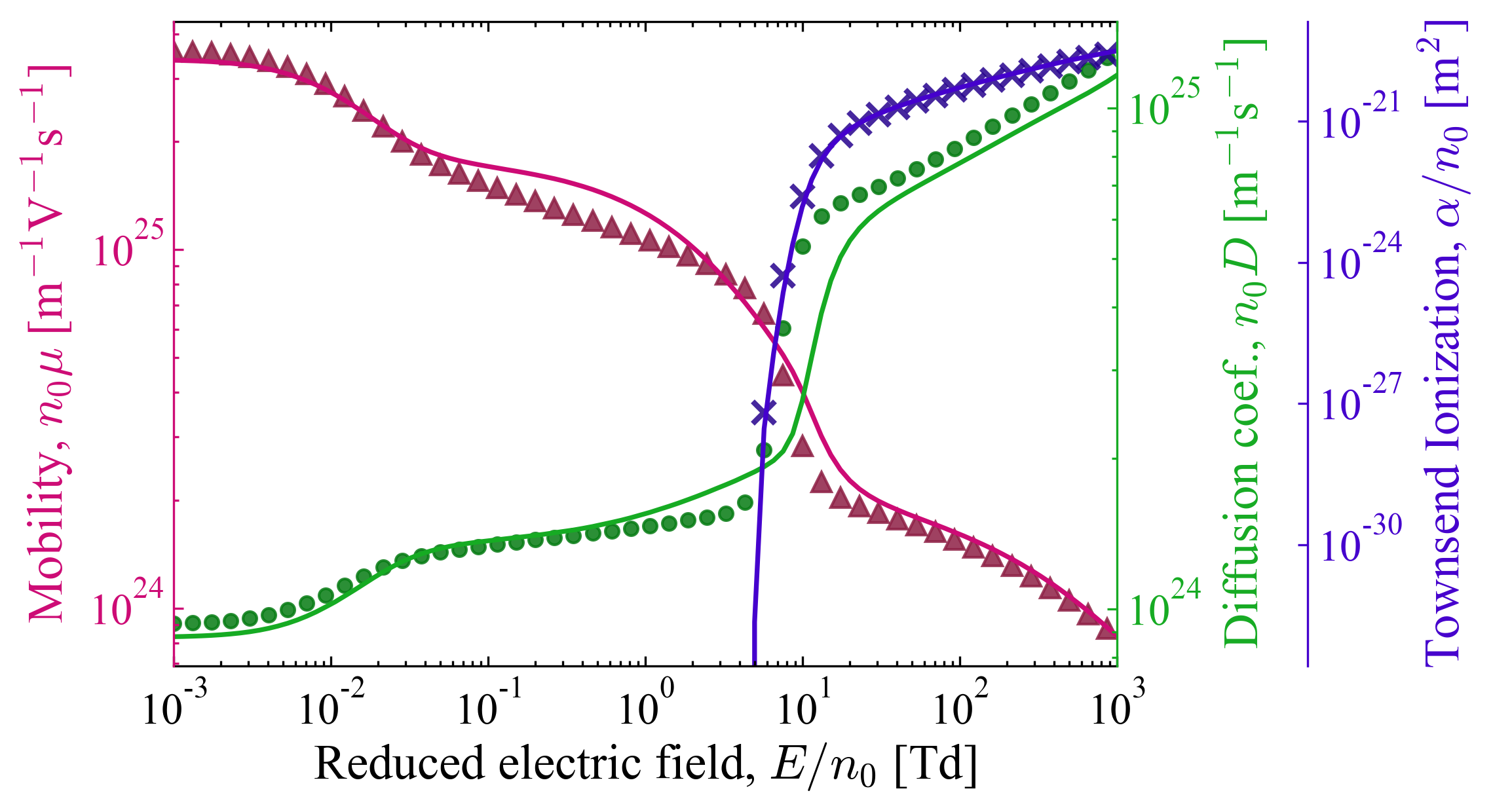}
        \caption{Oxygen (O\textsubscript{2})}
        \label{swarm_O2}
    \end{subfigure}
    \begin{subfigure}{0.5\textwidth}
        \centering
        \includegraphics[width=0.985\linewidth]{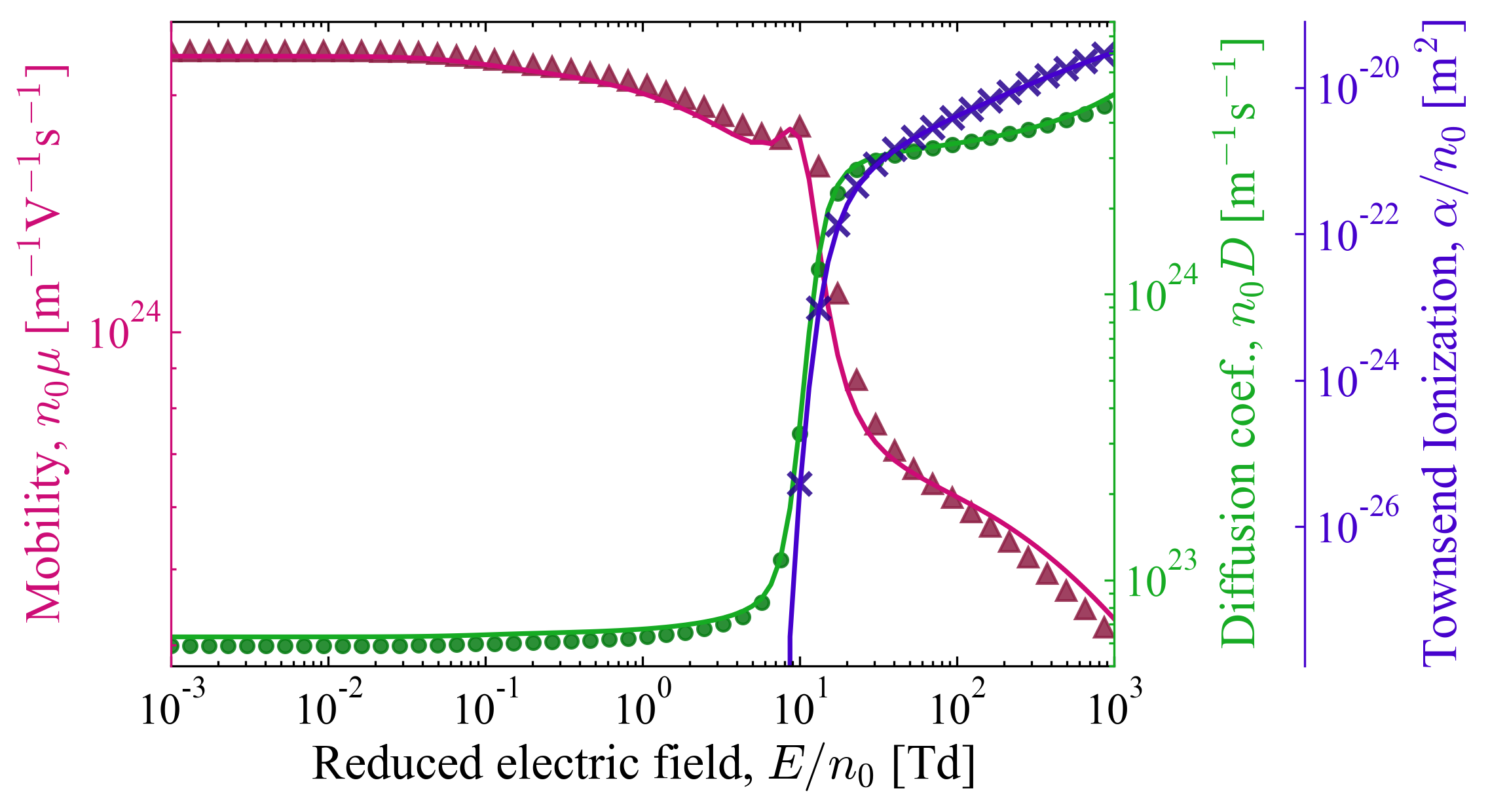}
        \caption{Sulfur hexafluoride (SF\textsubscript{6})}
        \label{swarm_SF6}
    \end{subfigure}
    \begin{subfigure}{0.5\textwidth}
        \centering
        \includegraphics[width=1\linewidth]{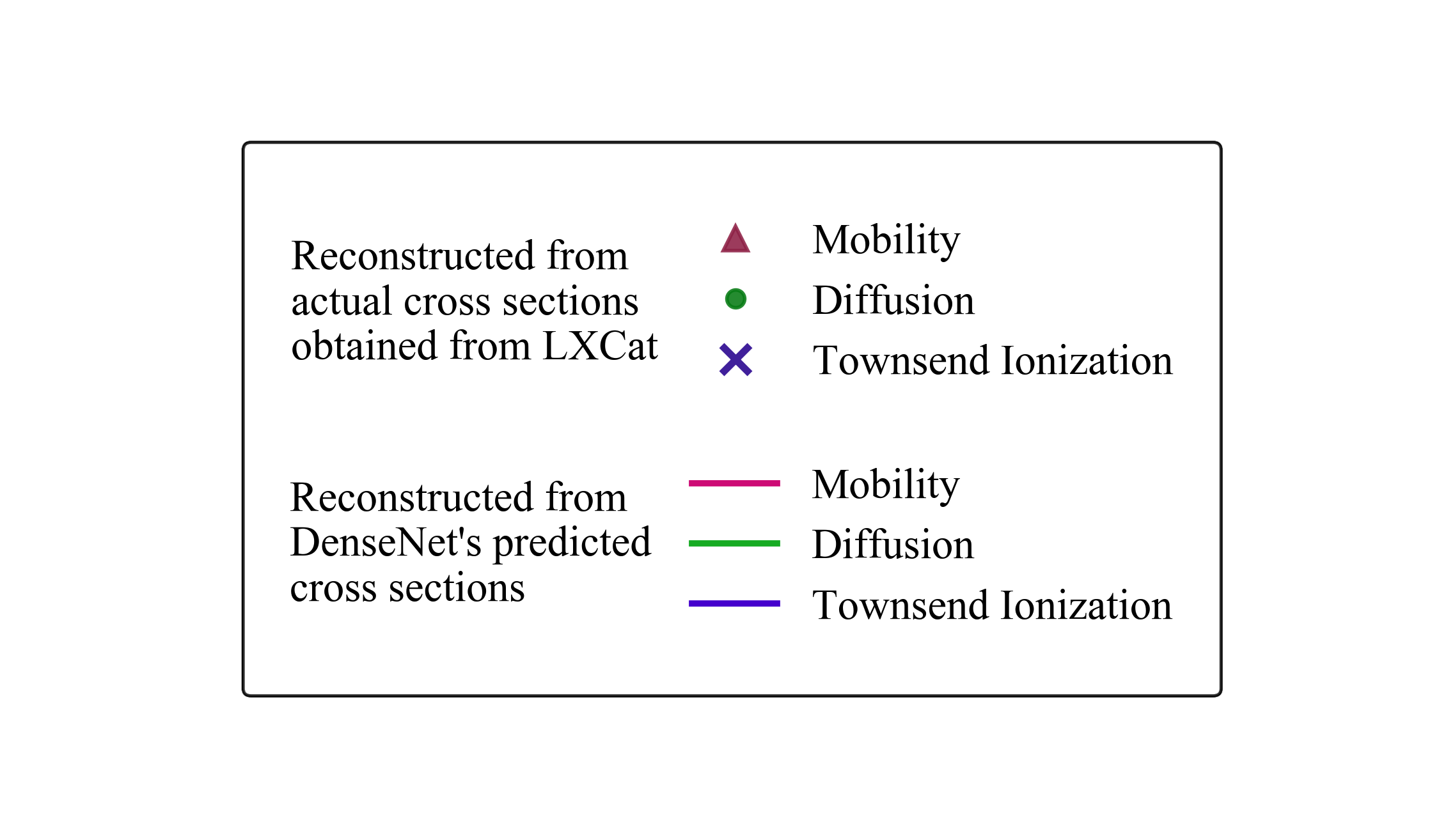}
        \label{swarm_legend}
    \end{subfigure}
    \caption{Comparison of swarm parameters reconstructed using - actual cross sections available on LXCat vs. DenseNet's predicted cross sections}
    \label{results_swarmimages}
\end{figure}

Fig.~\ref{results_elastic}, \ref{results_ionization} and \ref{results_excitation} shows the comparison between different architecture's estimate of elastic momentum transfer, ionization and excitation cross sections respectively, with the cross sections available on LXCat. For gas species N\textsubscript{2}, Ar, He, O\textsubscript{2} and SF\textsubscript{6}, cross sections (actual cs depicted in the Figs.~\ref{results_elastic}, \ref{results_ionization} and \ref{results_excitation} with black line) are sourced from the Biagi database~\cite{database2}, while those of F and CH\textsubscript{4} are taken from BSR~\cite{database4} and Hayashi~\cite{database9} database respectively. These cross sections were used to generate the simulated swarm data of these gas species using BOLSIG+, which were then used as the input to our trained model to predict the cross sections. Cross sections, sourced from other databases available on the LXCat, are plotted on the same graph just to give an estimate of the inherent variations in the values of cross sections available in the literature from past research works. Fig.~\ref{results_total} shows the similar comparison for the total cross sections, which is calculated by summing elastic momentum transfer, ionization and excitation cross sections. The predicted cross sections are again used to calculate the corresponding swarm coefficients using the BOLSIG+ solver and its comparison with the swarm coefficients calculated using the actual cross sections is shown in Fig.~\ref{results_swarmimages}

As evident from figures~\ref{results_elastic} and \ref{results_ionization}, prediction of both elastic momentum transfer and ionization cross sections, from all the three neural network architectures, agrees reasonably well over the entire energy range with the experimentally measured cross sections obtained from LXCat. Further, to quantitatively compare the performance of different architectures, we use three different metrics: mean absolute error (log-normalized scale), coefficient of determination ($R^2$) and Mean Absolute Relative Percentage Difference (\textit{MARPD})
\begin{equation}
    \mathit{MARPD} = \frac{1}{N}\sum_{i=1}^{N}\left| 100 \times \frac{y_i - \hat{y_i}}{|y_i|+|\hat{y_i}|}\right|
    \label{marpd}
\end{equation}
where $N$ is the number of data points, $y_i$ is predicted value and $\hat{y_i}$ is the true value. Mean absolute error on log-normalized scale depicts the error as seen by the model (test loss). Coefficient of determination ($R^2$) quantifies the degree of correlation between the actual and the predicted values. Its value lies between $(-\infty,\,1]$, with 1 representing complete dependency between the quantities being compared. Mean absolute relative percentage difference provides a standardized error value, which is not only comparable but also more interpretable even to those unfamiliar with the measurement scale of electron cross sections. These three metrics collectively provide a better understanding of network's performance compared to what a single metric alone provides. 

\newcommand\T{\rule{0pt}{3ex}}         
\newcommand\B{\rule[-1.2ex]{0pt}{0pt}}   
\renewcommand{\arraystretch}{1.05}
\begin{table}
    \centering
    \resizebox{\textwidth}{!}{%
    \begin{tabular}{|c|c|c c c|c c c|c c c|}
        \hline
        \multirow{2}{*}{Species} & \multirow{2}{*}{Cross section} & \multicolumn{3}{c|}{ANN*} & \multicolumn{3}{c|}{CNN} & \multicolumn{3}{c|}{DenseNet}\T\B\\
        
        \cline{3-11}
        
        & & MAE & R$^2$ & MARPD & MAE & R$^2$ & MARPD & MAE & R$^2$ & MARPD \B\T\\
        
        \hline

        \multirow{3}{*}{N\textsubscript{2}} 
        & Elastic & $0.0285$ & $0.578$ & $7.17\%$ & $0.0224$ & $0.615$ & $5.65\%$ & $0.0186$ & $0.637$ & $4.67\%$  \T\\
        & Ionization & $0.0164$ & $0.991$ & $5.65\%$ & $0.0083$ & $0.991$ & $3.04\%$ & $0.0080$ & $0.996$ & $2.99\%$ \\
        
        
        & Total & $0.0302$ & $0.468$ & $7.57\%$ & $0.0239$ & $0.504$ & $5.99\%$ & $0.0205$ & $0.567$ & $5.16\%$ \B\\
        
        \hline
        
        \multirow{3}{*}{Ar} & Elastic & $0.0661$ & $0.724$ & $15.95\%$ & $0.0584$ & $0.662$ & $14.23\%$ & $0.0315$ & $0.931$ & $7.90\%$  \T\\
        & Ionization & $0.0407$ & $0.867$ & $14.99\%$ & $0.0165$ & $0.968$ & $5.52\%$ & $0.0079$ & $0.994$ & $2.93\%$ \\
        
        
        & Total & $0.0597$ & $0.722$ & $14.46\%$ & $0.0551$ & $0.659$ & $13.46\%$ & $0.0274$ & $0.935$ & $6.88\%$ \B \\
        
        \hline
        
        \multirow{3}{*}{He} & Elastic & $0.0067$ & $0.986$ & $1.70\%$ & $0.0048$ & $0.997$ & $1.21\%$ & $0.0032$ & $0.999$ & $0.81\%$  \T\\
        & Ionization & $0.0125$ & $0.970$ & $4.67$\% & $0.0081$ & $0.989$ & $2.62\%$ & $0.0085$ & $0.975$ & $3.17\%$ \\
        
        
        & Total & $0.0062$ & $0.985$ & $1.58\%$ & $0.0043$ & $0.997$ & $1.09\%$ & $0.0045$ & $0.998$ & $1.15\%$ \B\\
        
        \hline
        
        \multirow{3}{*}{F} & Elastic & $0.0205$ & $0.803$ & $5.18\%$ & $0.0143$ & $0.931$ & $3.62\%$ & $0.0104$ & $0.986$ & $2.65\%$  \T\\
        & Ionization & $8.8971$ & $-964$ & $70.51\%$ & $0.1307$ & $-10.7$ & $35.65\%$ & $0.0411$ & $0.836$ & $12.97\%$ \\
        
        
        & Total & $0.0189$ & $0.814$ & $4.78\%$ & $0.0148$ & $0.929$ & $3.74\%$ & $0.0102$ & $0.987$ & $2.56\%$ \B\\
        
        \hline
        
        \multirow{3}{*}{CH\textsubscript{4}} & Elastic & $0.0293$ & $0.872$ & $7.39\%$ & $0.0198$ & $0.978$ & $5.01\%$ & $0.0165$ & $0.980$ & $4.17\%$  \T\\
        & Ionization & $0.0332$ & $0.902$ & $11.29\%$ & $0.0139$ & $0.995$ & $4.87\%$ & $0.0180$ & $0.953$ & $6.56\%$ \\
        
        
        & Total & $0.0519$ & $0.833$ & $12.85\%$ & $0.0276$ & $0.978$ & $6.95\%$ & $0.0183$ & $0.978$ & $4.64\%$ \B\\
        
        \hline
        
        \multirow{3}{*}{O\textsubscript{2}} & Elastic & $0.0129$ & $0.889$ & $3.28\%$ & $0.0104$ & $0.948$ & $2.61\%$ & $0.0079$ & $0.946$ & $2.01\%$  \T\\
        & Ionization & $0.0630$ & $0.719$ & $20.04\%$ & $0.4706$ & $0.991$ & $7.46\%$ & $0.0252$ & $0.980$ & $8.11\%$ \\
        
        
        & Total & $0.0165$ & $-0.281$ & $4.16\%$ & $0.0137$ & $0.619$ & $3.47\%$ & $0.0112$ & $0.773$ & $2.84\%$ \B\\
        
        \hline
        
        \multirow{3}{*}{SF\textsubscript{6}} & Elastic & $0.0178$ & $0.973$ & $4.50\%$ & $0.0169$ & $0.980$ & $4.27\%$ & $0.0125$ & $0.986$ & $3.17\%$  \T\\
        & Ionization & $0.0385$ & $0.951$ & $13.85\%$ & $0.0173$ & $0.967$ & $6.05\%$ & $0.0156$ & $0.975$ & $5.78\%$ \\
        
        
        & Total & $0.0154$ & $0.980$ & $3.90\%$ & $0.0169$ & $0.982$ & $4.82\%$ & $0.0122$ & $0.987$ & $3.08\%$ \B\\
        
        \hline
        
    \end{tabular}}
    \caption{Performance metrics of all architectures implemented in this study. *ANN architecture adopted from~\cite{stokes2019determining}}
    \label{table_results}
\end{table}

\renewcommand{\arraystretch}{1.05}
\begin{table}
    \centering
    \resizebox{0.85\textwidth}{!}{%
    \begin{tabular}{|c|c|c c|c c|c c|}
        \hline
        \multirow{2}{*}{Species} & \multirow{2}{*}{Swarm Coefficient} & \multicolumn{2}{c|}{ANN*} & \multicolumn{2}{c|}{CNN} & \multicolumn{2}{c|}{DenseNet} \T\B\\
        
        \cline{3-8}
        
        & & R$^2$ & MARPD & R$^2$ & MARPD & R$^2$ & MARPD \T\B\\
        
        \hline
        
        \multirow{3}{*}{N\textsubscript{2}} 
        & Mobility & $0.592$ & $16.61\%$ & $0.970$ & $6.38\%$ & $0.915$ & $6.41\%$  \T\\



        & Diffusion & $0.868$ & $19.15\%$ & $0.955$ & $5.71\%$ & $0.989$ & $6.18\%$ \\
        
        
        
        & Townsend Ionization & $0.998$ & $6.47\%$ & $0.998$ & $6.37\%$ & $0.999$ & $5.05\%$ \B\\
        
        
        \hline
        
        \multirow{3}{*}{Ar} 
        & Mobility & $0.877$ & $10.02\%$ & $0.901$ & $10.99\%$ & $0.944$ & $3.44\%$  \T\\



        & Diffusion & $0.673$ & $8.54\%$ & $0.726$ & $8.34\%$ & $0.901$ & $4.77\%$ \\
        
        
        
        & Townsend Ionization & $0.967$ & $20.93\%$ & $0.989$ & $19.41\%$ & $0.999$ & $14.97\%$ \B\\
        
        
        \hline
        
        \multirow{3}{*}{He} 
        & Mobility & $0.999$ & $0.78\%$ & $0.999$ & $0.74\%$ & $0.999$ & $0.72\%$  \T\\



        & Diffusion & $0.990$ & $1.85\%$ & $0.998$ & $1.35\%$ & $0.997$ & $1.49\%$ \\
        
        
        
        & Townsend Ionization & $0.988$ & $14.48\%$ & $0.985$ & $13.12\%$ & $0.996$ & $2.51\%$ \B\\
        
        
        \hline
        
        \multirow{3}{*}{F} 
        & Mobility & $0.906$ & $6.54\%$ & $0.996$ & $3.88\%$ & $0.999$ & $6.043\%$  \T\\



        & Diffusion & $-0.75$ & $17.96\%$ & $0.356$ & $12.03\%$ & $0.987$ & $7.78\%$ \\
        
        
        
        & Townsend Ionization & $0.098$ & $32.24\%$ & $0.658$ & $26.67\%$ & $0.982$ & $17.58\%$ \B\\
        
        
        \hline
        
        \multirow{3}{*}{CH\textsubscript{4}} 
        & Mobility & $0.712$ & $12.08\%$ & $0.716$ & $8.01\%$ & $0.966$ & $7.26\%$  \T\\



        & Diffusion & $0.001$ & $21.83\%$ & $0.662$ & $9.97\%$ & $0.933$ & $7.41\%$ \\
        
        
        
        & Townsend Ionization & $0.989$ & $12.43\%$ & $0.999$ & $6.25\%$ & $0.995$ & $9.29\%$ \B\\
        
        
        \hline
        
        \multirow{3}{*}{O\textsubscript{2}} 
        & Mobility & $0.907$ & $8.38\%$ & $0.967$ & $8.26\%$ & $0.984$ & $4.39\%$  \T\\



        & Diffusion & $0.877$ & $7.24\%$ & $0.862$ & $9.58\%$ & $0.948$ & $4.92\%$ \\
        
        
        
        & Townsend Ionization & $0.967$ & $12.88\%$ & $0.998$ & $12.78\%$ & $0.995$ & $3.13\%$ \B\\
        
        
        \hline
        
        \multirow{3}{*}{SF\textsubscript{6}} 
        & Mobility & $0.923$ & $5.47\%$ & $0.972$ & $3.09\%$ & $0.990$ & $1.71\%$  \T\\



        & Diffusion & $0.978$ & $5.69\%$ & $0.990$ & $3.79\%$ & $0.998$ & $2.60\%$ \\
        
        
        
        & Townsend Ionization & $0.996$ & $6.02\%$ & $0.999$ & $5.17\%$ & $0.998$ & $0.98\%$ \B\\
        
        
        \hline
        
    \end{tabular}}
    \caption{Performance metrics of reproduced swarm coefficients by predictions of all architectures implemented in this study. *ANN architecture adopted from~\cite{stokes2019determining} 
    }
    \label{results_swarm}
\end{table}

From the performance metrics (shown in Table~\ref{table_results}), we can safely conclude that the DenseNet architecture performs significantly better compared to CNN, which in turn yield better results than ANN architecture for predicting the elastic momentum cross sections over the entire energy domain considered, of all the gas species considered in our study. A common trend across all the gas species in prediction of elastic MTCS is that all the three architectures predict the cross section with significantly higher accuracy for the range $30-100$ eV. To further comment on the accuracy of the architectures, we analysed the prediction trends of individual gas species in detail. Nitrogen's elastic MTCS has a characteristic peak between $2-2.5$ eV, which is not present in any of the other gas species used in the training data and thus, both ANN and CNN fail to predict this peak. This is due to a quantum mechanical effect specific to $N_2$ in this energy range and it may be difficult to teach the network about the same. DenseNet, on the other hand, does notably better in predicting the presence of this peak, yet, its estimate of the energy at which it occurs is off by $\sim0.5$ eV. Likewise, Argon has Ramsauer-Townsend minimum whose value is significantly lower than all other gas species considered in the training data (another quantum mechanical effect). Still, DenseNet is able to predict the presence of Ramsauer-Townsend minimum at the correct energy value, only erring slightly in determining its magnitude, whereas both CNN and ANN fail to even determine the presence of this minimum. Such trends are observed in prediction of elastic MTCS of all other gas species too, wherein DenseNet is able to determine the characteristic local maximum/minimum values and its locations with remarkably higher accuracy compared to ANN or CNN architecture. 
We believe our use of convolution kernels of varying sizes allowed the DenseNet architecture to better understand the trends of swarm data which in turn lead to this enhanced performance. Also, layers in the DenseNet architecture receive additional supervision from the loss function through shorter connections, alleviating the vanishing gradient problem and improving the flow of information and gradients throughout the network. This deep supervision provided by the DenseNet could also be one of the reasons for this improved accuracy of the predicted cross sections.

For predicting the ionization cross sections, both DenseNet and CNN gives equally good results compared to ANN over the entire energy domain, according to the performance metrics. Moreover, even though no prior information about the threshold energy of ionization cross sections was provided to any neural network, both CNN and DenseNet were able to predict the threshold energy of all gas species, with an accuracy upto one decimal place. Specifically for the case of fluorine, all the three models somewhat struggle to determine the ionization cross section with accuracy as compared to other  predictions. This is purely due to the fact that the ionization cross section of fluorine is unusually lower than the ionization cross sections of the other gas species in the training data and thus can be considered as an outlier.

Prediction of excitation cross sections by all the architectures differ substantially from the actual cross sections. A possible reason for the same is that the swarm data themselves provide less information about the excitation cross sections compared to elastic momentum transfer and ionization cross sections. This assumption is backed up by a comparison of the two sets of swarm parameters, as depicted in Fig.~\ref{results_swarmimages}. The first being calculated from predicted elastic momentum transfer, ionization and excitation cross sections, while the second set of swarm parameters is calculated using the actual elastic momentum transfer, ionization and excitation cross sections. Another point to be noted here is that only the lowest  threshold processes is used in the training and for many cases, this is far less than the sum of all the excitation cross sections.
Although the predicted excitation cross sections differed substantially from the actual cross section, the same is not replicated in the comparison of swarm parameters, whose metrics (Table~\ref{results_swarm}) are almost consistent with those predicted of elastic momentum transfer and ionization cross sections' predictions. Thus, we can attribute lack of information content about excitation cross sections in swarm coefficients as one of the possible reasons behind the inaccuracy of predicted excitation cross sections. However, this requires a more detailed investigation in future.

\subsection{Uncertainty Quantification}

Solutions obtained using deep learning methods have some inherent uncertainty. Quantifying this uncertainty would assist us in determining the reliability of the predictions. Moreover, the mapping of swarm coefficients to cross sections is non-unique - there exist multiple cross sections which map to the same swarm coefficient and the probability distribution of the cross sections generated by the uncertainty quantification (UQ) allows us to sample all these plausible solutions.

\begin{figure}
    \begin{subfigure}{0.5\textwidth}
        \centering
        \includegraphics[width=0.927\linewidth]{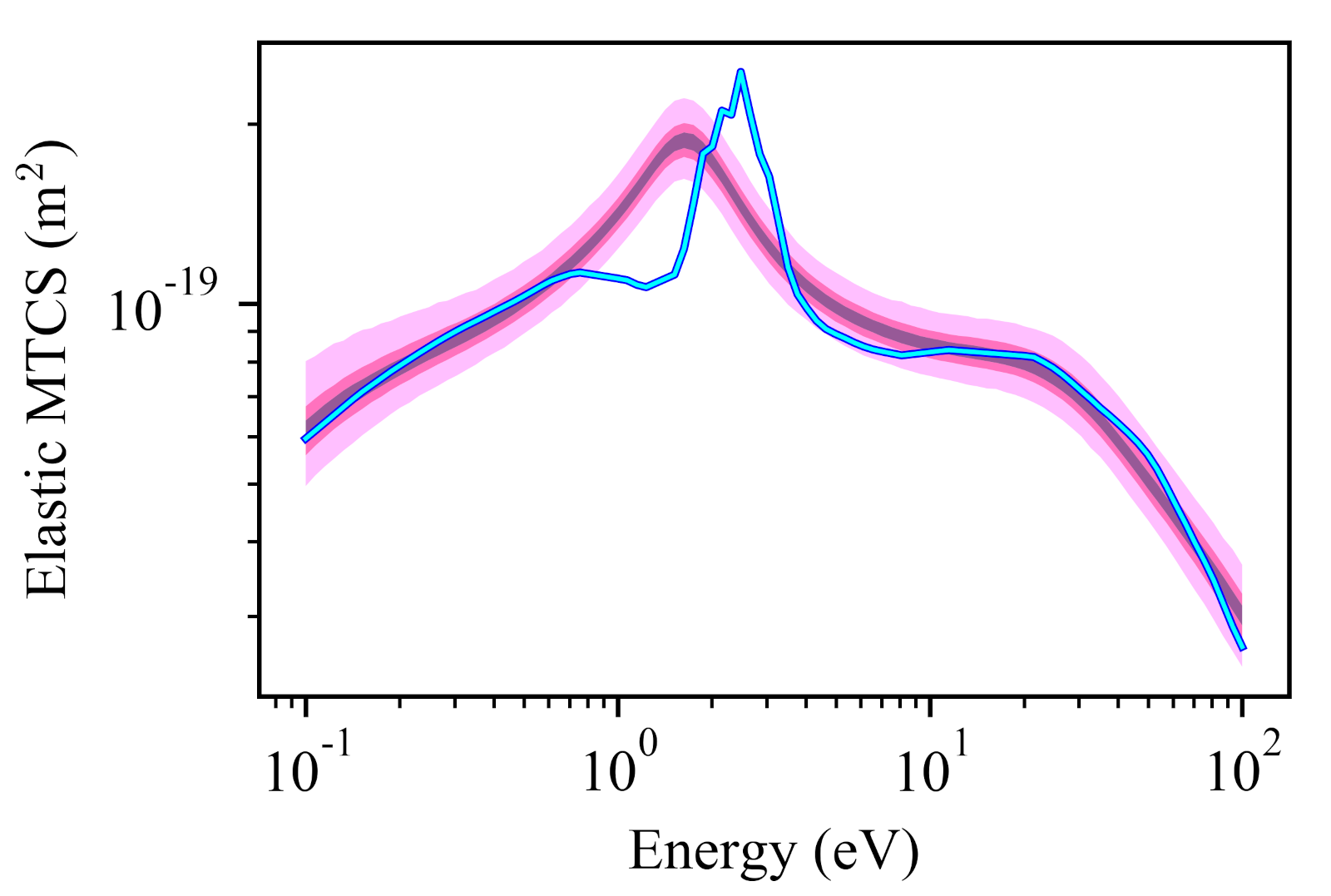}
        \caption{Nitrogen (N\textsubscript{2})}
        \label{elastic_uncertainty_N2}
    \end{subfigure}
    \begin{subfigure}{0.5\textwidth}
        \centering
        \includegraphics[width=0.927\linewidth]{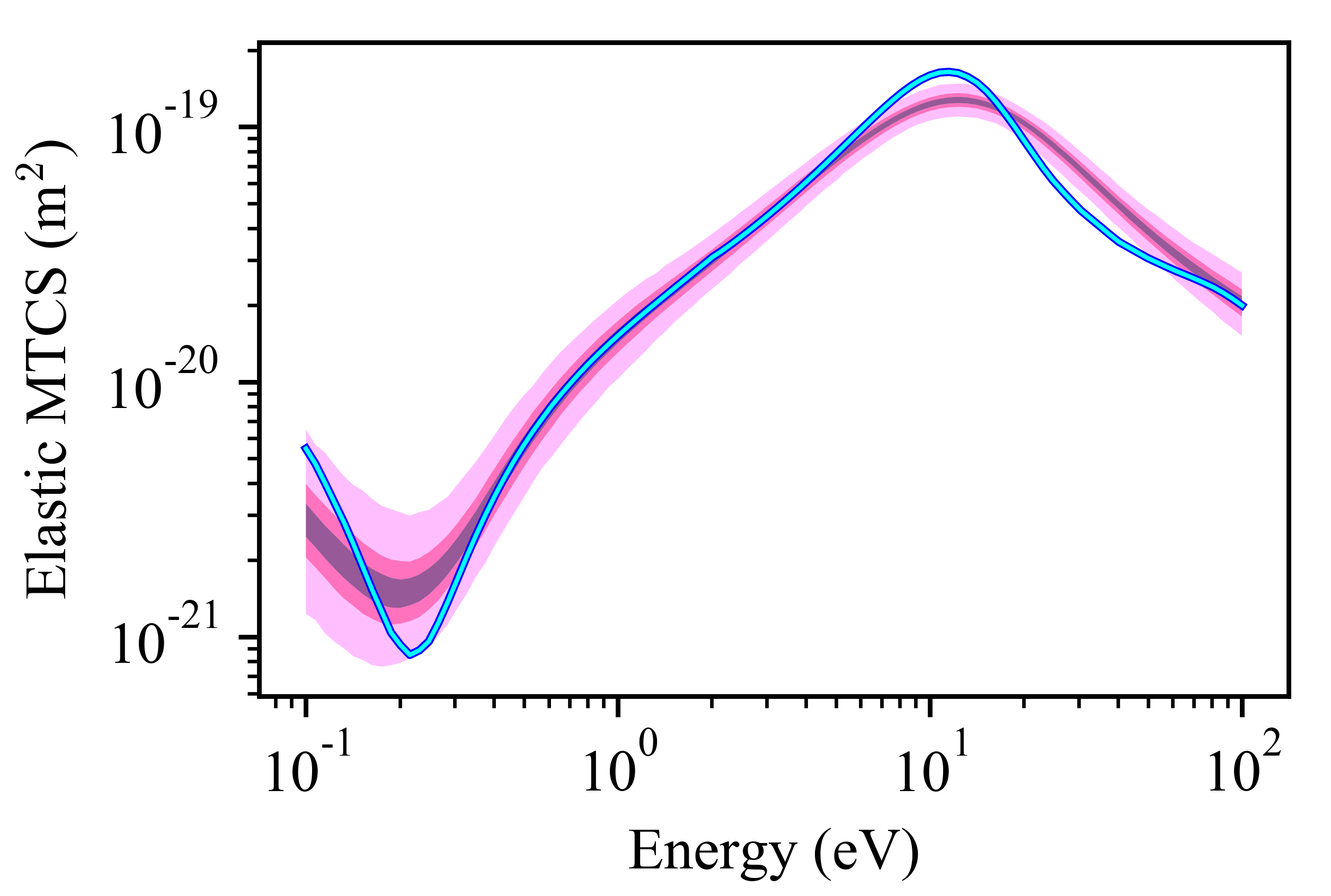}
        \caption{Argon (Ar)}
        \label{elastic_uncertainty_Ar}
    \end{subfigure}
    \begin{subfigure}{0.5\textwidth}
        \centering
        \includegraphics[width=0.927\linewidth]{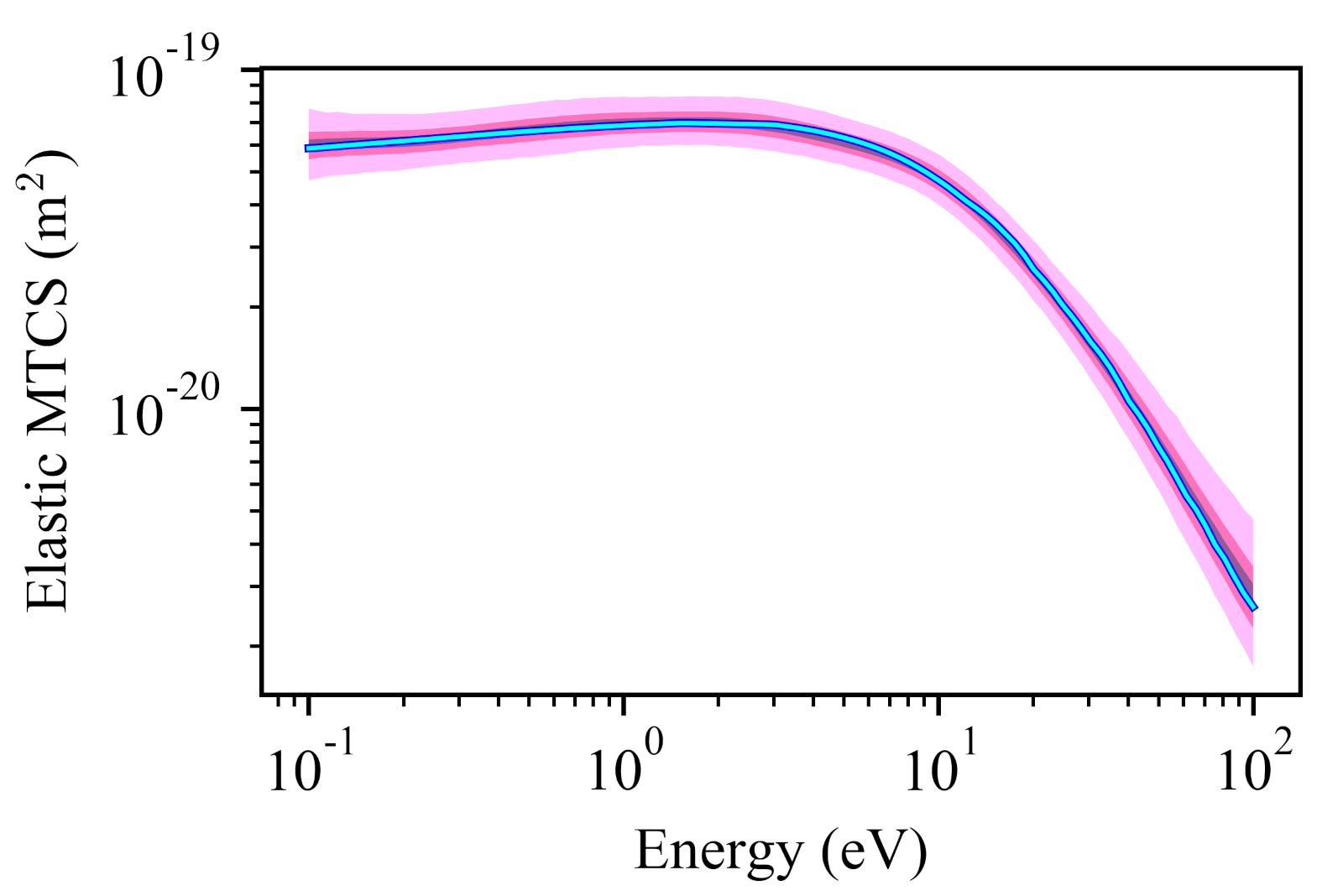}
        \caption{Helium (He)}
        \label{elastic_uncertainty_He}
    \end{subfigure}
    \begin{subfigure}{0.5\textwidth}
        \centering
        \includegraphics[width=0.927\linewidth]{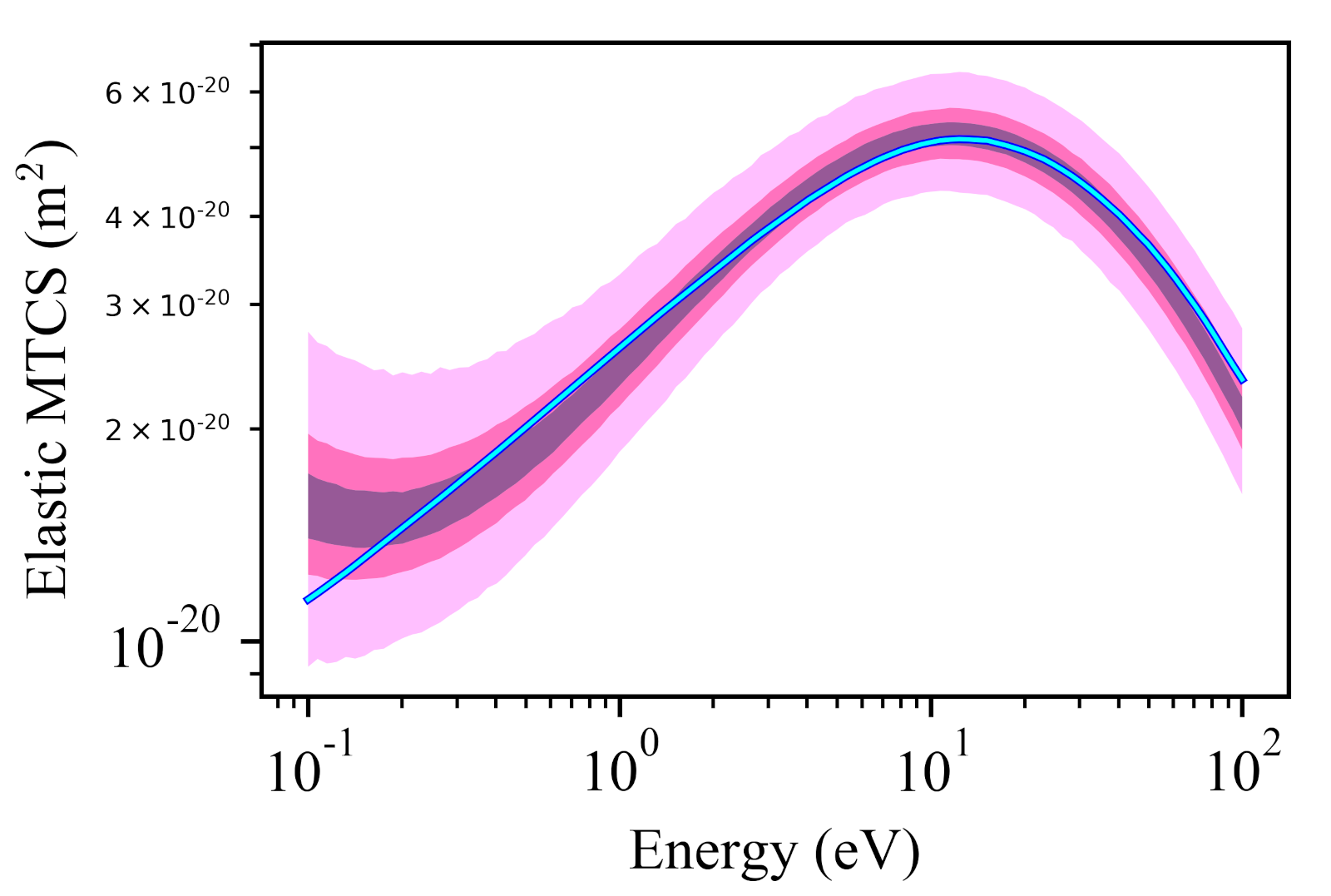}
        \caption{Fluorine (F)}
        \label{elastic_uncertainty_F}
    \end{subfigure}
    \begin{subfigure}{0.5\textwidth}
        \centering
        \includegraphics[width=0.927\linewidth]{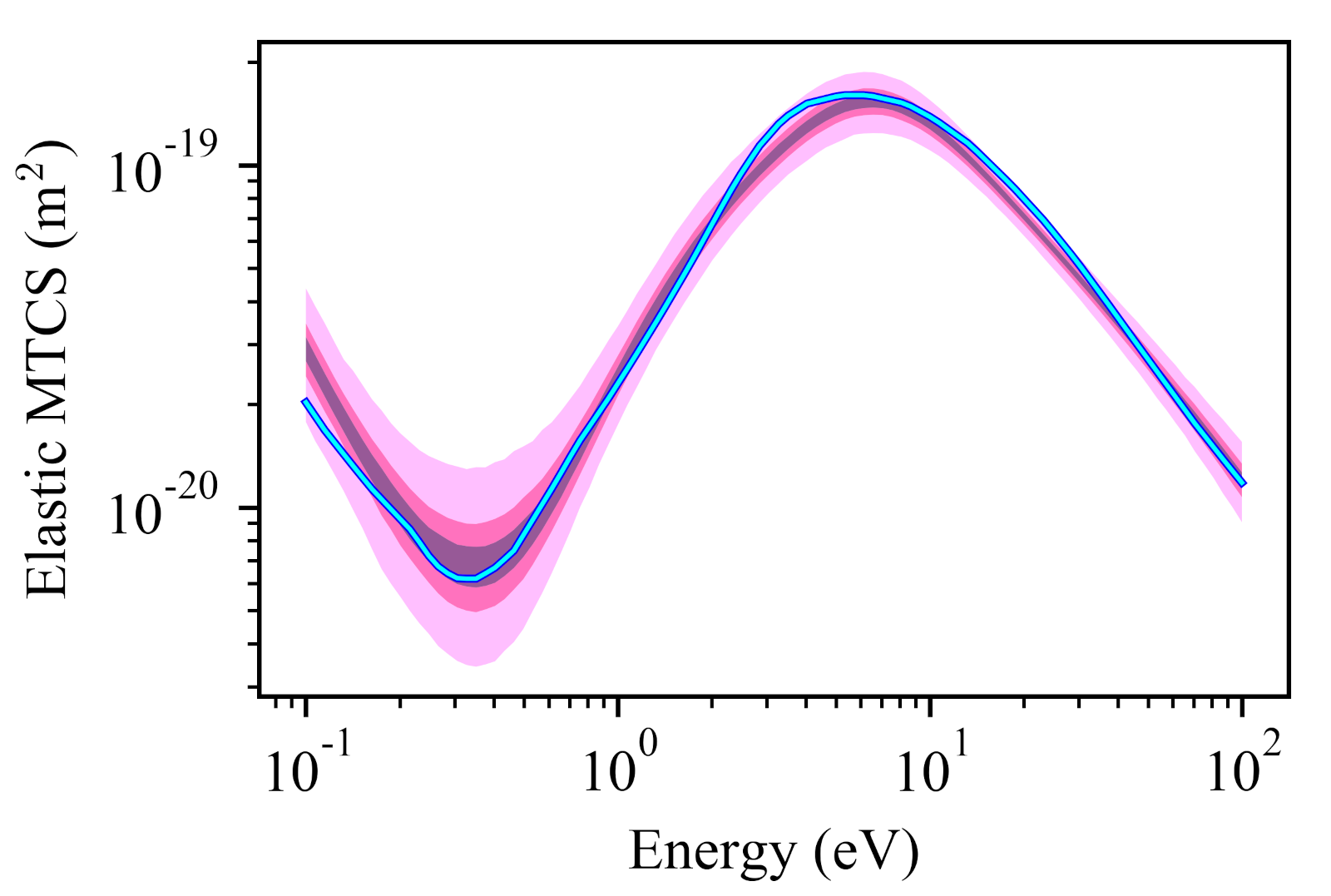}
        \caption{Methane (CH\textsubscript{4})}
        \label{elastic_uncertainty_CH4}
    \end{subfigure}
    \begin{subfigure}{0.5\textwidth}
        \centering
        \includegraphics[width=0.927\linewidth]{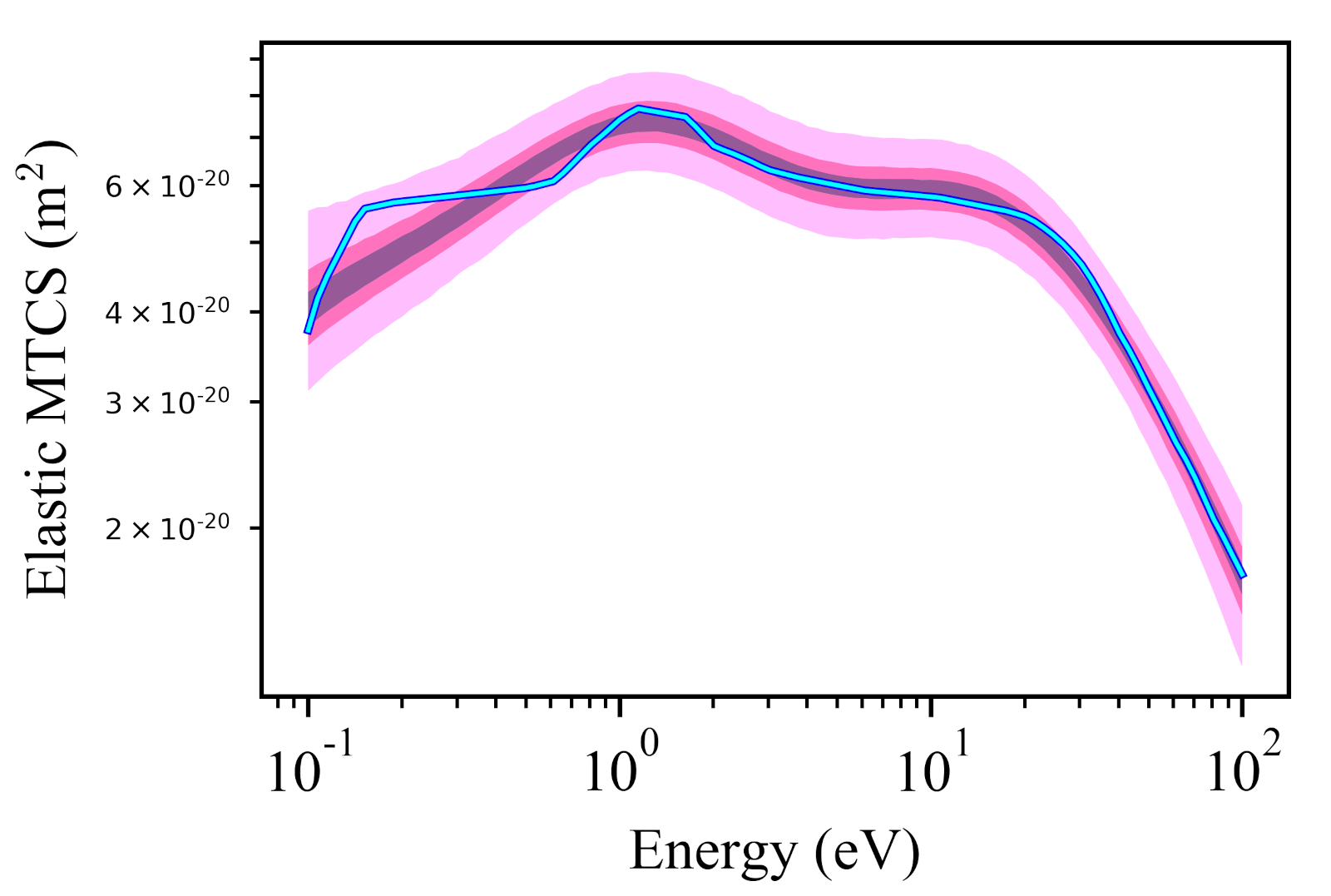}
        \caption{Oxygen (O\textsubscript{2})}
        \label{elastic_uncertainty_O2}
    \end{subfigure}
    \begin{subfigure}{0.5\textwidth}
        \centering
        \includegraphics[width=0.927\linewidth]{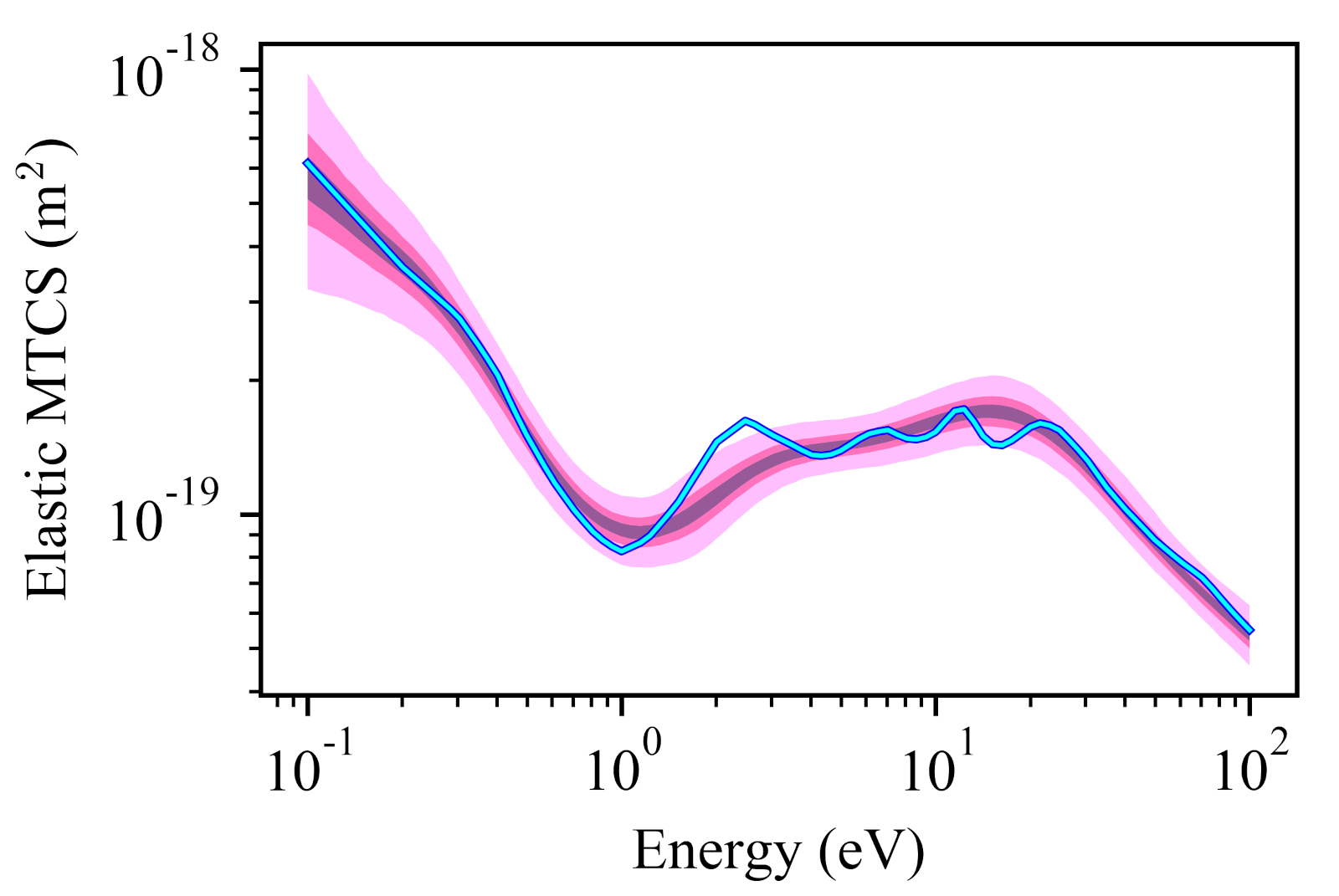}
        \caption{Sulfur hexafluoride (SF\textsubscript{6})}
        \label{elastic_uncertainty_SF6}
    \end{subfigure}
    \begin{subfigure}{0.5\textwidth}
        \centering
        \includegraphics[width=0.875\linewidth]{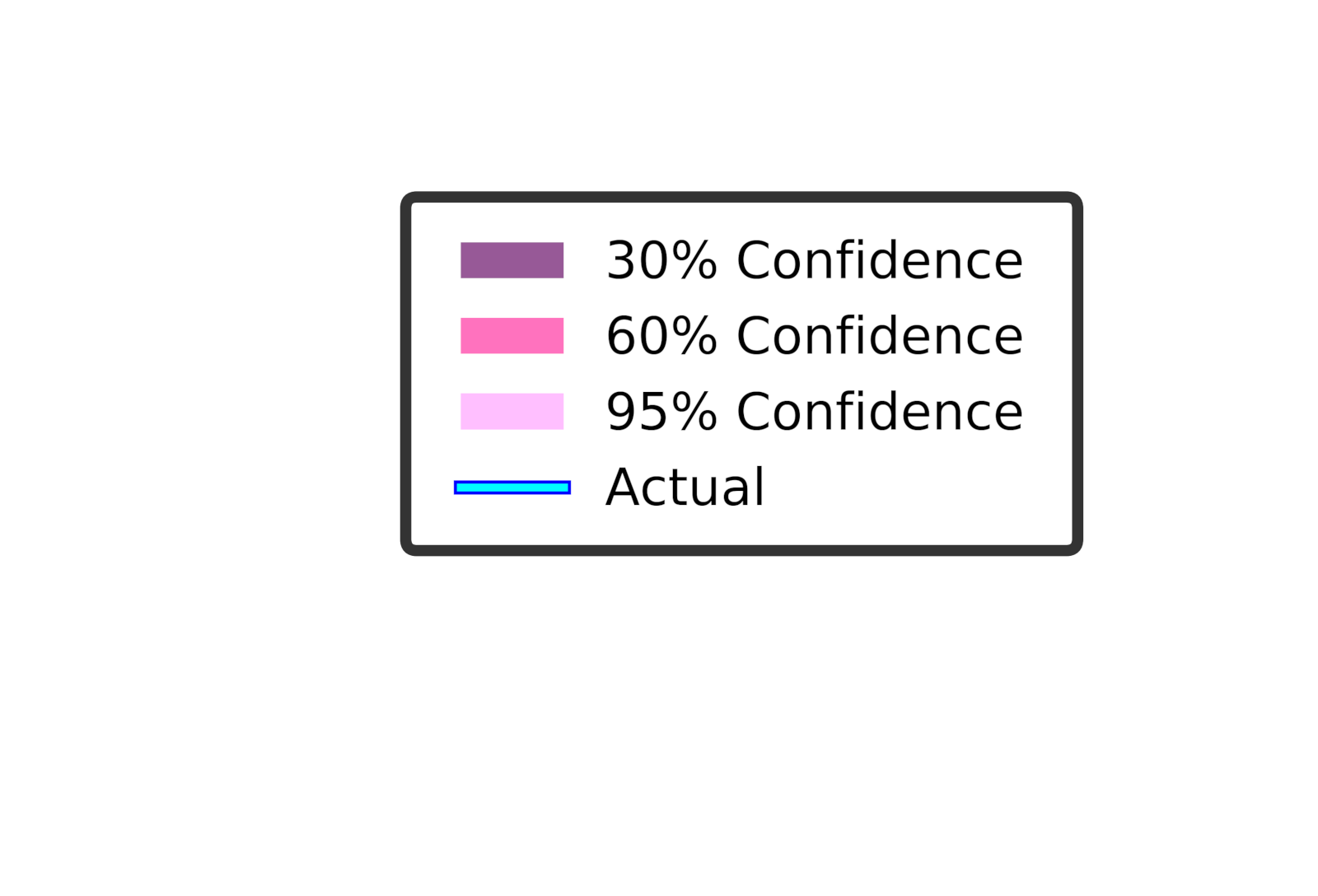}
        \label{elastic_uncertainty_legend}
    \end{subfigure}
    \caption{Uncertainty in prediction of elastic cross sections of various gas species}
    \label{results_elastic_uncertainty}
\end{figure}

\begin{figure}
    \begin{subfigure}{0.5\textwidth}
        \centering
        \includegraphics[width=0.927\linewidth]{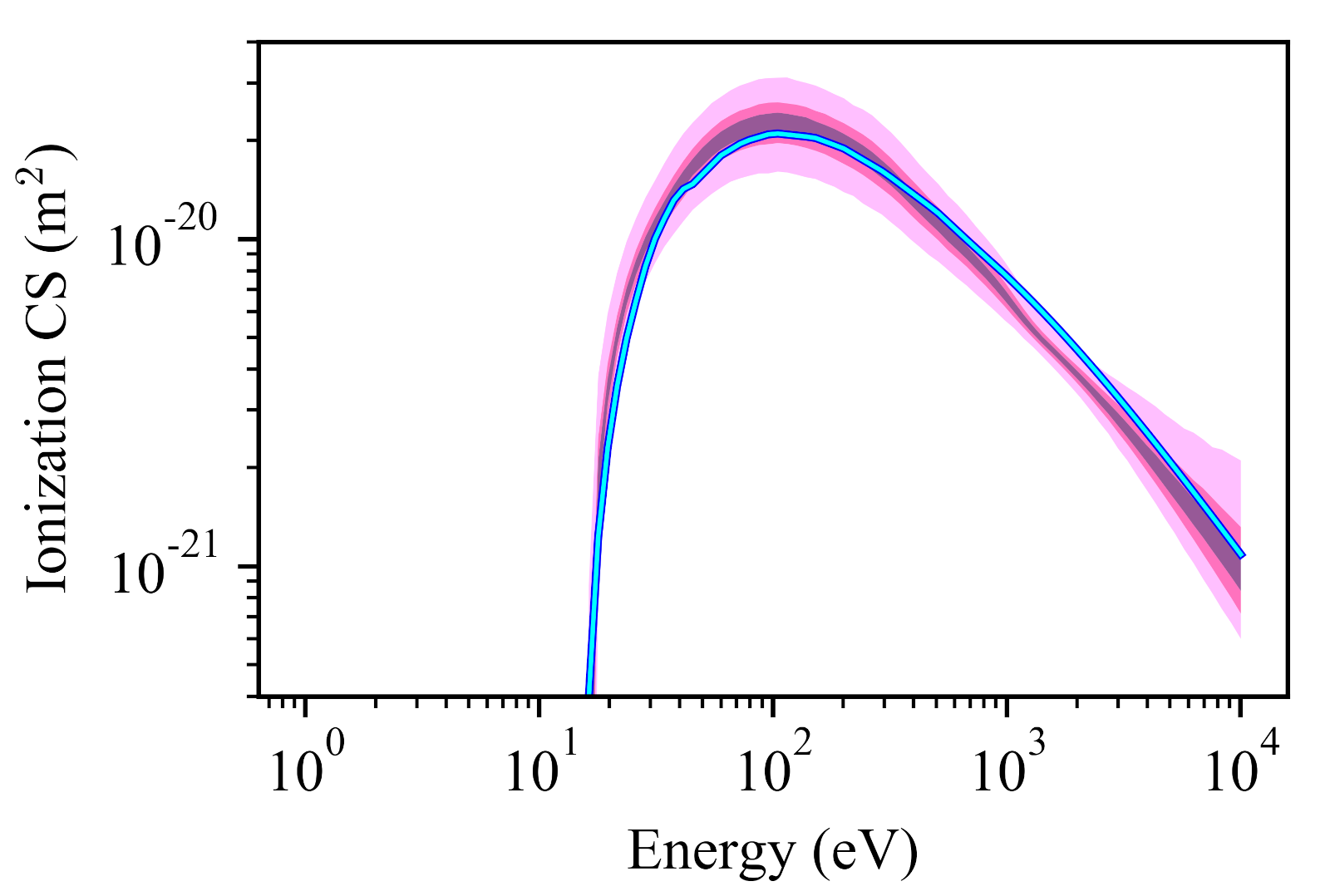}
        \caption{Nitrogen (N\textsubscript{2})}
        \label{ionization_uncertainty_N2}
    \end{subfigure}
    \begin{subfigure}{0.5\textwidth}
        \centering
        \includegraphics[width=0.927\linewidth]{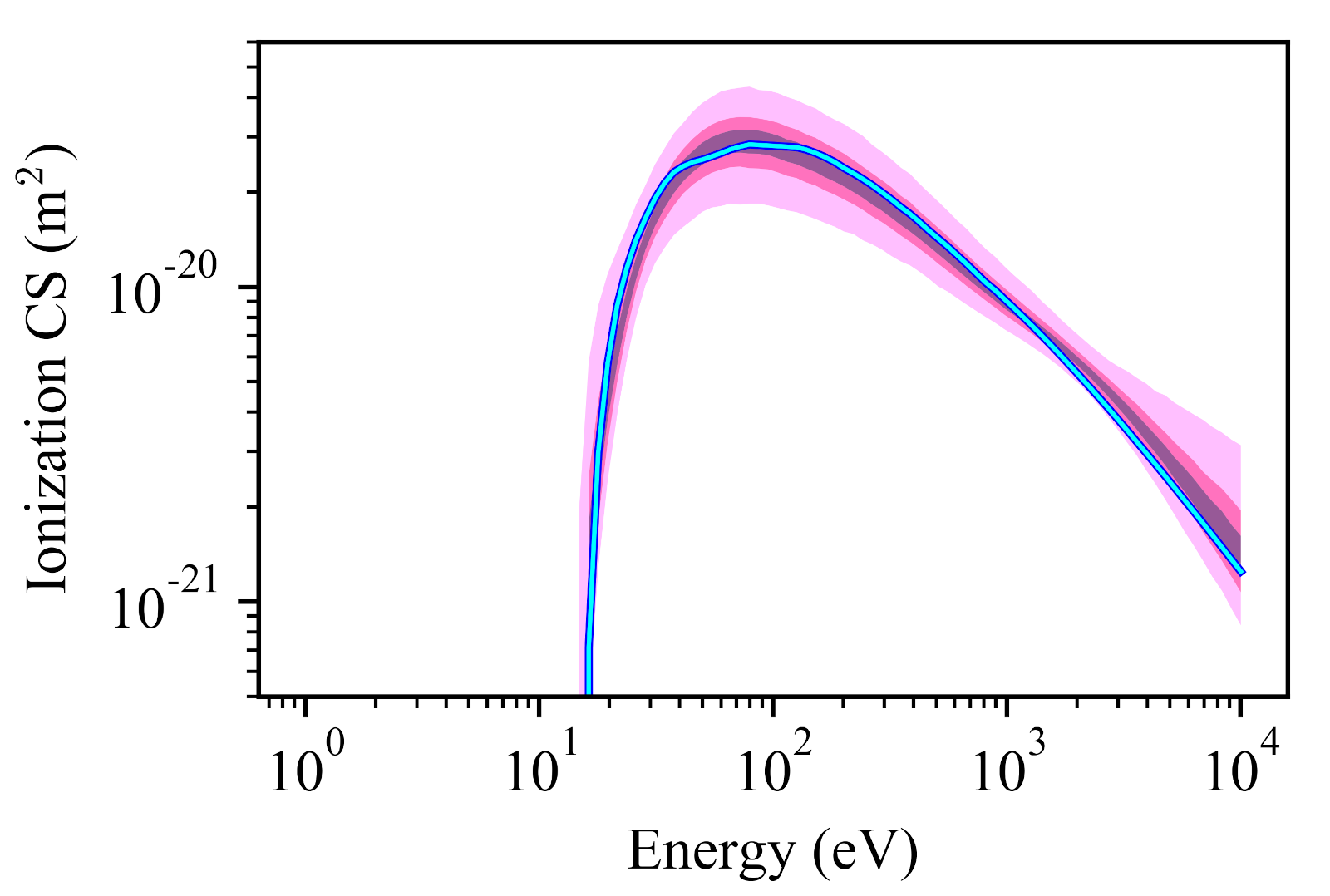}
        \caption{Argon (Ar)}
        \label{ionization_uncertainty_Ar}
    \end{subfigure}
    \begin{subfigure}{0.5\textwidth}
        \centering
        \includegraphics[width=0.927\linewidth]{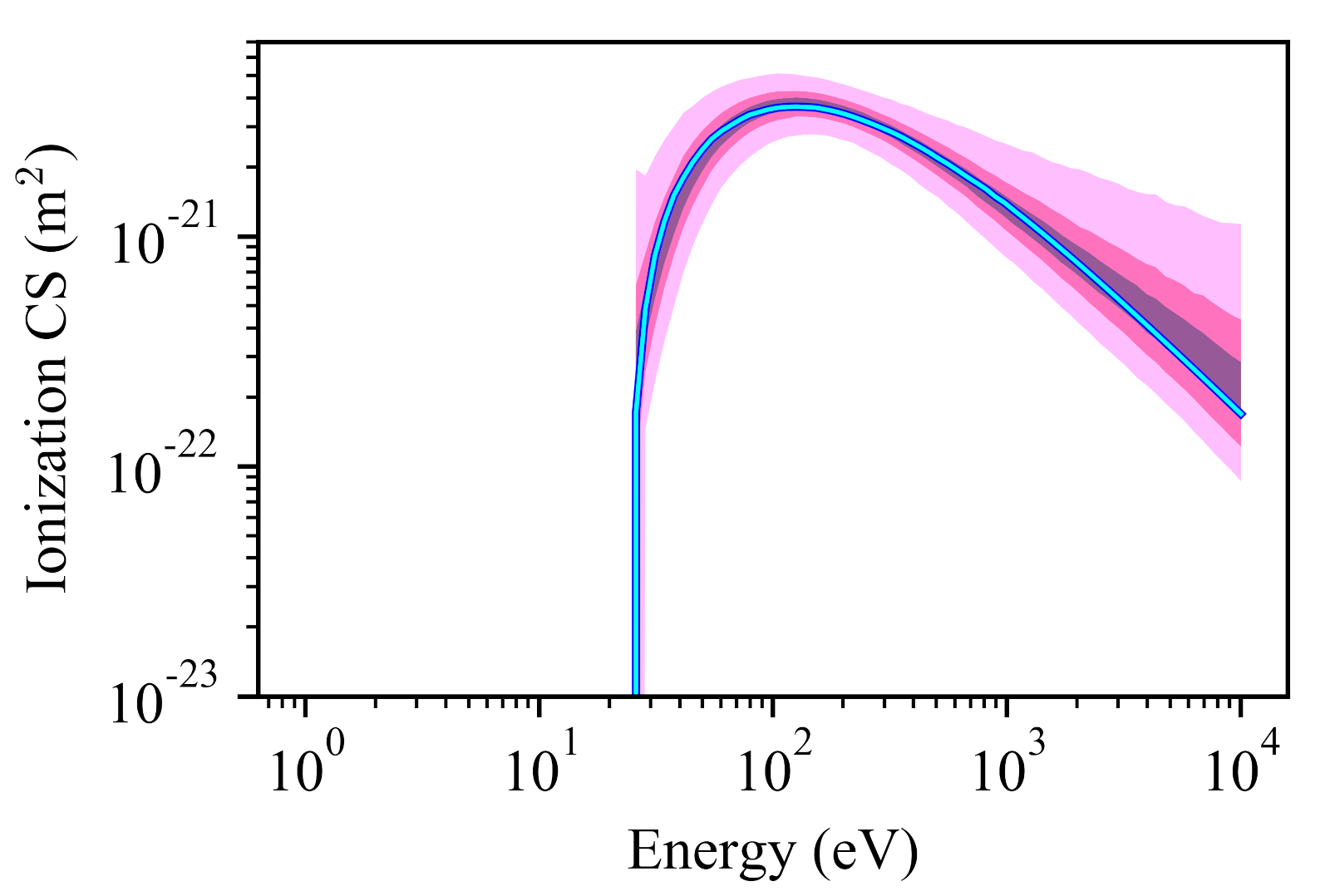}
        \caption{Helium (He)}
        \label{ionization_uncertainty_He}
    \end{subfigure}
    \begin{subfigure}{0.5\textwidth}
        \centering
        \includegraphics[width=0.927\linewidth]{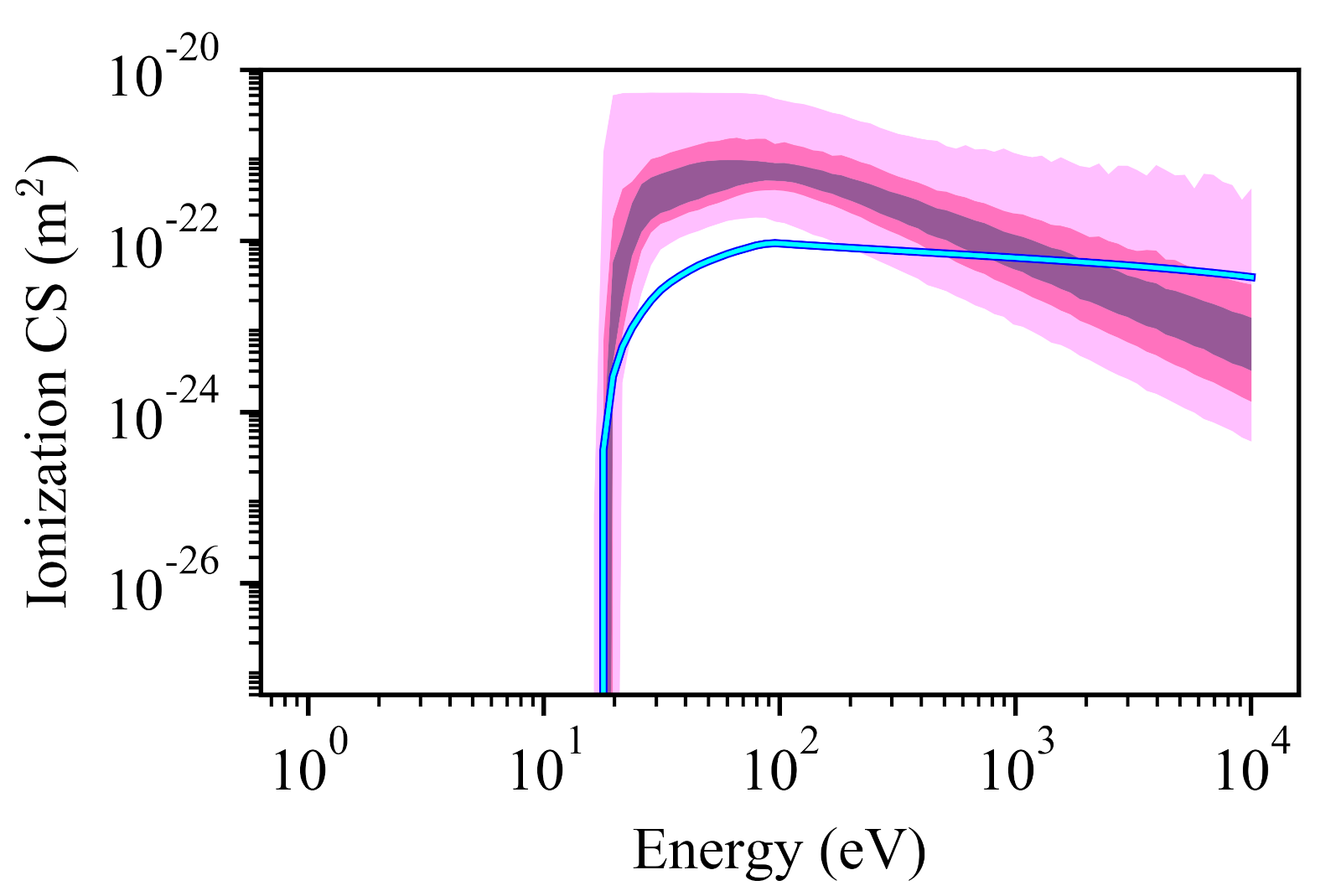}
        \caption{Fluorine (F)}
        \label{ionization_uncertainty_F}
    \end{subfigure}
    \begin{subfigure}{0.5\textwidth}
        \centering
        \includegraphics[width=0.927\linewidth]{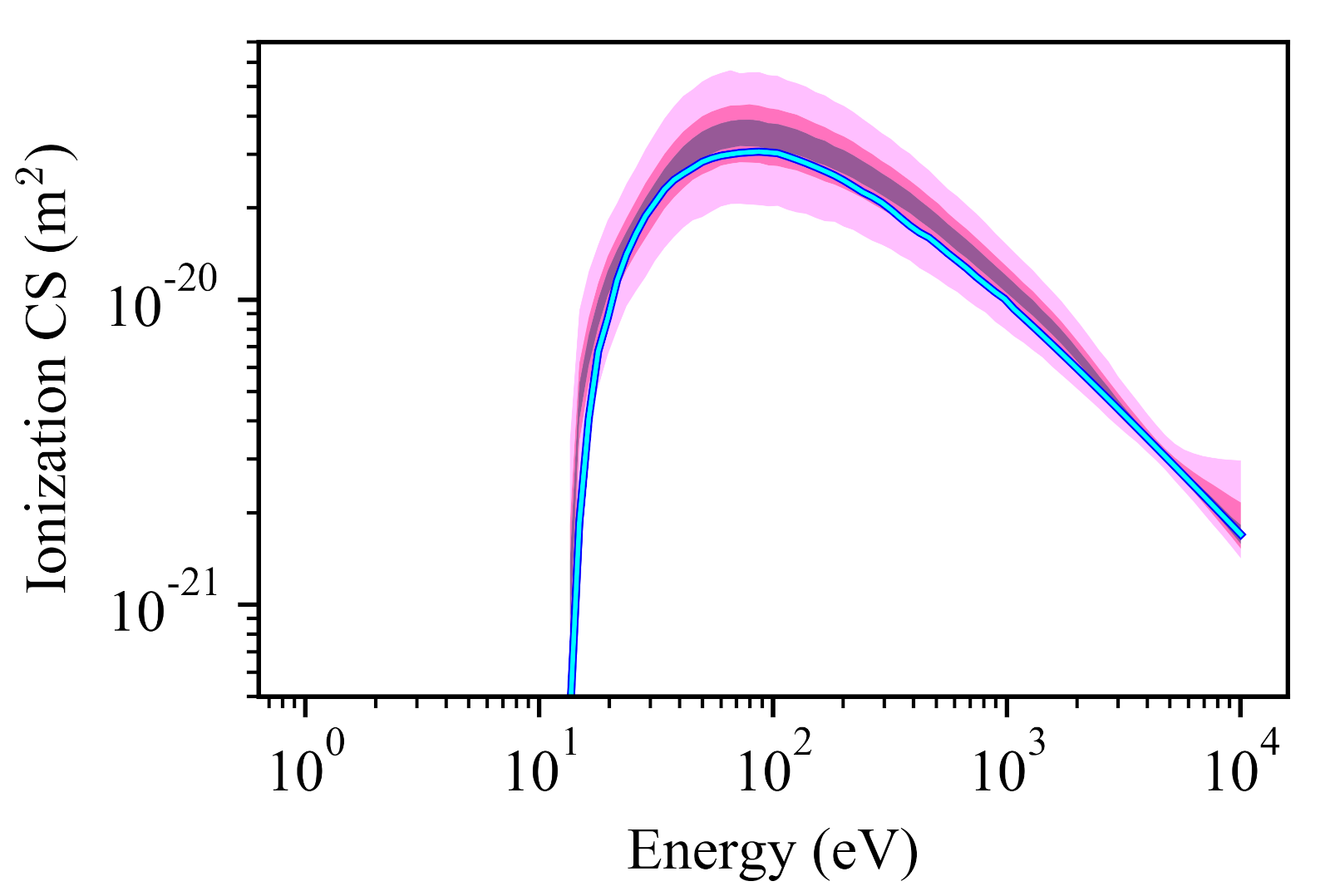}
        \caption{Methane (CH\textsubscript{4})}
        \label{ionization_uncertainty_CH4}
    \end{subfigure}
    \begin{subfigure}{0.5\textwidth}
        \centering
        \includegraphics[width=0.927\linewidth]{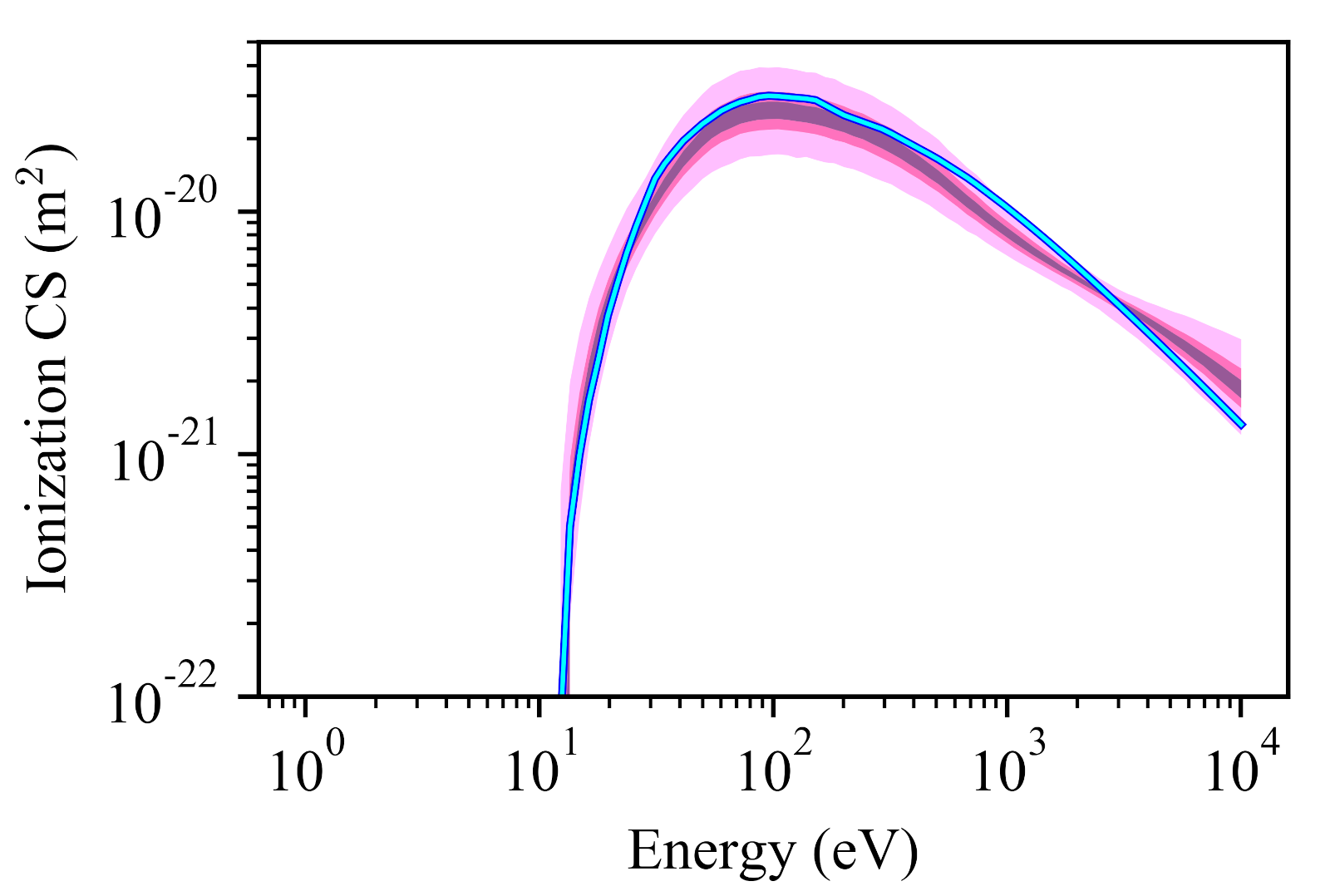}
        \caption{Oxygen (O\textsubscript{2})}
        \label{ionization_uncertainty_O2}
    \end{subfigure}
    \begin{subfigure}{0.5\textwidth}
        \centering
        \includegraphics[width=0.927\linewidth]{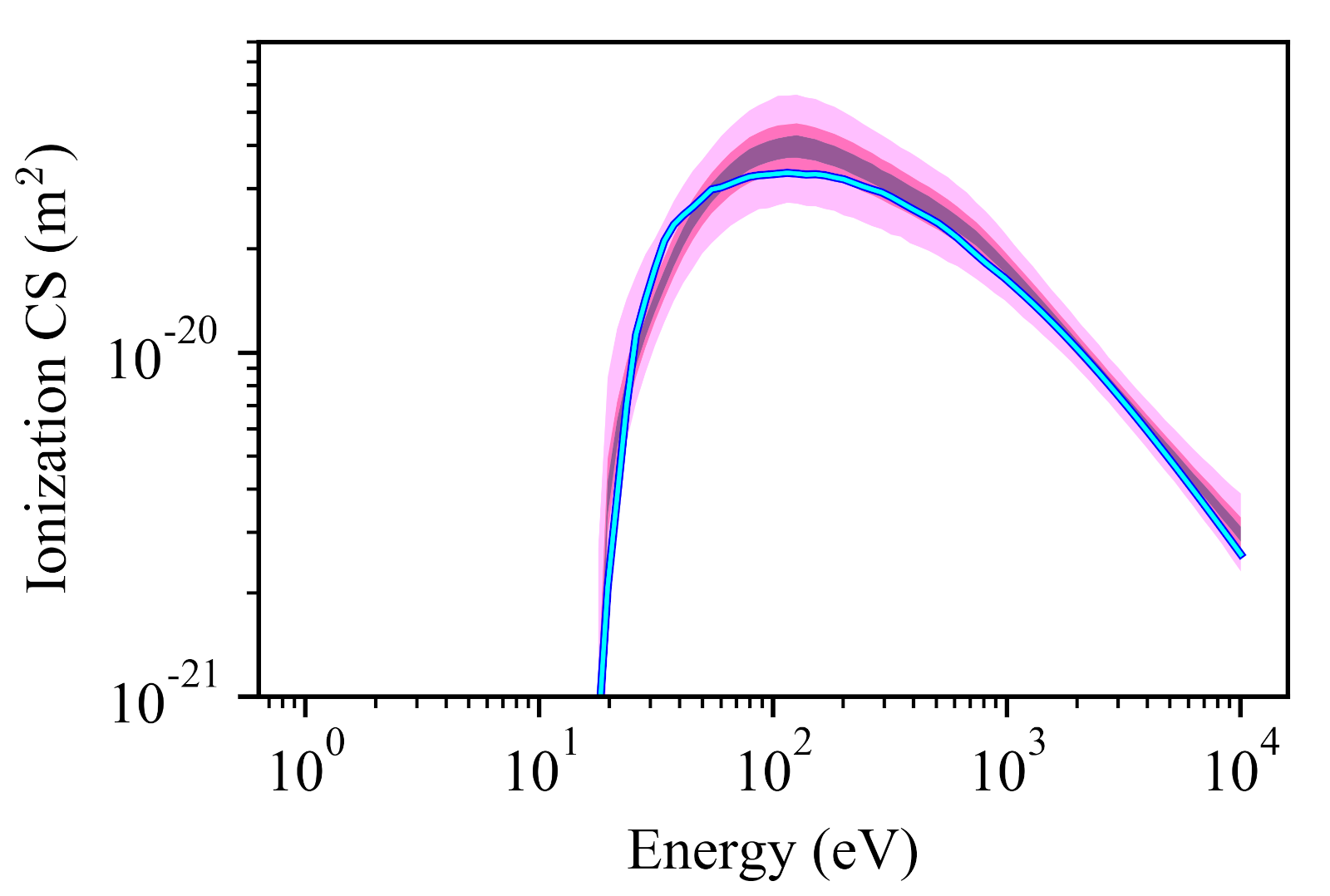}
        \caption{Sulfur hexafluoride (SF\textsubscript{6})}
        \label{ionization_uncertainty_SF6}
    \end{subfigure}
    \begin{subfigure}{0.5\textwidth}
        \centering
        \includegraphics[width=0.875\linewidth]{images/results/legend_uncertainty.png}
        \label{ionization_uncertainty_legend}
    \end{subfigure}
    \caption{Uncertainty in prediction of ionization cross sections of various gas species}
    \label{results_ionization_uncertainty}
\end{figure}

\begin{figure}
    \begin{subfigure}{0.5\textwidth}
        \centering
        \includegraphics[width=0.927\linewidth]{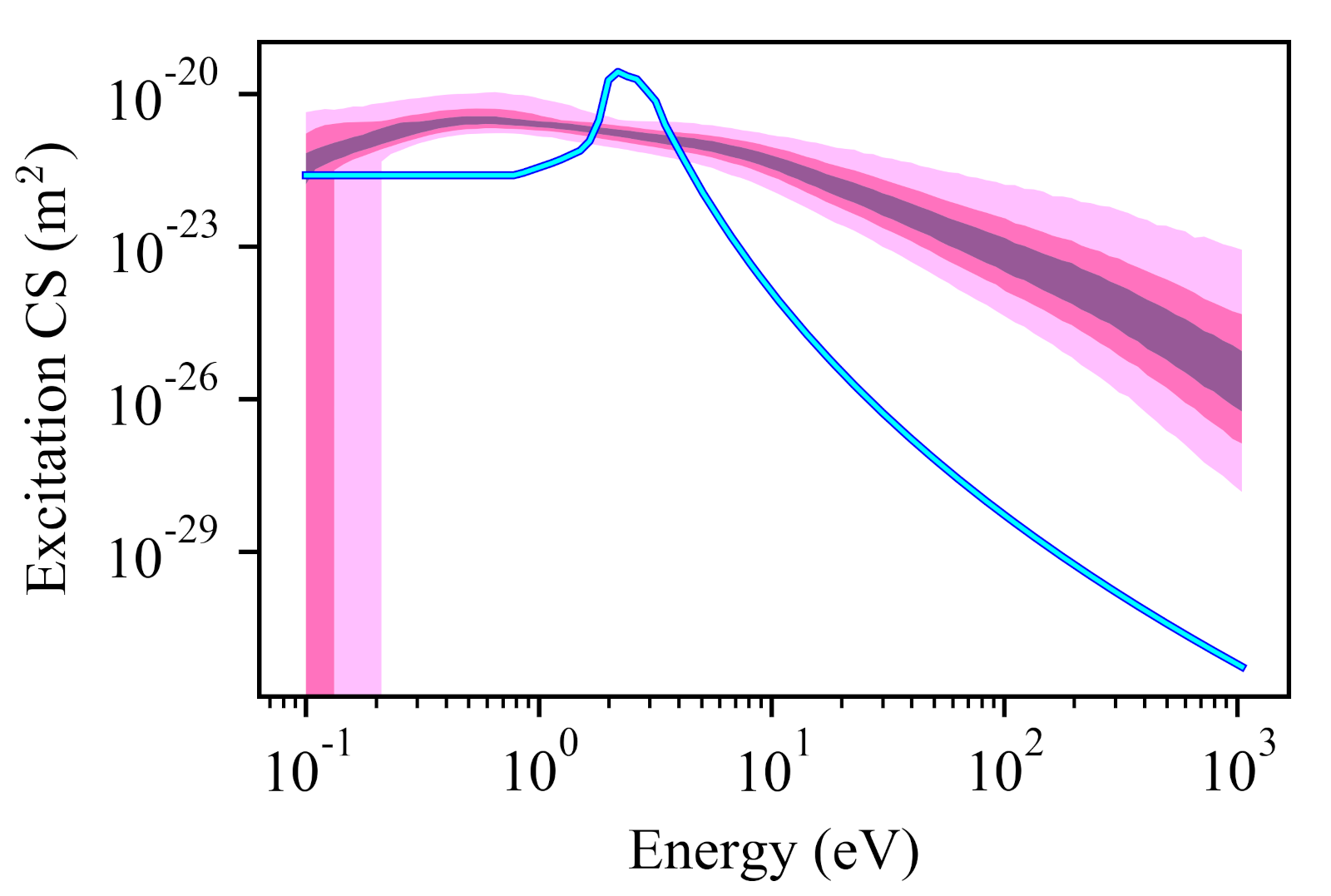}
        \caption{Nitrogen (N\textsubscript{2})}
        \label{excitation_uncertainty_N2}
    \end{subfigure}
    \begin{subfigure}{0.5\textwidth}
        \centering
        \includegraphics[width=0.927\linewidth]{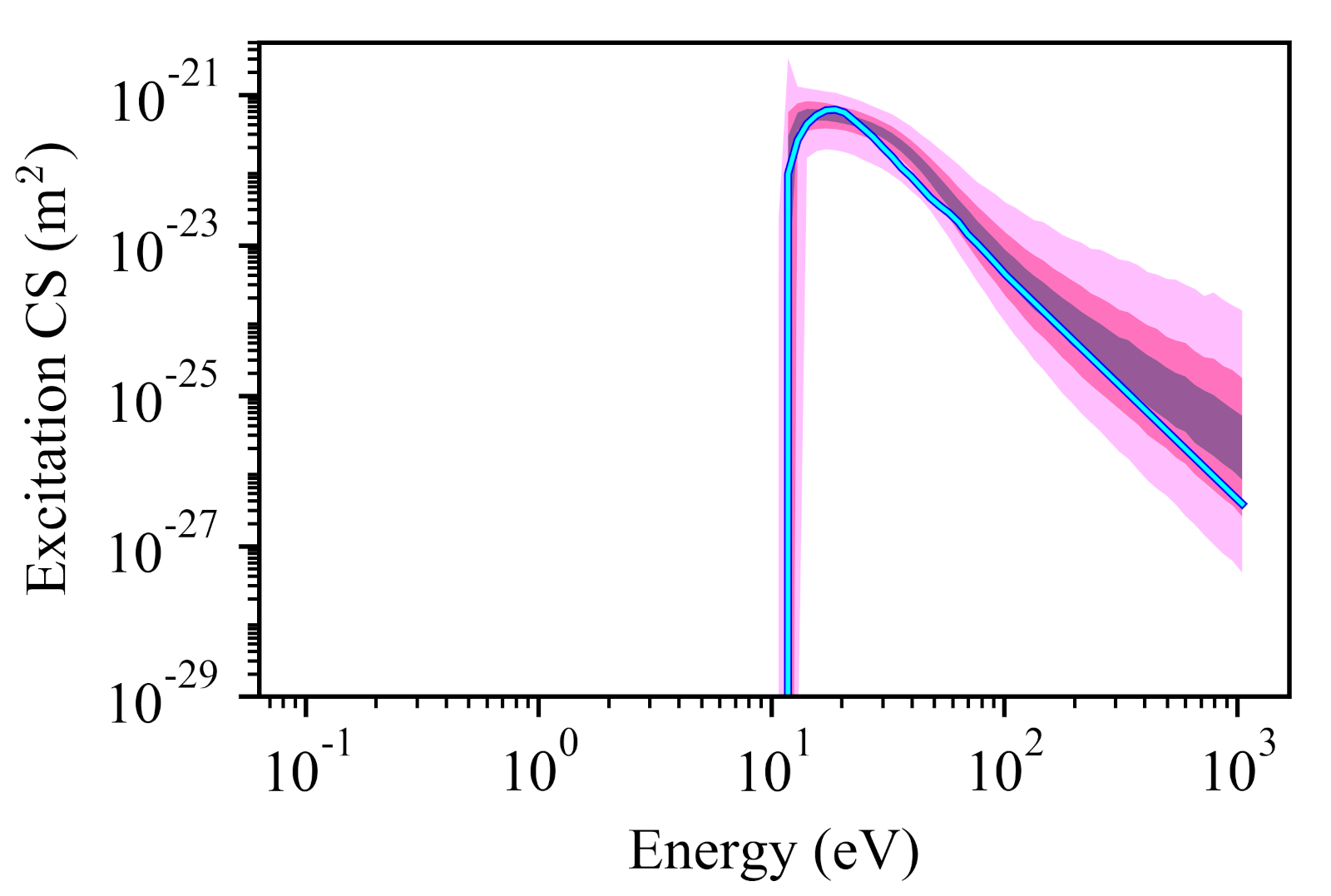}
        \caption{Argon (Ar)}
        \label{excitation_uncertainty_Ar}
    \end{subfigure}
    \begin{subfigure}{0.5\textwidth}
        \centering
        \includegraphics[width=0.927\linewidth]{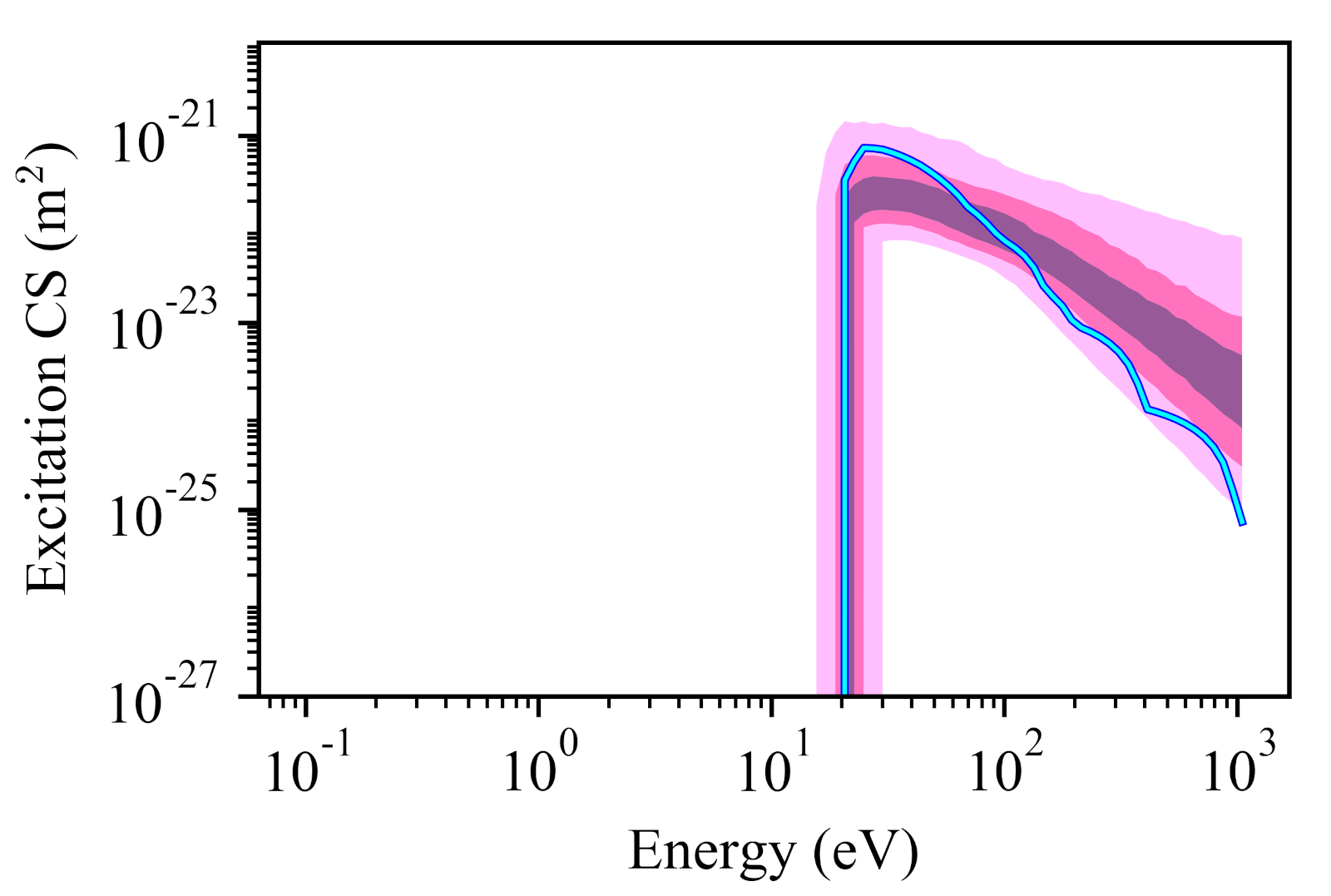}
        \caption{Helium (He)}
        \label{excitation_uncertainty_He}
    \end{subfigure}
    \begin{subfigure}{0.5\textwidth}
        \centering
        \includegraphics[width=0.927\linewidth]{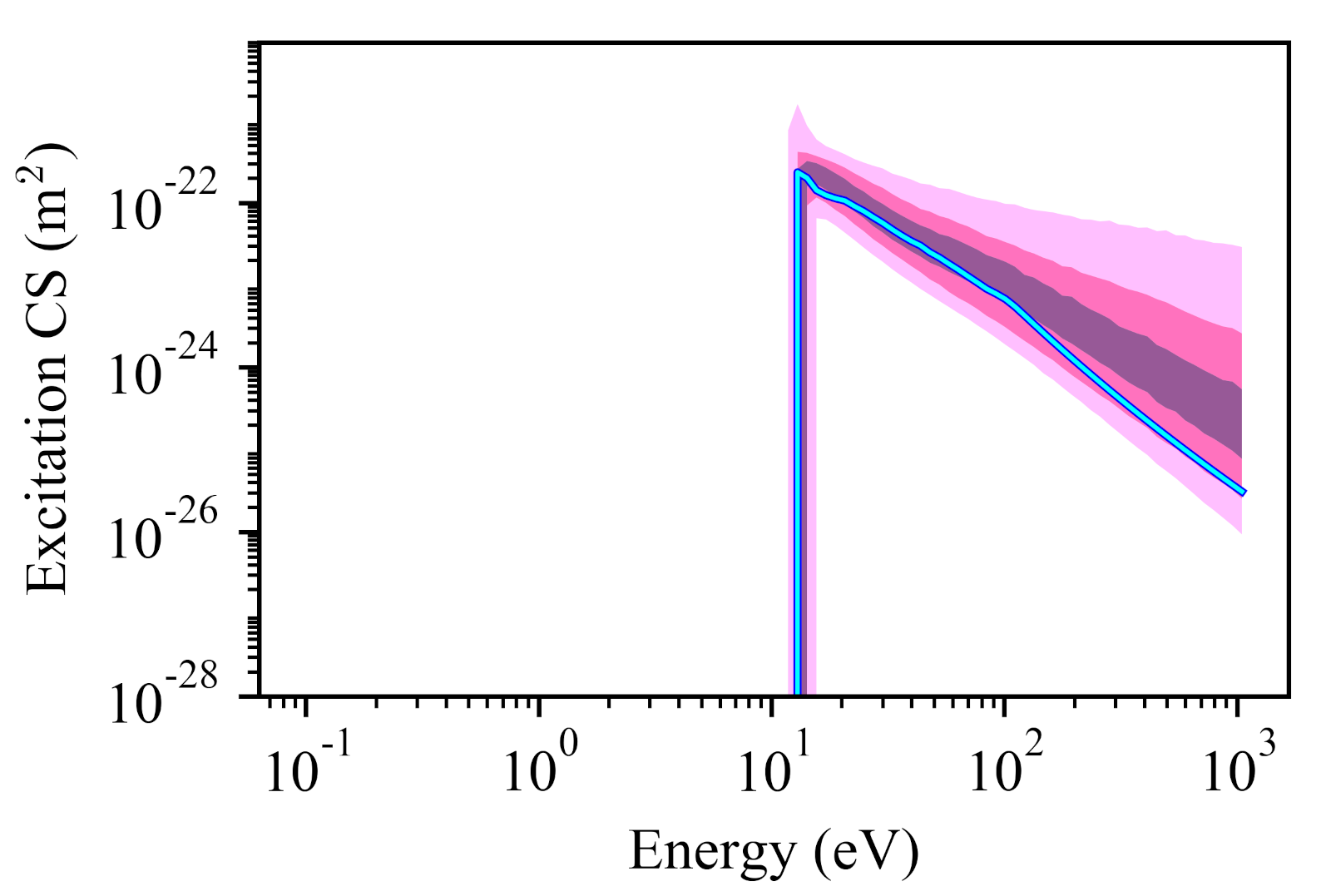}
        \caption{Fluorine (F)}
        \label{excitation_uncertainty_F}
    \end{subfigure}
    \begin{subfigure}{0.5\textwidth}
        \centering
        \includegraphics[width=0.927\linewidth]{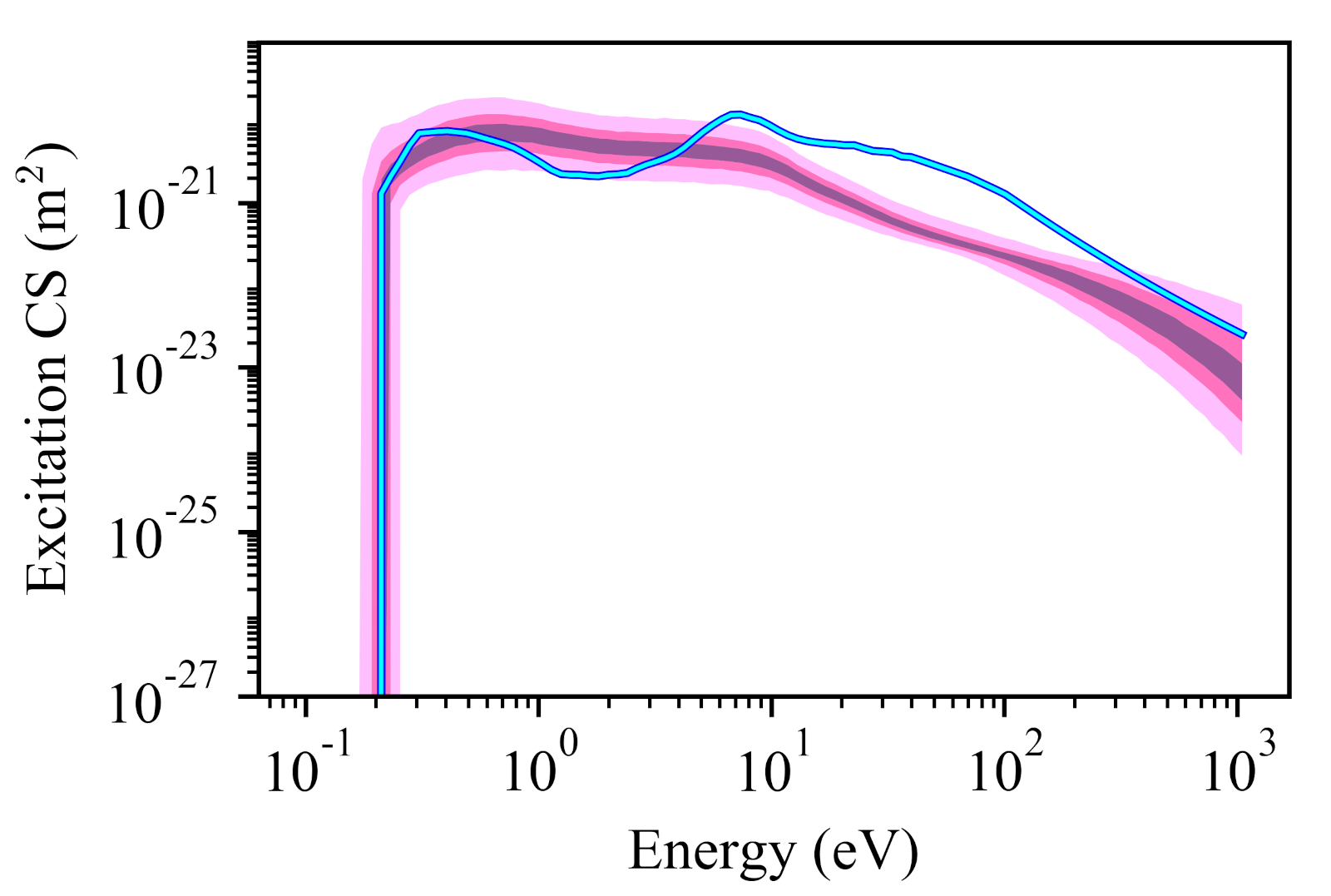}
        \caption{Methane (CH\textsubscript{4})}
        \label{excitation_uncertainty_CH4}
    \end{subfigure}
    \begin{subfigure}{0.5\textwidth}
        \centering
        \includegraphics[width=0.927\linewidth]{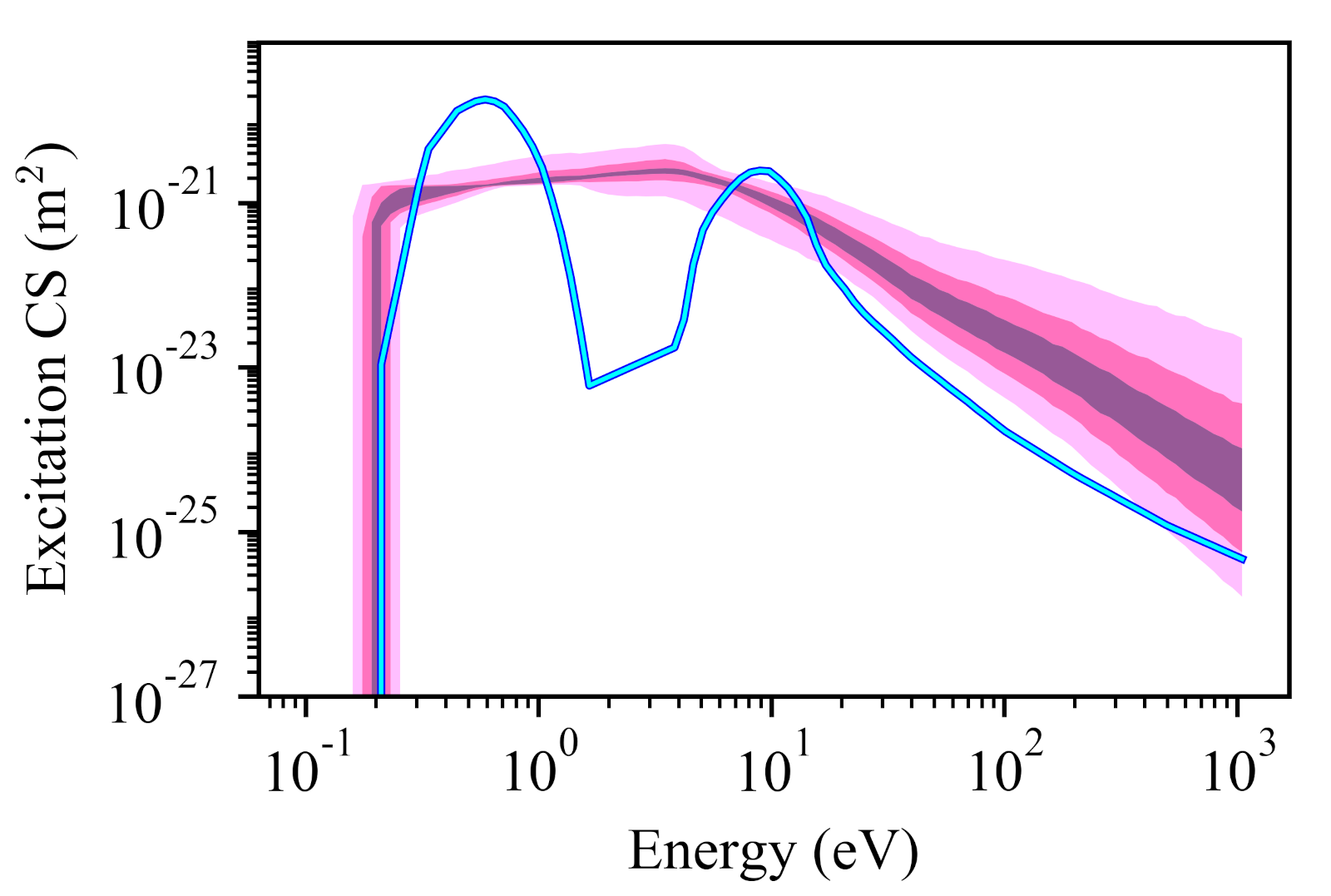}
        \caption{Oxygen (O\textsubscript{2})}
        \label{excitation_uncertainty_O2}
    \end{subfigure}
    \begin{subfigure}{0.5\textwidth}
        \centering
        \includegraphics[width=0.927\linewidth]{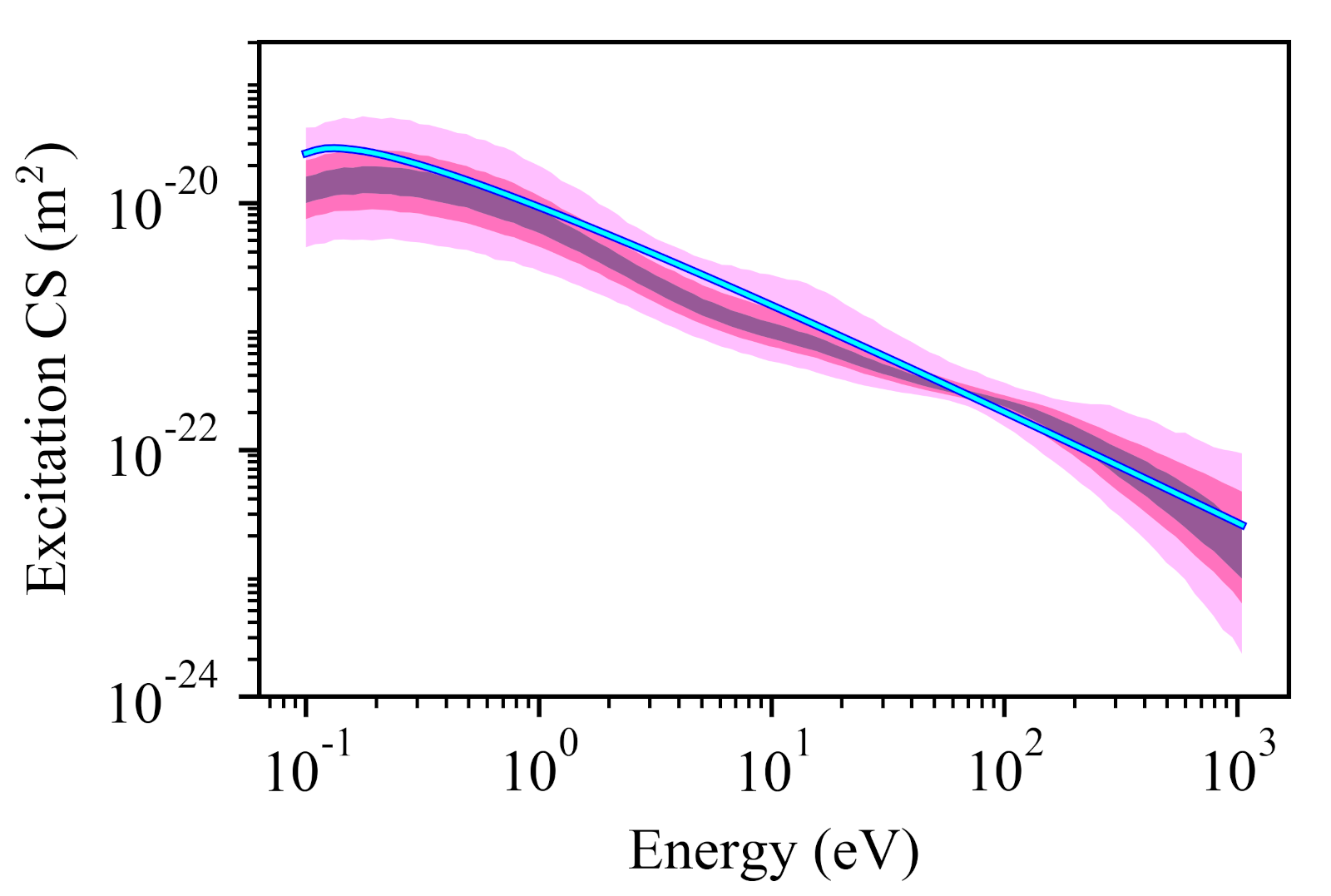}
        \caption{Sulfur hexafluoride (SF\textsubscript{6})}
        \label{excitation_uncertainty_SF6}
    \end{subfigure}
    \begin{subfigure}{0.5\textwidth}
        \centering
        \includegraphics[width=0.875\linewidth]{images/results/legend_uncertainty.png}
        \label{excitation_uncertainty_legend}
    \end{subfigure}
    \caption{Uncertainty in prediction of excitation cross sections of various gas species. ``Actual" curves shown here correspond to only a part (only the lowest energy process) of real excitation.}
    \label{results_excitation_uncertainty}
\end{figure}

Bayesian neural networks (BNNs)~\cite{mackay1992practical} predicts the complete probability distribution of the output variable and hence are most suited to determine the model uncertainty. Yet, BNNs are not frequently used due to its high computational cost. Thus, Monte Carlo Dropout~\cite{gal2016dropout} is generally used as an approximate Bayesian inference and we apply this to quantify the uncertainty in our inverse solution. It is implemented by first replicating the DenseNet architecture outlined previously. Subsequently, a dropout of $20\%$ is introduced in the dense layers. These neurons are disabled randomly during both the training and the testing phase. Therefore, every time an input value is passed to the model, different values are predicted which are sampled from some probabilistic distribution. We deduce this distribution by sampling a total of $10^4$ estimated cross sections, and the results are shown in Fig.~\ref{results_elastic_uncertainty}, \ref{results_ionization_uncertainty} and \ref{results_excitation_uncertainty}, which depict the confidence intervals in which the cross section value might lie. We observe a general trend for all gas species except Helium, that the model has a higher uncertainty in determining elastic MTCS at low energies ($0.1-0.8$ eV) compared to that at high energy values. Conversely, the model has higher uncertainty in predicting the ionization cross section at higher energies ($>4000$ eV). Additionally, we find that the model is absolutely certain about the predicted ionization threshold energy but is less certain in determining the peak value of the ionization cross section, even though it gives almost accurate results for both of these quantities. Further, the uncertainty in predicting the excitation cross sections is more compared to both elastic momentum transfer and ionization cross section and as suggested earlier, the lack of information content about the excitation cross sections in swarm data, could be one of the possible reasons for this higher uncertainty of excitation cross sections.

\section{Conclusion}
We have presented a data-driven approach, to obtain cross sections from the corresponding swarm data using different deep learning models which are trained upon the synthetic data generated from cross sections available on the LXCat. We have demonstrated the feasibility and the robustness of this deep learning based approach, by testing the trained networks to predict the elastic momentum transfer, ionization and excitation cross sections of various gas species, having diverse physical and chemical properties, and found the predicted cross sections to be consistent with the cross sections for elastic momentum transfer and ionization. Also, the swarm coefficients calculated using the predicted cross sections agrees reasonably well with those calculated using the cross sections sets for each species from LXCat (considering only the lowest energy excitation process). We have quantitatively analysed the performance of three different neural network architectures (ANN, CNN and DenseNet) in finding the solution to the inverse swarm problem and found that the Dense Convolutional Neural Network (DenseNet), due to its ability to effectively extract both long and short term trends from the swarm data, significantly outperformed artificial neural network used in previous works, as indicated by the ensemble of metrics used to access the accuracy of the architecture. In summary, we have tested our models on a wide range of gas species, used more performance metrics for statistical analysis and determined cross sections over a greater energy range compared to previous works based on ANNs. Finally, the uncertainty quantification of the model provides us a good estimate of the probability distribution of the cross sections from which all the physically plausible solutions of this inverse swarm problem can be sampled. 
Based on our results, we can conclusively say that CNN based models, particularly DenseNet, are better compared to ANN models in accurate determination of cross sections from swarm data. Interestingly, unlike ANNs, DenseNet could also predict characteristic peaks in specific energy ranges present in some gas species such as Nitrogen and Argon; these peaks are due to quantum mechanical effects and require domain expertise for such analysis. These significant improvements in prediction accuracy and pattern recognition while using DenseNet will provide the required confidence to the LTP community to accept such data driven approaches. However, additional work is needed before using actual swarm measurements (experimental) as input to such models. Many real gas species have multiple excitation cross sections and they all have an effect on the corresponding swarm coefficients but our proposed model is trained upon the swarm data which is computed using only a single excitation cross sections. Future works should address this issue.

The performance of deep learning models is highly dependent upon the training data fed to it. In this work, we have generated synthetic training data by interpolating the actual cross sections which have been categorized based on the characteristics of elastic momentum transfer cross sections. This approach is sufficient to provide the model with a large amount of data to train upon but clearly limits new trends in the synthetic data. Thus, we believe the performance of these neural networks would further improve if we actually use a sophisticated synthetic data generation scheme which can provide artificial cross sections which are physically-plausible, yet have unique trends of their own. One such possible approach is to use Generative Adversarial Networks (GANs), which is a machine learning framework used to extract complex features from a dataset and based on it, generate completely new data with random noise as input. Work on improving the quality of synthetic data with the use of GANs is currently underway.

\section*{Acknowledgment}
The authors would like to thank Dr. Leanne Pitchford, emeritus senior research scientist at LAPLACE Laboratory, CNRS, Toulouse, France, for discussions about the inverse problem in the context of swarm parameters, and for her valuable comments after careful reading of this manuscript.

\section*{References}
\bibliographystyle{ieeetr}
\bibliography{main.bib}

\end{document}